\documentclass[sigconf, authorversion]{acmart}
\usepackage{enumitem}
\usepackage{graphicx}  % another package that works for figures
\usepackage{booktabs} % For formal tables
\usepackage{cleveref} % for better references
\usepackage{caption,subcaption}
\usepackage{csquotes}% Recommended
\usepackage{siunitx}
\usepackage{tabularx,acmart-taps}
\usepackage{xcolor}
\usepackage{rotating}
\usepackage{lscape}
\usepackage{afterpage}
\usepackage{hyperref}
\hypersetup{breaklinks=true}
\usepackage{wrapfig}
\usepackage{float}

\usepackage{framed} % replacing \usepackage{mdframed}
\usepackage{fontawesome} % for the icon

\usepackage[ruled]{algorithm2e} % For algorithms

\SetAlFnt{\small}
\SetAlCapFnt{\small}
\SetAlCapNameFnt{\small}
\SetAlCapHSkip{0pt}
\IncMargin{-\parindent}

\usepackage{url}
\definecolor{lightgray}{gray}{0.9}
\definecolor{darkgray}{gray}{0.4}
\definecolor{darkred}{rgb}{0.6, 0, 0}

\newcommand{\bracketellipsis}{[\ldots]\xspace}

\renewenvironment{displayquote}
  {\list{}{\small\leftmargin=0.5em\rightmargin=0em\topsep=0.75ex}\item\relax\itshape\color{darkgray}}
  {\endlist}

\newcommand{\quoteby}[1]{%
    \\[-0.7ex]
    \begin{minipage}[t]{\linewidth}
        \hfill \textnormal{\faCommentsO}~\texttt{#1}
    \end{minipage}
}

\newcommand{\smallquote}[2]{
    {\color{darkgray}\textup{\texttt{#1}}~\faCommentsO~\textit{#2}}
}

\definecolor{lapislazuli}{rgb}{0.15, 0.38, 0.61}

\newcommand{\change}[1]{{\color{black}{#1}}}

\copyrightyear{2025}
\acmYear{2025}
\setcopyright{cc}
\setcctype{by-nc}
\acmConference[CHI '25]{CHI Conference on Human Factors in Computing Systems}{April 26-May 1, 2025}{Yokohama, Japan}
\acmBooktitle{CHI Conference on Human Factors in Computing Systems (CHI '25), April 26-May 1, 2025, Yokohama, Japan}\acmDOI{10.1145/3706598.3714193}
\acmISBN{979-8-4007-1394-1/25/04}

\AtBeginMaketitle{%
\let\oldAtBeginDocument\AtBeginDocument%
\renewcommand\AtBeginDocument[1]{#1}
\setcopyright{cc}
\setcctype{by-nc}
\copyrightyear{2025}
\acmYear{2025}
\acmDOI{10.1145/3706598.3714193}
\acmISBN{979-8-4007-1394-1/25/04}
\acmConference[CHI '25]{CHI Conference on Human Factors in Computing Systems}{April 26--May 01, 2025}{Yokohama, Japan}
\acmBooktitle{CHI Conference on Human Factors in Computing Systems (CHI '25), April 26--May 01, 2025, Yokohama, Japan}
\let\AtBeginDocument\oldAtBeginDocument%
}

\begin{document}

% \title[QV vs Likert]{``\textellipsis I can show what I really like.'': 
% Comparing Quadratic Voting with Likert Surveys at aligning respondents' preferences}

\title{Organize, Then Vote: Exploring Cognitive Load in Quadratic Survey Interfaces}


%% Author list
\author{Ti-Chung Cheng}
\orcid{0000-0001-7647-338X}
\affiliation{\institution{Computer Science \\ University of Illinois at Urbana-Champaign}
\city{Urbana}
\state{Illinois}
\country{USA}}
\email{tcheng10@illinois.edu}
\author{Yutong Zhang}
\authornotemark[1]
\orcid{0009-0007-6027-3122}
\affiliation{\institution{Computer Science \\ Stanford University}
\city{Stanford}
\state{California}
\country{USA}}
\email{yutongz7@stanford.edu}
\author{Yi-Hung Chou}
\authornote{Both authors contributed equally to this research.}
\orcid{0000-0001-8537-0200}
\affiliation{\institution{University of California, Irvine}
\city{Irvine}
\state{California}
\country{USA}}
\email{yihungc1@uci.edu}
\author{Vinay Koshy}
\orcid{0000-0002-1410-3911}
\affiliation{\institution{Computer Science \\ University of Illinois at Urbana Champaign}
\city{Urbana}
\state{Illinois}
\country{USA}}
\email{vkoshy2@illinois.edu}
\author{Tiffany Wenting Li}
\orcid{0000-0002-0954-5627}
\affiliation{\institution{Computer Science \\ University of Illinois at Urbana-Champaign}
\city{Urbana}
\state{Illinois}
\country{USA}}
\email{wenting7@illinois.edu}
\author{Karrie Karahalios}
\orcid{0000-0001-8788-3405}
\affiliation{\institution{Computer Science \\ University of Illinois at Urbana-Champaign}
\city{Urbana}
\state{Illinois}
\country{USA}}
\email{kkarahal@illinois.edu}
\author{Hari Sundaram}
\orcid{0000-0003-3315-6055}
\affiliation{\institution{Computer Science \\ University of Illinois}
\city{Urbana}
\state{Illinois}
\country{USA}}
\email{hs1@illinois.edu}

% %%
% %% By default, the full list of authors will be used in the page
% %% headers. Often, this list is too long, and will overlap
% %% other information printed in the page headers. This command allows
% %% the author to define a more concise list
% %% of authors' names for this purpose.
\renewcommand{\shortauthors}{Ti-Chung Cheng et al.}

\begin{abstract}
    Quadratic Surveys (QSs) elicit more accurate preferences than traditional methods like Likert-scale surveys. However, the cognitive load associated with QSs has hindered their adoption in digital surveys for collective decision-making. We introduce a two-phase ``organize-then-vote'' QS to reduce cognitive load. As interface design significantly impacts survey results and accuracy, our design scaffolds survey takers' decision-making while managing the cognitive load imposed by QS. In a 2x2 between-subject in-lab study on public resource allotment, we compared our interface with a traditional text interface across a QS with 6 (short) and 24 (long) options. Two-phase interface participants spent more time per option and exhibited shorter voting edit distances. We qualitatively observed shifts in cognitive effort from mechanical operations to constructing more comprehensive preferences. We conclude that this interface promoted deeper engagement, potentially reducing satisficing behaviors caused by cognitive overload in longer QSs. This research clarifies how human-centered design improves preference elicitation tools for collective decision-making.
\end{abstract}




%% Main Text
\maketitle
\section{Introduction}
% par 1: Introduction to the Problem
% Purpose: Define the problem and explain its significance.
%  What is the problem, what is the challenge, and why is it important

Designing intuitive survey interfaces is crucial for accurately capturing respondents' preferences, which directly impact the quality and reliability of the data collected. Recent Human-Computer Interaction (HCI) studies highlight how certain survey response formats can increase errors~\cite{pielotDidYouMisclick2024, kimComparingDataChatbot2019} and influence survey effectiveness~\cite{ugur2015evaluating}. In this paper, our goal is to introduce an effective interface for a~\textbf{Quadratic Survey (QS)}, a survey tool designed to elicit preferences more accurately than traditional methods~\cite{chengCanShowWhat2021}. Despite the promise of QSs, there has been no research on designing interfaces to support their unique quadratic mechanisms~\cite{grovesOptimalAllocationPublic1977}, where participants must rank and rate items --- a task that poses significant cognitive challenges. To popularize QSs and ensure high-quality data, this paper addresses the question: \textit{How can we design interfaces to support participants in completing Quadratic Surveys (QSs) more effectively?}

\begin{figure*}[p]
    \centering
    \includegraphics[width=0.96\textwidth]{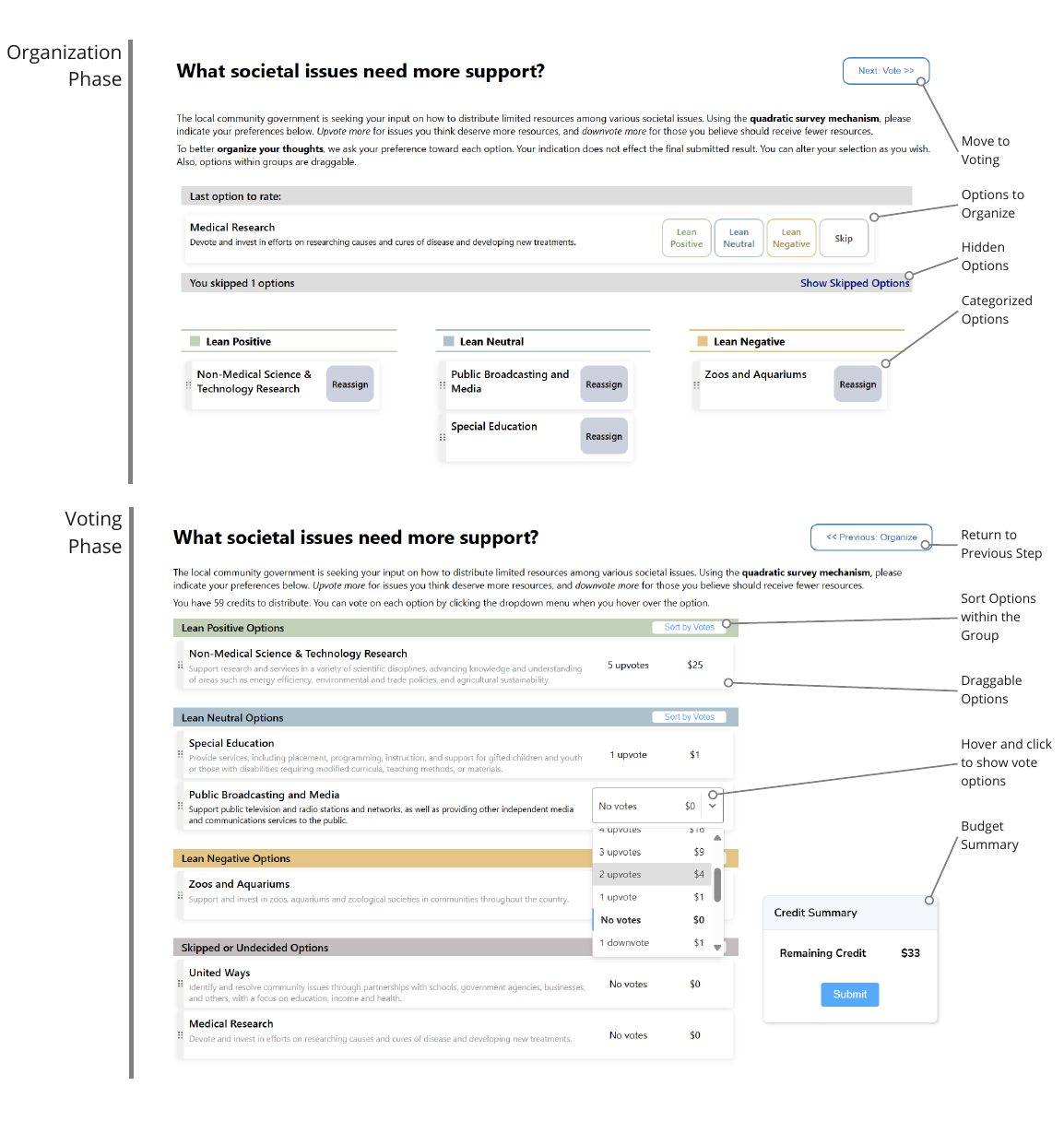}
    \caption{The Two-Phase Interface: The interface consists of two phases. Survey respondents can navigate between phases using the top right button. In the organization phase, the interface presents one option at a time to the respondents, and they chose one of four positional choices: ``Lean Positive'', ``Lean Neutral'', ``Lean Negative'', or ``Skip''. Skipped options are hidden and can be evaluated later. The chosen options then appear below. Items can be dragged and dropped across categories or returned to the stack. In the voting phase, options are listed in the order of the four categories. When hovering over each option, respondents can select a vote for that option using a dropdown menu. Each dropdown menu contains the cost associated with the vote. A sort button allows ascending sorting within each category. A summary box tracks the remaining credit balance.}
    \label{fig:interactiveInterface}
    \Description{
    This image shows two screen captures: the Organization Phase at the top and the Voting Phase at the bottom. The Organization Phase screen contains a question titled "What societal issues need more support?" with two sections. One section shows a block with descriptions of an option, and to the right of the block are four choices: "Lean Positive," "Lean Neutral," "Lean Negative," or "Skip." The interface also shows a skipped option. Below the block, three columns contain options inside, each showing the option title and a reassign button. In the Voting Phase, the same title and instructions are displayed, but now options are listed by their previously assigned categories (columns). The image shows the mouse hovering over one of the options, revealing a dropdown menu to allocate votes, along with the associated cost in credits. Each category box has a sorting button on the right, allowing users to reorder options within the category. Dots on the left side of the options indicate that drag-and-drop functionality is available for rearranging options. In the lower right corner, a summary box titled "Credit Summary" displays the remaining credit balance for voting. A button in the top left corner allows users to return to the previous Organization Phase.
    }

\end{figure*}

We envision an effective interface that navigates participants through the complex mechanism and preference construction process\change{, tailored to a QS.} A QS improves accuracy in individual preference elicitation compared to traditional methods like Likert scales by requiring participants to make trade-offs using a fixed budget of credits, where purchasing $k$ votes for an option in QS costs $k^2$ credits~\cite{quarfoot2017quadratic,chengCanShowWhat2021}. This quadratic cost structure forces respondents to carefully evaluate their preferences, balancing the strength of their support or opposition against the limited budget. However, the process of making these thoughtful trade-offs introduces challenges. As individual preferences are often constructed when presented with the options~\cite{lichtensteinConstructionPreference2006}, the act of weighing costs, evaluating options, and constructing rankings increases cognitive load. Moreover, QSs, often referred to as Quadratic Voting (QV) in voting scenarios, can involve hundreds of options~\cite{rogersColoradoTriedNew2019, teamTaiwanDigitalMinister}, increasing the risk of cognitive overload and the taking of mental shortcuts~\cite{simonBehavioralModelRational1955, payneAdaptiveStrategySelection1988, tverskyJudgmentsRepresentativeness}.

% ================================ %
% par 2: Approaches to Address the Challenges
% Purpose: Describe the existing approaches related to the problem.
% Key Questions:
%  - What are some broad approaches to addressing these challenges?
%  - Do not go into detail about related work but give an idea of the major themes in related work.

To date, existing quadratic mechanism-powered applications simply present options, allow vote adjustments and automatically calculate votes, costs, and budget usage. Such designs focused heavily on the mechanics operating the tool, rather than supporting possible challenges these application users faced. Survey interface literature, while addressing decision-making and usability, focuses on traditional surveys that do not share the unique option-to-option trade-offs that a QS introduces~\cite{engstrom2020politics, weijtersEffectRatingScale2010, kierujVariationsResponseStyle2010, toepoelSmileysStarsHearts2019, farzandAestheticsEvaluatingResponse2024, pielotDidYouMisclick2024}. Prior research in HCI and beyond explored techniques to manage cognitive load~\cite{paula2023, oviatt2006human, toepoelSmileysStarsHearts2019, softwareBrad2021, reis2012towards} and scaffold challenging tasks~\cite{task2014, moderate2021, ibiliEffectAugmentedReality2019, amyChatSensing2018} showing promise in supporting preference construction with a QS. Thus, this study aims to bridge this gap.

% ================================ %
% par 3: Your Proposal
% Purpose: Present your main ideas and proposed solution.
% Key Question:
%  - What are you proposing? Provide a sketch of the major ideas.

We propose a novel interactive two-phase ``organize-then-vote'' QS interface (referred to as the two-phase interface for short, Figure~\ref{fig:interactiveInterface}), which was developed through multiple iterations. It aims to facilitate preference construction and reduce cognitive load when making trade-offs through three key elements. First, the interface scaffolds the preference construction process by having participants initially categorize the survey options into ``Lean Positive,'' ``Lean Neutral,'' or ``Lean Negative.'' This serves as a cognitive warm-up, easing participants into the more complex QS voting task. Second, the interface arranges the options according to these categorizations, providing a structured visual layout. Third, participants can refine the positions of these options using drag-and-drop functionality, giving them greater control and agency in the preference-construction process. %These design features are aligned with preference construction theory and build upon prior research in interface design to reduce cognitive load and enhance user engagement.

To explore how these interface elements affect cognitive load and support preference construction in QSs, we pose the following research questions:
\begin{itemize}
    \item [\textbf{RQ1.}] How does the number of options in Quadratic Surveys impact respondents' cognitive load?
    \item [\textbf{RQ2a.}] How does the two-phase interface impact respondents' cognitive load compared to a single-phase text interface?
    \item [\textbf{RQ2b.}] What are the similarities and differences in sources of cognitive load across the two interfaces?
    \item [\textbf{RQ3.}] What are the differences in Quadratic Survey respondents' behaviors when coping with long lists of options across the two-phase interface and the single-phase text interface?
\end{itemize}

% ================================ %
% par 4: Main Findings
% Purpose: Summarize the key findings from your work.
% Key Question:
%  - What are the main findings?

We invited 41 participants to a lab study comparing our two-phase interface with a baseline to understand how different interface designs and option lengths (6 options or 24 options) impact cognitive load. 

Self-reported cognitive load using the NASA Task Load Index (NASA-TLX) and semi-structured interviews identified common challenges in QS, such as preference construction and budget management, while highlighting differences between text and two-phase interfaces. The two-phase interface fostered more strategic engagement with survey options, encouraging consideration of broader impacts in the long QS, reducing time pressure in the short QS, and eliciting greater affirmative satisfaction (e.g., "feeling good"). Quantitative results support these observations: participants in the two-phase interface—particularly in long surveys—traversed the list less frequently but maintained the same number of edits while spending more time per option. This suggests that reduced traversal did not diminish engagement. Together, these findings highlight the organizing phase's role in fostering deeper engagement with survey options.

% ================================ %
% par 5: Main Contributions
% Purpose: Identify and explain the primary contributions of your work.
% Key Structure:
%  1. Line 1: Identify your contribution—conceptualization, framework, interface, algorithm, etc.
%  2. Line 2: Contrast your contribution with prior work.
%  3. Line 3: Explain how you accomplished your contribution.
%  4. Line 4: Emphasize the impact of the contribution—why should anyone care?

\paragraph{Contributions}
We contribute to the body of knowledge in the HCI community by proposing the first interface specifically designed for QS and QV-like applications, which aims to reducing cognitive challenges and scaffolding preference construction through a two-phase interface with direct manipulation. Before our work, no research had explored QS interfaces. This is particularly important for long QSs, which are prone to cognitive overload. Few HCI studies have addressed interfaces for surveys and questionnaires. Our study demonstrates how user interfaces can facilitate preference construction in situ and promote deeper engagement with survey options through interface elements. Additionally, this paper offers the first in-depth qualitative analysis of user experiences with Quadratic Mechanism applications, identifying usability challenges and key factors contributing to cognitive load. The impact of our contribution extends beyond QSs, offering design implications for other preference-elicitation tools used in multi-option scenarios. By making QSs easier to use and more accurate, our design encourages wider adoption among researchers and practitioners. Finally, our work lays the groundwork for future Quadratic Mechanism interface design to facilitate individuals expressing their preferences.
\section{Related Work}
\label{sec:relatedWorks}
This research lies at the intersection of three core areas:~\change{quadratic surveys, existing QV interfaces and choice overload along with cognitive challenges.} In this section, we review the related works in each of these areas.

\begin{figure*}[!t]
    \centering
    \includegraphics[width=0.85\textwidth]{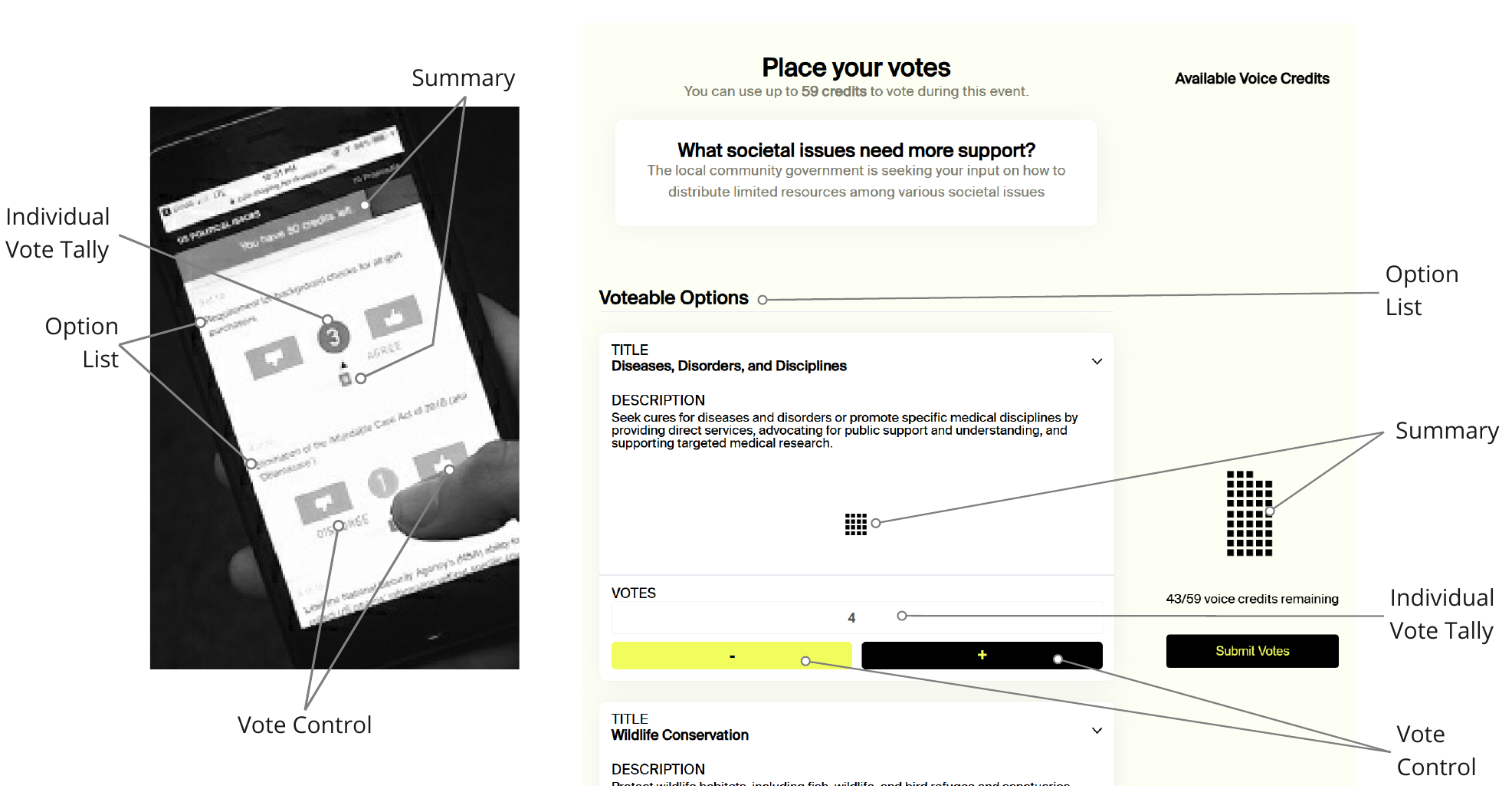}
    \caption{A selection of two QV interfaces. The interface on the left was used in the first empirical QV research~\cite{quarfoot2017quadratic}. Little information is available about the software, except for an image from~\citet{posner2018radical}. The interface on the right is an open-sourced QV interface~\cite{RadicalxChangeQuadraticvoting2024} forked from GitCoin~\cite{gitcoinReadWhitepaperGitcoin}, used by the RadicalxChange community~\cite{radicalxchange}. Both interfaces share the common elements with different visual representations.}
    \Description{There are two smaller figures in this figure. On the left is a black and white image of a mobile phone displaying a voting interface from software by WeDesign, used in empirical QV research. A hand holds the phone, and the screen shows a prompt related to background checks for gun purchases. There are thumbs-up and thumbs-down icons labeled "Agree" and "Disagree," with numbers indicating the current number of votes (e.g., 3 votes for "Agree"). A remaining vote budget is displayed at the top as a progress bar, indicating "50 choices left." The user is interacting with the interface, selecting either agree or disagree on the prompts. On the right is A screenshot of a QV interface designed for voting on societal issues that need more support. The screen displays two options: "Diseases, Disorders, and Disciplines" and "Wildlife Conservation," with a brief description under each. Users can adjust votes with plus and minus buttons, and the current vote count (e.g., 3 votes for Diseases, -2 votes for Wildlife Conservation) is displayed. The total available credits are shown on the right side as a grid of small blocks, with 46 out of 59 credits remaining. There is also a "Submit Votes" button. A menu on the left allows users to jump to different societal issues. Both images are annotated to locate the elements described in the main text.}
    \label{fig:rcx_interface_annotated}
\end{figure*}

\subsection{Quadratic Survey and the Quadratic Mechanism}
We introduce the term \textbf{Quadratic Survey (QS)} to describe surveys that utilize the quadratic mechanism to collect individual attitudes. The~\textbf{quadratic mechanism} is a theoretical framework designed to encourage the truthful revelation of individual preferences through a quadratic cost function~\cite{grovesOptimalAllocationPublic1977}. This framework gained popularity through~\textbf{Quadratic Voting (QV)}, also known as plural voting, which uses a quadratic cost function in a voting framework to facilitate collective decision-making~\cite{lalley2016quadratic}.

To illustrate how QS works, we formally define the mechanism: each survey respondent is allocated a fixed budget, denoted by $B$, to distribute among various options. Participants can cast $n$ votes for or against option~$k$. The cost~$c_k$ for each option $k$ is derived as:
\begin{equation*}
c_k = n_k^2 \quad \text{where}\quad n_k \in \mathbb{Z}
\end{equation*}
The cost of all votes must not exceed the participant's budget:
\begin{equation*}
\sum_k c_k \leq B
\end{equation*}
Survey results are determined by summing votes for each option:
\begin{equation*}
\text{Total Votes for Option } k = \sum_{i=1}^{S} n_{i,k}    
\end{equation*}
where $S$ represents the total number of participants, and~$n_{i,k}$ is the number of votes cast by participant~$i$ for option~$k$. Each additional vote for each option increases the marginal cost linearly, encouraging participants to vote proportionally to their level of concern for an issue~\cite{posner2018radical}.

QS adapts these strengths of the quadratic mechanism in \textit{voting} to encourage truthful expression of preferences in \textit{surveys}. Unlike traditional surveys that elicit either rankings~\textit{or} ratings, QS allows for~\textit{both}, enabling participants to cast multiple votes for or against options, incurring a quadratic cost.~\citet{chengCanShowWhat2021} showed that this mechanism aligns individual preferences with behaviors more accurately than Likert Scale surveys, particularly in resource-constrained scenarios like prioritizing user feedback on user experiences.

In recent years, empirical studies on QV have expanded into various domains~\cite{naylor2017first, cavaille2024cares}. Applications based on the quadratic mechanism have also grown, including Quadratic Funding, which redistributes funds based on outcomes from consensus made using the quadratic mechanism~\cite{buterinFlexibleDesignFunding2019a, freitasQuadraticFundingIncomplete2024}. Recent work by \citet{southPluralManagement2024} applies the quadratic mechanism to networked authority management, later used in Gov4git~\cite{Gov4gitDecentralizedPlatform2023}. Despite the increasing breadth and depth of applications utilizing the quadratic mechanism, little attention has been paid to user experience and interface design, which support individuals' preference intensity elicitation. Our work aims to address this by designing interfaces for quadratic mechanisms.

\subsection{Existing QV Interfaces}
\label{sec:related_qv}

Since QS shares QV's underlying mechanism, we used snowball sampling to identify publicly available QV applications mentioned in news and academic sources. Currently, no widely adopted QV interface is tied to a single vendor or platform.~\Cref{fig:rcx_interface_annotated} shows two variations of existing QV interfaces, with both employing a single-step approach with different visual representations of common elements~\cite{Gov4gitDecentralizedPlatform2023, yehjxraymondYehjxraymondQvapp2024, chengCanShowWhat2021, cavaille2024cares}. All QV interfaces generally include:

\begin{itemize}[leftmargin=*]
    \item Option list: A list of options for voting.
    \item Vote controls: Buttons to add or remove votes for each option.
    \item Individual vote tally: A display of the votes cast per option.
    \item Summary: A summary of costs and the remaining budget.
\end{itemize}

These components let users operate QV mechanically, providing little understanding of voters' usability needs nor offering cognitive support. In addition, HCI research on survey interfaces is limited~\cite{nobarany2012design, van2007design} with most efforts focusing on alternative input modalities like bots, voice interfaces, or virtual reality~\cite{voiceWei2022, khullar2021, kimComparingDataChatbot2019, feick2020virtual}.

\subsection{Cognitive Challenges and Choice Overload}
The challenge of respondents making difficult decisions using quadratic mechanisms remains unexplored in the literature.~\citet{lichtensteinConstructionPreference2006} identified three key elements that make decisions difficult. \change{These elements include making decisions in unfamiliar contexts, quantifying the value of one's opinions, and being forced to make trade-offs due to conflicting choices. QS fits at least two of the three elements: participants may encounter a selection of unfamiliar options by the survey designer; they are asked to quantify the difference between option preferences through a numerical vote; and the budget constraint enforces trade-offs under a non-linear function, which means that a vote decrease for one option is not necessary equivalent to an increase for another, making iterative adjustment and evaluating tradeoffs difficult. Thus, we believe QS introduces a high cognitive load.}

Cognitive load refers to the demands placed on a user's working memory during the interaction process, which significantly influences the usability of the system~\cite{cooper1998research, seppCognitiveLoadTheory2019}. Cognitive overload can adversely affect performance~\cite{drommi2001interface}, leading individuals to rely on heuristics rather than deliberate, logical decision-making~\cite{daniel2017thinking}. When presented with excessive information, such as too many options, individuals 'satisfice', settling for a 'good enough' solution rather than an optimal one~\cite{simonBehavioralModelRational1955, payneAdaptiveStrategySelection1988, tverskyJudgmentsRepresentativeness}. Subsequently, too many options can overwhelm individuals, resulting in decision paralysis, demotivation, and dissatisfaction~\cite{iyengarWhenChoiceDemotivating2000}.

Additionally,~\citet{alwinMeasurementValuesSurveys1985} highlighted that the use of ranking techniques in surveys can be time-consuming and potentially more costly to administer. These challenges are compounded when ranking numerous items, requiring substantial cognitive sophistication and concentration from survey respondents \cite{featherMeasurementValuesEffects1973}.

Notable applications of QV include the 2019 Colorado House, which considered 107 bills~\cite{coyNewWayVoting2019}, and the 2019 Taiwan Presidential Hackathon, which featured 136 proposals~\cite{QuadraticVotingFrontend2022}; both used a single QV question with hundreds of options. These empirical applications of QV suggest the importance of understanding QS with many options' impact on cognitive load and support developing interfaces for practical uses.
\section{Quadratic Survey Interface Design}
\label{sec:interfaceDesign}
This section presents our QS interface. \change{Drawing on existing QV interfaces described in Section~\ref{sec:related_qv} and prior literature, we iterated through paper prototypes and three design pre-tests, detailed in Appendix~\ref{apdx:design}. Initially,} participants struggled to~\textit{rank} relative preferences among options and~\textit{rate} the degree of trade-offs between them. In this study, we focus on addressing the former challenge, which pertains to preference construction.

\subsection{`Organize-then-Vote': The Two-Phase Interface}
\label{sec:finalInterfaceDesign}

\subsubsection{Justifying a two-phase approach}
The main objective of the two-phase interface is to facilitate preference construction and reduce cognitive load. As shown in Figure~\ref{fig:interactiveInterface}, the interface consists of two steps: an organization phase and a voting phase. In both phases, survey respondents can drag and drop options across the presented list.

\paragraph{A two-phase approach}
Preferences are shaped through a series of decision-making processes~\cite{lichtensteinConstructionPreference2006}. Two decision-making theories~\change{inspired this} two-step interaction interface design:~\citet{montgomeryDecisionRulesSearch1983}'s Search for a Dominance Structure Theory (Dominance Theory) and~\citet{svensonDifferentiationConsolidationTheory1992}'s Differentiation and Consolidation Theory (Diff-Con Theory). The former suggested that decision-makers prioritize creating dominant choices to minimize cognitive effort by focusing on evidently superior options~\cite{montgomeryDecisionRulesSearch1983}. The latter described a two-phase process where decisions are formed by initially~\textit{differentiating} among alternatives and then~\textit{consolidating} these distinctions to form a stable preference~\cite{svensonDifferentiationConsolidationTheory1992}. Pre-tests showed participants puzzled by ranking all options before voting. These theories suggest decisions emerge by eliminating choices, not by fully ranking them. Therefore, the organize-then-vote design makes this natural process more explicit. Phase one focused on differentiating and identifying dominant options, enabling survey respondents to preliminarily categorize and prioritize their choices. Phase two presented these categorized options in a comparable manner, with drag-and-drop functionality, enhancing one's ability to consolidate preferences. This structured approach aimed to construct a clear decision-making procedure that reduced cognitive load and enhanced clarity and confidence in the decisions made.

\paragraph{Phase 1: Organization Phase}
The goal of the organization phase was to support participants in identifying clearly superior options or partitioning choices into distinguishable groups. In this section, we first describe how the interaction works, then we detail the reasons for the implemented design decisions.

The organizing interface, depicted on the top half of Figure~\ref{fig:interactiveInterface}, sequentially presents each survey option. Participants select a response among three ordinal categories -- ``Lean Positive'', ``Lean Negative'', or ``Lean Neutral''. Once selected, the system moves that option to the respective category. Participants can skip the option if they do not want to indicate a preference. Options within the groups are draggable and rearrangeable to other groups should the participants wish.

To support preference formation, respondents are shown one option at a time, allowing them to either recall a prior judgment or construct a new one based on the presented choices~\cite{strackThinkingJudgingCommunicating1987}. Limiting the information presented this way also helps reduce cognitive load by preventing overload from too many options~\cite{swellerCognitiveLoadTheory2011}. This incremental process ensures that participants form opinions on individual options.

The three possible options --- Lean Positive, Lean Neutral, and Lean Negative --- aim to scaffold participants in constructing their own choice architecture~\cite{munscherReviewTaxonomyChoice2016, thalerNudgeImprovingDecisions2008a}, which strategically segments options into diverse and alternative choice presentations while avoiding biases from defaults. We believed that these three categories were sufficient for participants to segment the options. We do not limit the number of options one can place in each category to prioritize user agency, allowing participants full control over how they organize their preferences~\cite{norman2013design}. Immediate feedback displays the placement of options and allows participants to rearrange them via drag-and-drop, adhering to key interface design principles~\cite{norman2013design}. It also allows finer-grain control for individuals to surface dominating options and create differentiating groups of options.

\paragraph{Phase 2: Interactive Voting Phase}

The objective of the voting phase is to facilitate the consolidation of differentiated options through interactive elements while reinforcing the differentiation across options constructed by participants in the previous phase. This facilitation is achieved by retaining the drag-and-drop functionality for direct manipulation of position and enabling sorting within each category.

Options are displayed as they are categorized within each category from the previous step and in the following section --- Lean Positive, Lean Neutral, Lean Negative, and Skipped or Undecided --- as detailed on the bottom half of Figure~\ref{fig:interactiveInterface}. The Skipped or Undecided category contains options left in the organization queue, possibly because survey respondents have a pre-existing preference or chose not to organize their thoughts further. The original order within these categories is preserved to maintain and reinforce the differentiated options. This ordering sequence mitigated early prototype concerns where uncategorized options were left at the top of the voting interface confusing survey respondents. Respondents have the flexibility to return to the organization interface at any point during the survey to revise their choices.

In the voting interface, options are draggable, allowing participants to modify or reinforce their preference decisions as needed. Each category features a sort-by-vote function for reordering within the group, which, although it doesn't affect the final outcome, supports information organization and consolidation. Both features aim to group similar options automatically and emphasize proximity, reducing cognitive load by following the proximity compatibility principle to enhance decision-making~\cite{wickens1990proximity}.

While multiple interaction mechanisms exist, drag-and-drop has been extensively explored in rank-based surveys. For instance,~\citet{krosnick2018measurement} demonstrated that replacing drag-and-drop with traditional number-filling rank-based questions improved participants' satisfaction with little trade-off in their time. Similarly,~\citet{timbrook2013comparison} found that integrating drag-and-drop into the ranking process, despite potentially reducing outcome stability, was justified by the increased satisfaction and ease of use reported by respondents. The trade-off was deemed worthwhile as QSs did not use the final position of options as part of the outcome if it significantly enhanced user satisfaction and usability~\cite{rintoulVisualAnimatedResponse}. Together, these design decisions led to our belief that a two-phase interface with direct interface manipulation could reduce the cognitive load for survey respondents to form preference decisions when completing QSs.

\begin{figure}[ht!]
    \centering
    \includegraphics[width=0.45\textwidth]{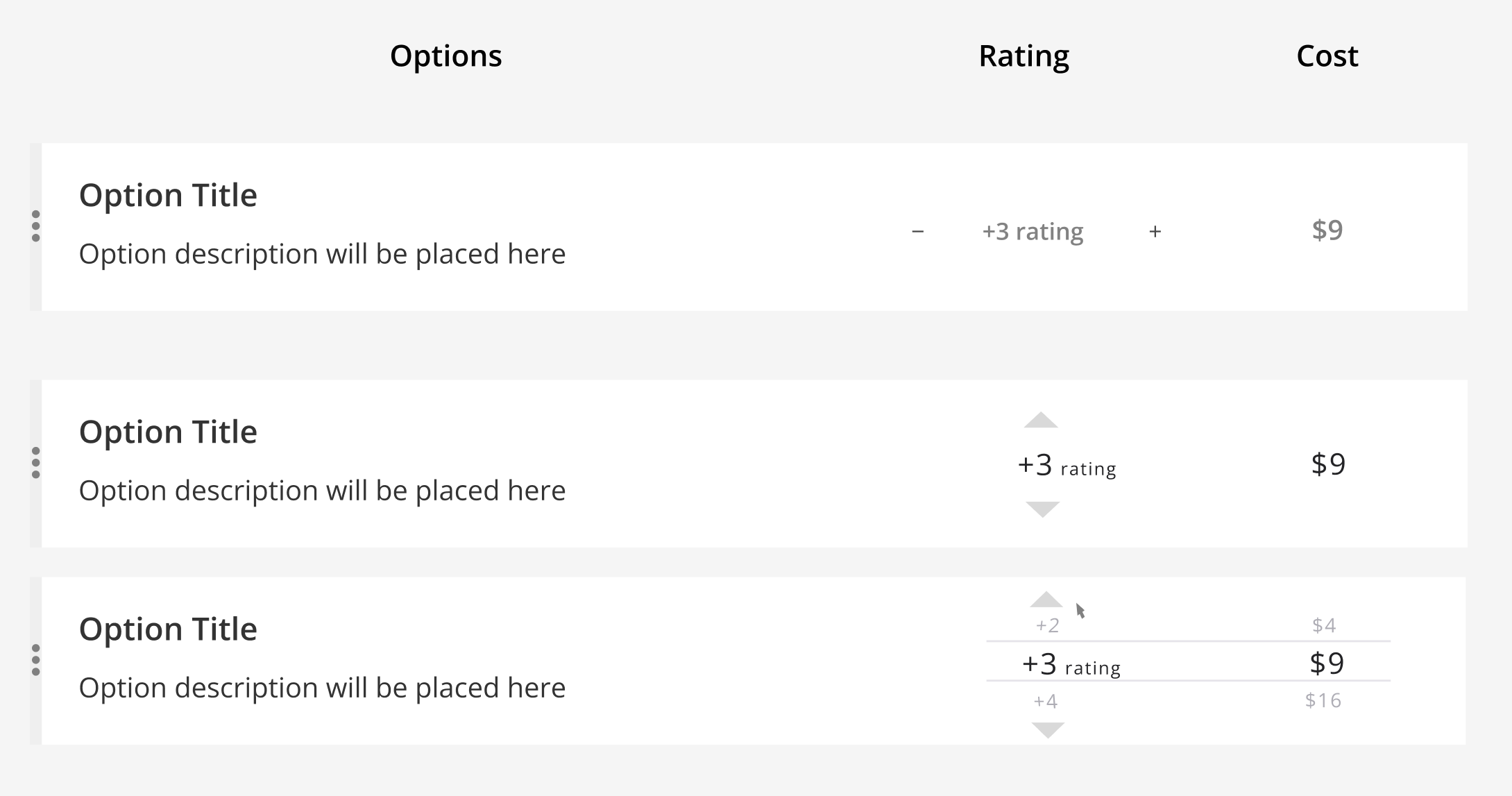}
    \caption{Alternative vote control. The click-based design (upper) mirrors traditional vote control used in other QV interfaces, where each click controls one vote. The wheel-based design (the latter two) allows control through both clicks and mouse wheel rotation.}
    \Description{Three voting control interfaces are displayed. Each row represents a different interface. The first row shows a traditional click-based voting interface with options to decrease, increase, or maintain a rating of +3. The second and third rows show a wheel-based voting interface with mouse wheel functionality. In these, the middle row indicates a current rating of +3, with +2 and +4 ratings also visible. The cost for each option is listed on the right, ranging from 4 to 16. The last row mirrors the previous one with a rating of +3 and a cost of 9.}

    \label{fig:btn_design}
\end{figure}

In addition, we made three aesthetic design decisions~\change{considering existing QV-based interfaces}. First, we removed visual elements like icons, emojis, progress bars, and vote visualizations, as prior research indicated that emojis could influence survey interpretations and reduce user satisfaction~\cite{herringGenderAgeInfluences2020, toepoelSmileysStarsHearts2019}. While effective visualizations can aid decision-making, this study does not aim to address that question. Second, all options are visible on the screen simultaneously.~\change{Prior research recommends placing all items on the voting screen to prevent overlooked votes~\cite{centerforcivicdesignCenterCivicDesign}.} This echoes the proverb ``out of sight, out of mind,'' reducing where individuals might be biased toward visible options, and additional effort is required for individuals to retrieve specific information if options are hidden. Last, use a dropdown positioned to the right of each survey option for ease of access to the budget summary when determining the votes. The layout of the votes and cost was inspired by online shopping cart checkout interfaces where quantities are supplied next to the itemized costs followed by the total checkout amount.~\change{Figure~\ref{fig:btn_design} shows the two alternatives—click-based buttons (participants disliked multiple clicks) and a wheel-based design (unfamiliar to some)—and settled on the dropdown.}

\subsection{Baseline Interface: Single-Phase Text Interface}

\begin{figure*}[!t]
    \centering
    \includegraphics[width=\textwidth]{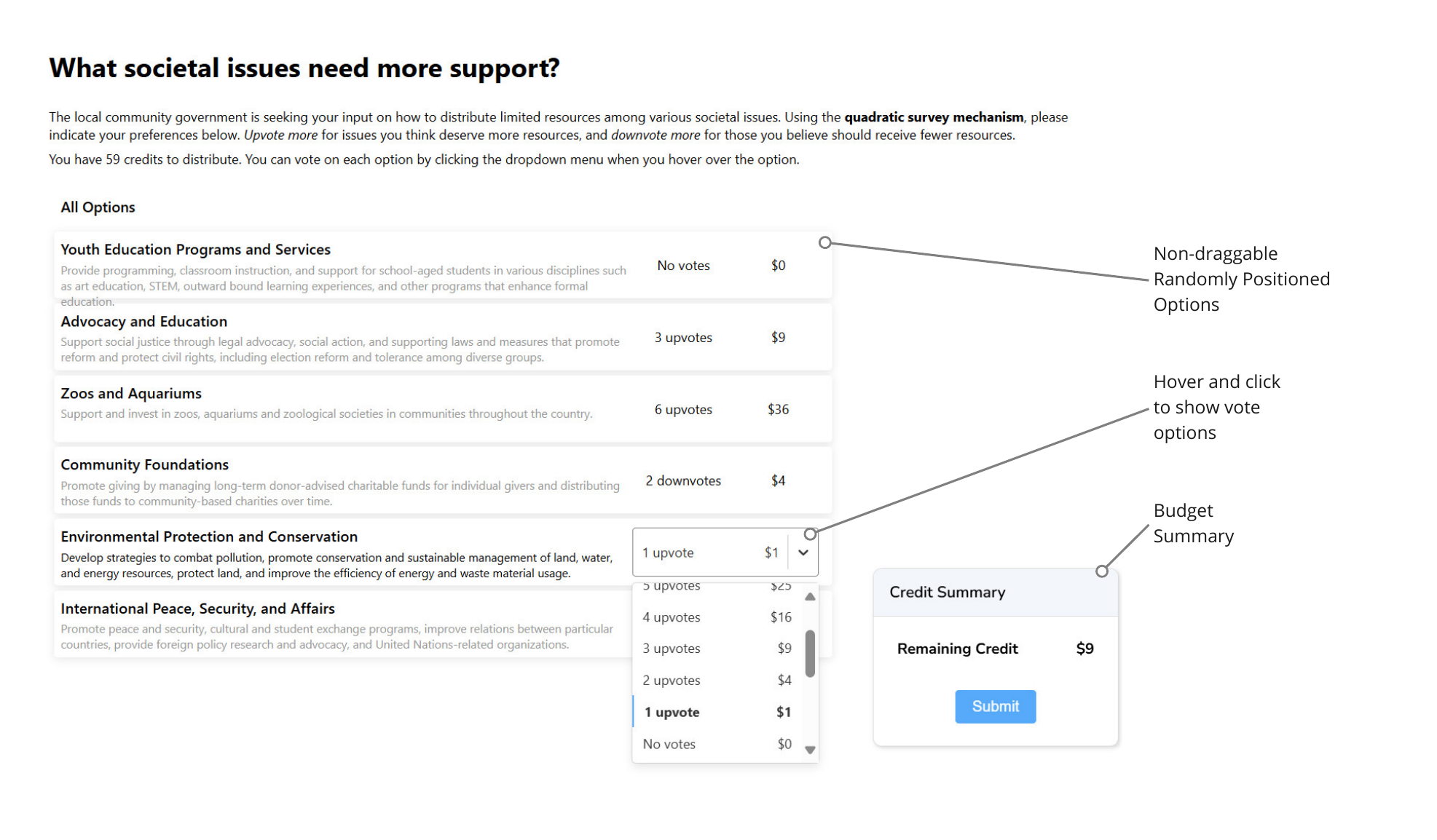}
    \caption{The text-based interface: This interface is based on the two-phase version but does not include the organization phase and lacks the drag-and-drop functionality. Options are randomly positioned.}
    \Description{An image of a voting interface asking users to select societal issues needing support. The title reads, "What societal issues need more support?" with a brief explanatory paragraph underneath. Below, a list of six options is displayed, including "Youth Education Programs and Services," "Advocacy and Education," "Zoos and Aquariums," "Community Foundations," "Environmental Protection and Conservation," and "International Peace, Security, and Affairs." Each option has a description, a current vote count, and a dollar amount. The right side of the image shows an expanded dropdown menu for one of the options with selectable voting choices, such as "1 upvote" and "2 upvotes." A separate box labeled "Credit Summary" shows the remaining credit of 9 and a "Submit" button below it.}
    \label{fig:textInterface}
\end{figure*}

We created a single-phase text interface (referred to text interface for short, Figure~\ref{fig:textInterface}) as a control, enabling us to see how organizational features affect cognitive load and behavior. Like existing interfaces, it uses static lists, a summary box, and a vote control. To ensure a fair comparison, we applied the same design principles: no extraneous visuals, all options on one screen, and dropdown-based voting. The prompt appears at the top, followed by a randomly ordered list to prevent ordering bias~\cite{ferberOrderBiasMail1952, couperWebSurveyDesign2001}. Costs and the credits summary appear on the right.

Both experimental interfaces were developed with a ReactJS frontend and a NextJS backend powered by MongoDB. We open-source both interfaces.\footnote{https://github.com/CrowdDynamicsLab/Quadratic-Survey-Frontend}
\begin{figure*}[ht!]
    \centering
    % Top figure (full-width)
    \begin{subfigure}{0.84\textwidth} % Full width for the first subfigure
        \centering
        \includegraphics[width=\textwidth, trim=0 13 0 13, clip]{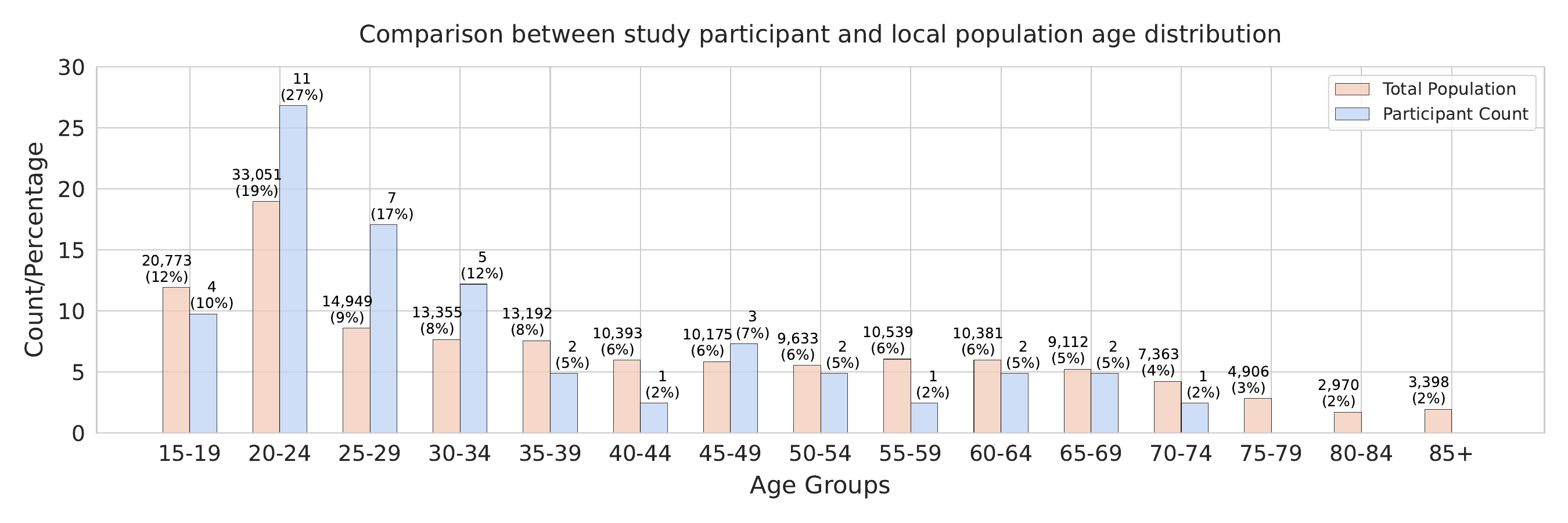}
        \caption{Age distribution of the study participants were similar to the locale's demographic profile.}
        \Description{A bar chart comparing the age distribution between study participants and the local population. The x-axis represents age groups from 15-19 to 85+, and the y-axis shows count/percentage values ranging from 0 to 30. Each age group has two bars: one for the total population (in peach) and one for participant count (in light blue). Some key differences are visible, such as the 20-24 age group, where the total population is 33,051 (19\%) and participants are 11 (27\%). The 25-29 group shows 14,949 (9\%) for the population and 7 (17\%) for participants. Percentage and count data are displayed above each bar. Other age groups had similar bars. The chart title reads "Comparison between study participant and local population age distribution," and the legend distinguishes the two categories (Total Population and Participant Count).}
        \label{fig:demoAge}
    \end{subfigure}
    
    \vspace{0.25cm} % Add some vertical space between the subfigures

    % Bottom figures (two subfigures side by side)
    \begin{subfigure}{0.44\textwidth} % First subfigure on the second row
        \centering
        \includegraphics[width=\textwidth]{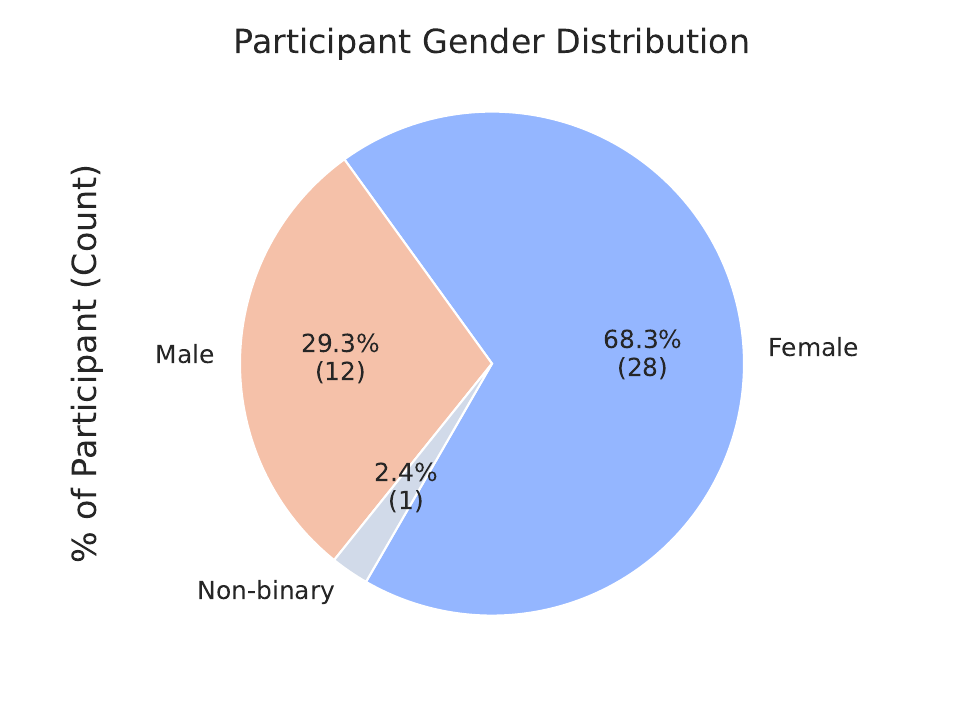}
        \caption{Gender distribution of our participants skewed towards female participants.}
        \Description{A pie chart displaying the gender distribution of study participants. The chart is divided into three sections: 68.3\% (28 participants) are labeled as female and shown in blue, 29.3\% (12 participants) are labeled as male and shown in peach, and 2.4\% (1 participant) are labeled as non-binary, represented by a small gray slice. The title reads "Participant Gender Distribution," and the y-axis on the left is labeled "\% of Participant (Count)."}
        \label{fig:demoGender}
    \end{subfigure}
    \hfill
    \begin{subfigure}{0.44\textwidth} % Second subfigure on the second row
        \centering
        \includegraphics[width=\textwidth]{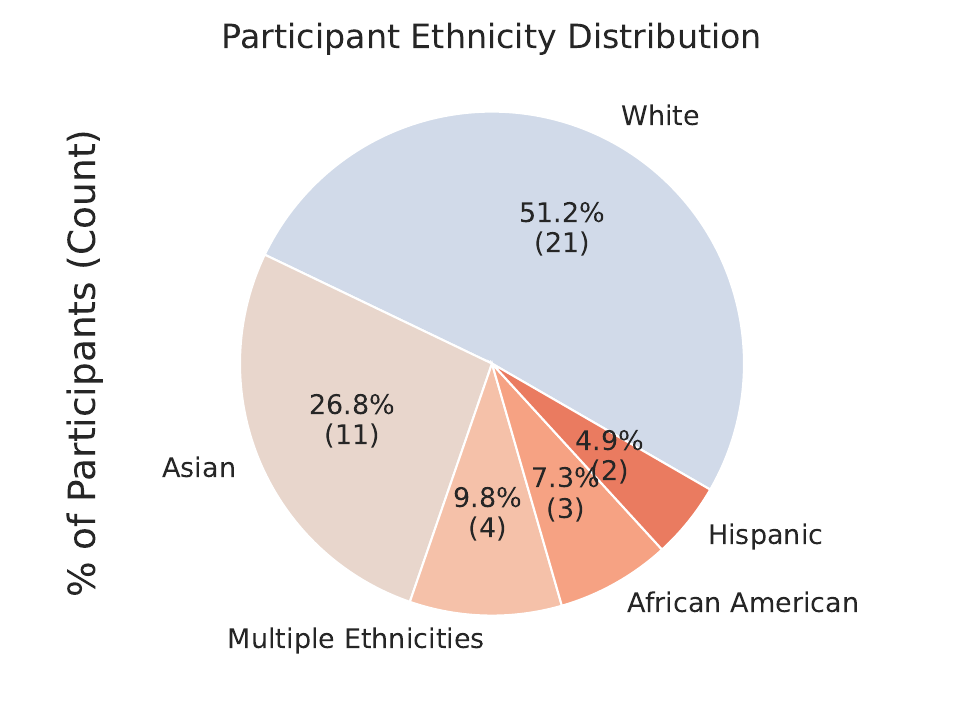}
        \caption{Ethnicity distribution remains diverse with fewer Hispanic and African American participants.}
        \Description{A pie chart showing the ethnicity distribution of participants. The largest segment, representing 51.2\% (21 participants), is labeled White and shown in light blue. Other segments include 26.8\% (11 participants) for Asian, 9.8\% (4 participants) for Multiple Ethnicities, 7.3\% (3 participants) for African American, and 4.9\% (2 participants) for Hispanic. The percentages and counts are displayed within each section. The y-axis on the left is labeled "\% of Participants (Count)," and the title reads "Participant Ethnicity Distribution."}
        \label{fig:demoEthnicity}
    \end{subfigure}

    \caption{Demographic distributions: Age, Gender, and Ethnicity}
    \Description{A collection of three demographic graphs demonstrating age, gender, and ethnicity distribution.}
    \label{fig:Demographics}
\end{figure*}

\section{Experiment Design}
\label{sec:experiment}
In this section, we describe our experiment design. The study was approved by the university's Institutional Review Board (IRB).

\subsection{Recruitment and Participants}
We recruited 41 participants from a United States college town using online ads, digital bulletins, social media posts, email newsletters, and physical flyers in public spaces beyond campus. We described the study as exploring societal attitudes to reduce response bias. One participant was excluded due to data quality concerns\footnote{The participant reported not completing the survey seriously, as they believed the experiment was fake.}.

To ensure diversity, we prioritized non-students by selectively accepting them and monitoring demographic distribution. The mean participant age was 34.63 years, with an age distribution similar to the county's demographic profile (Figure~\ref{fig:demoAge}), although there was a slightly higher representation of younger adults. Gender and race demographics are presented in Figures~\ref{fig:demoGender} and~\ref{fig:demoEthnicity}. Demographic differences between groups were reasonably balanced, although participants using the short text interface skewed slightly younger ($\mu$=32.1), and those in the long two-phase interface group had a broader age range ($\mu$=38.8, $\sigma$=19.6). Appendix \ref{sec:apdx:demo} contains full details.

\subsection{Experiment Design}
We implemented a between-subject design to avoid learning effects and minimize participants' fatigue from potential complexity of QSs. The experiment focused on public resource allotment, following the methodology of~\citet{chengCanShowWhat2021}, in which participants expressed preferences across societal issues. These issues are relevant to all citizens and effectively highlight the need to prioritize limited public resources. Participants received a survey with options randomly drawn from the 31 societal topics evaluated by Charity Navigator~\cite{charitynavigatorCharityNavigator2023}, an organization that assesses over 20,000 charities in the United States (see Appendix~\ref{sec:charityList} for the full list).
Randomly selecting the options each participant saw helped control for potential systematic content biases that specific voting options might introduce across surveys of different lengths. Participants were randomly assigned to one of four groups, each with 10 participants:
\begin{itemize}[leftmargin=*]
    \item Short Text (ST): A text interface with 6 options.
    \item Short Two-Phase (S2P): A two-phase interface 6 options.
    \item Long Text (LT): A text-based interface 24 options.
    \item Long Two-Phase (L2P): A two-phase interface with 24 options.
\end{itemize}

Prior research informed the choice of 6 and 24 options, representing short and long lists. These studies recommend fewer than 10 options for constant-sum surveys~\cite{moroneyQuestionnaireDesignHow2019} and fewer than 7 for the Analytic Hierarchy Process~\cite{saatyPrinciplesAnalyticHierarchy1987}. Classic cognitive load research~\cite{millerMagicalNumberSeven1956, saaty2003magic} suggests the use of 7$\pm$2 items. A meta-analysis by~\citet{chernevChoiceOverloadConceptual2015} identified 6 and 24 as common values for short and long lists in choice overload studies, which are rooted in the original choice overload experiment by~\citet{iyengarWhenChoiceDemotivating2000}.

\subsection{Experiment Procedure}
\begin{figure*}[ht!]
    \centering
    \includegraphics[width=\textwidth, trim=13 13 13 13, clip]{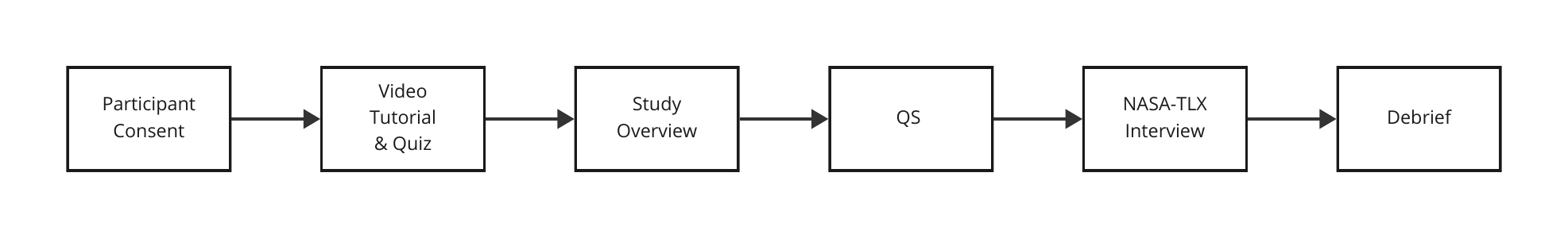}
    \caption{Study protocol: Participants are asked to learn about the mechanism of QSs after consenting to the study. The researcher explained the study overview and asked participants to complete the QS. A NASA-TLX survey followed by interviews to understand participants' cognitive load. We debriefed participants after the study.}
    \Description{A flowchart depicting the study protocol, consisting of six stages. Each stage is represented by a rectangular box connected by right-facing arrows. The boxes, from left to right, are labeled: "Participant Consent," "Video Tutorial \& Quiz," "Study Overview," "QS," "NASA-TLX Interview," and "Debrief." The arrows between the boxes indicate the sequence of the study process.}

    \label{fig:studyProtocol}
\end{figure*}

Participant's spent on average 40 minutes (range:~27$-$68,~$\sigma$=9) in the lab. Figure~\ref{fig:studyProtocol} visually represents the study protocol detailed in the following subsections.

\subsubsection{Consent, Instructions, and Quiz}
Participants were invited to the lab to control for external influences and used a 32-inch vertical monitor to display all options. After consenting, participants watched a video explaining the quadratic mechanism without any mention of the interface's operation, followed by a quiz to ensure understanding. Participants rewatched the video or consulted the researcher until they successfully selected the correct answers. Each participant's screen was captured throughout the study.

\subsubsection{Quadratic Survey}
The researcher informed participants that the study aimed to help local community organizers understand preferences on societal issues to improve resource allocation. Aware that their screens were being recorded, participants completed the survey independently inside a semi-enclosed space in the lab, without the researcher's presence. Once they completed the survey, participants notified the researcher.

\subsubsection{NASA-TLX Survey and Interview}
The researcher joins study participant and administer a paper-based weighted NASA Task Load Index (NASA-TLX), followed by a semi-structured interview after being informed that the researcher would begin audio recording with their laptop. We adopted the paper-based weighted NASA-TLX, a widely used multidimensional tool that averages six subscale scores to measure overall workload after task completion~\cite{hart1988development, hartNasaTaskLoadIndex2006, cain2007review}. NASA-TLX is favored for its low cost and ease of administration~\cite{gaoMentalWorkloadMeasurement2013}, and it exhibits less variability compared to one-dimensional workload scores~\cite{rubioEvaluationSubjectiveMental2004}, making it suitable for our study. 

While cognitive load can be assessed through psychophysiological, performance, subjective, and analytical measures~\cite{gaoMentalWorkloadMeasurement2013}, the length and complexity of QSs make some of these impractical. Performance and analytical measures require task switching or interruptions, which risk increasing overall cognitive load and experiment time. Psychophysiological measures, such as pupil size~\cite{palinkoEstimatingCognitiveLoad2010} and ECG~\cite{haapalainenPsychophysiologicalMeasuresAssessing2010}, are costly, sensitive to external factors, and often require participants to wear additional equipment.

\subsubsection{Demographic, Debrief, and Compensation}
After the interview, the researcher collected participant's demographics and debriefed them, explaining that the study's goal was to understand interface design and cognitive load. Participants received a \$15 cash compensation.

\newpage

\subsection{Quantitative Measures: Clickstream Data}
\label{subsec:measures}

Besides using NASA-TLX and interviews to capture cognitive load, we also recorded participants' clickstream data from the interface (i.e., each click and the corresponding UI component). These log data enabled us to analyze how participants navigated and engaged with the survey options.

\paragraph{Edit Distance} We introduce three related metrics---edit distance per option, edit distance per action, and cumulative edit distance---to quantify the distance participants traveled across the interface. Edit distance per option sums the total number of options traversed while modifying a single vote option, edit distance per action measures the distance traversed during each individual adjustment, and cumulative edit distance captures the total distance traversed throughout the entire survey. The formal definitions and modeling approach are provided in~\Cref{sec:dist}.

\paragraph{Time Spent per Option} In addition, we computed the total time participants spent interacting with each specific option by aggregating the time spent on that specific option during the survey. We describe and discuss these findings in~\Cref{sec:timeAnalysis}.

\begin{figure*}[!h]
    \centering
    \begin{subfigure}[t]{0.47\textwidth}
        \centering
        \includegraphics[width=0.95\textwidth]{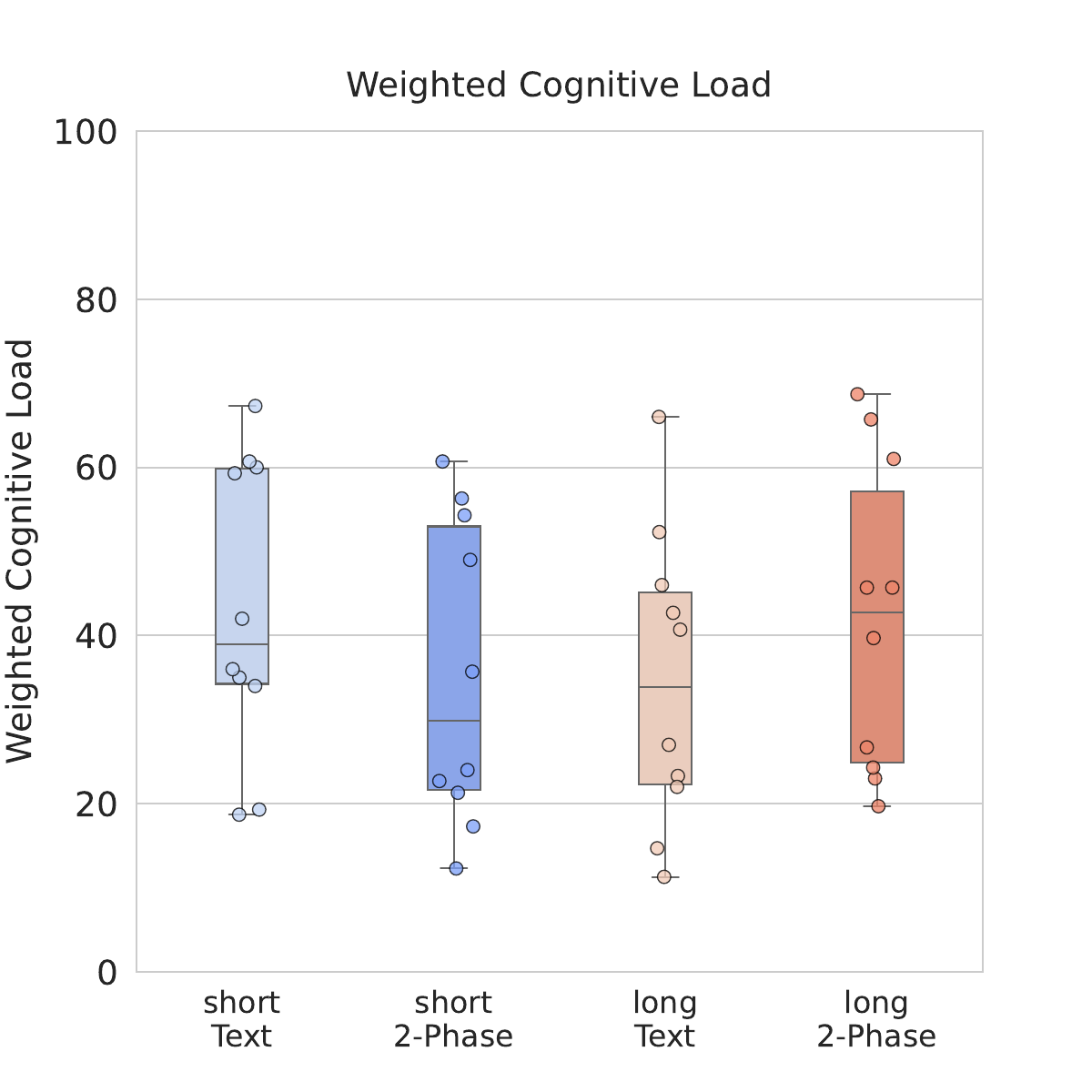}
        \caption{NASA-TLX Weight Score: The Long Two-Phase Interface exhibits the highest weighted cognitive load with a median of $42.70$, a mean of $42.02$. This is higher than the long text interface, which has a median cognitive load of $33.85$ and a mean of $34.60$. However, the short text interface demonstrates a higher cognitive load with a median of $39.00$, a mean of $43.23$, compared to the short two-phase interface, which has a median of $29.85$, a mean of $35.36$. The standard deviation is similar across groups at around $18$.}
        \Description{A box plot comparing weighted cognitive load scores across four interface conditions: short text, short two-phase, long text, and long two-phase. The y-axis is labeled "Weighted Cognitive Load" and ranges from 0 to 100. Each condition is represented by a box with whiskers, and individual data points are plotted as circles. The short text and long two-phase interfaces show higher cognitive load distributions, while the short two-phase and long text interfaces show lower distributions. The title reads "Weighted Cognitive Load."}
        \label{fig:nasatlx-final1}
    \end{subfigure}
    \hfill
    \begin{subfigure}[t]{0.49\textwidth}
        \centering
        \includegraphics[width=1\textwidth]{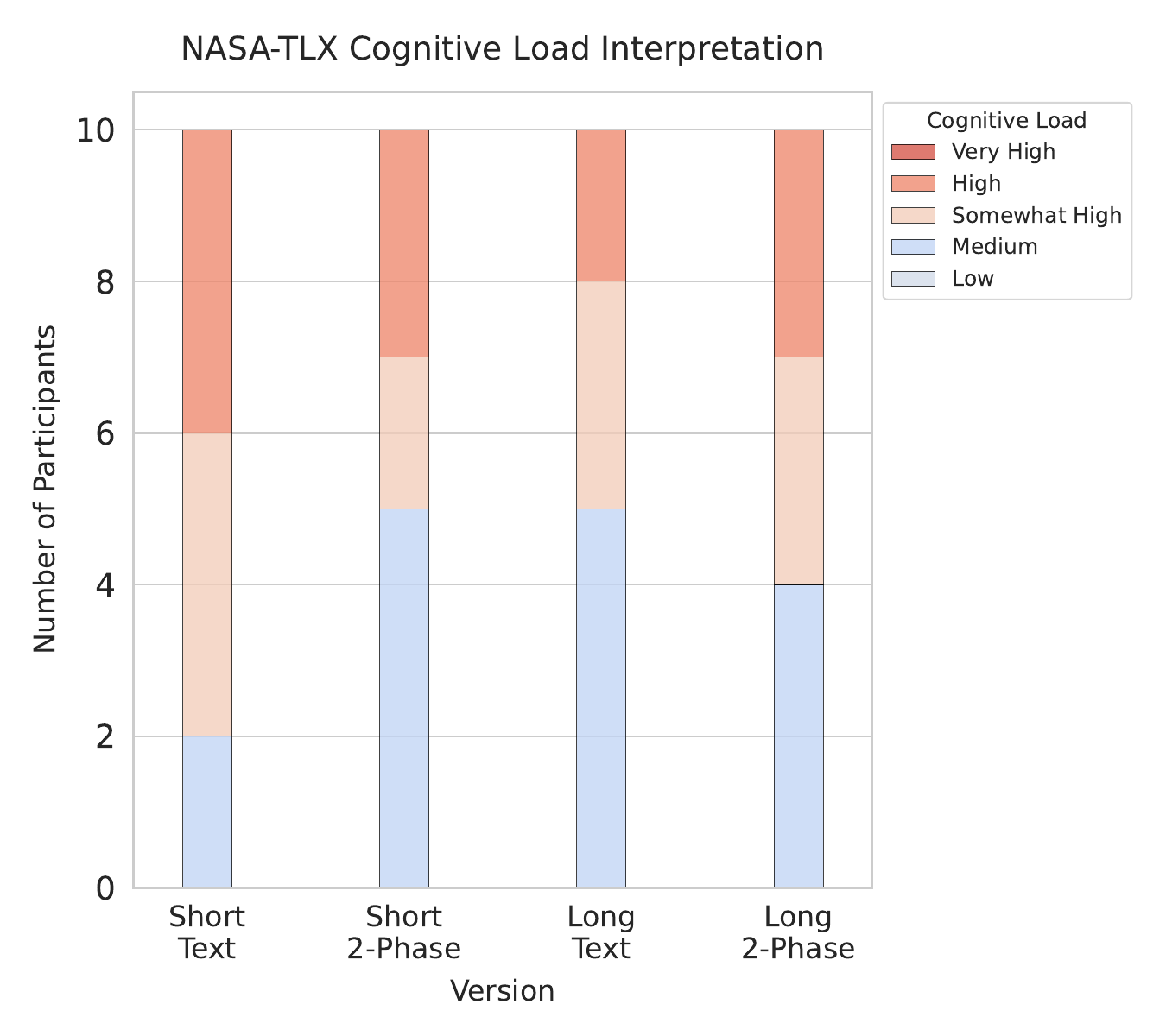}
        \caption{NASA-TLX Cognitive Interpretation: More participants in the short text interface, totaling $8$, reported a somewhat high or above cognitive load, which is significantly higher compared to the $5$ participants who reported similarly for the short two-phase interface. However, the long two-phase interface saw slightly more participants, $6$ in total, reporting somewhat high or above cognitive load compared to the long text interface.}
        \Description{Stacked bar chart showing the number of participants experiencing different levels of cognitive load across four interface versions: Short Text, Short 2-Phase, Long Text, and Long 2-Phase. The y-axis represents the number of participants (from 0 to 10), and the legend differentiates five cognitive load levels: Low, Medium, Somewhat High, High, and Very High. The chart highlights more participants reporting higher cognitive loads for short text-based interfaces.}
        \label{fig:nasatlx-final2}
    \end{subfigure}
    \caption{This figure shows the box plot results for weighted NASA-TLX scores across experiment groups and participant counts based on individual score interpretations. In~\ref{fig:nasatlx-final1}, we observe a downward trend in cognitive load for the short QS, while the long QS shows an upward trend. Interestingly, there is a counterintuitive downward trend between short and long text interfaces. In~\ref{fig:nasatlx-final2}, these trends are clearer when NASA-TLX scores are grouped into five tiers.}
    \Description{Two figures side by side, a box plot to the left and a stacked bar chart to the right, representing the weighted NASA-TLX distributions and the interpretations respectively.}
    \label{fig:nasatlx-final}
\end{figure*}

\section{Result: Self-Reported Cognitive Load in Quadratic Surveys}
\label{sec:cog}
This section presents findings on cognitive load in QSs, focusing on how the number of options and different interfaces influence it (\textbf{RQ1}, \textbf{RQ2a}). We analyze similarities and differences in cognitive load sources across conditions (\textbf{RQ2b}).

Qualitative findings are based on an inductive thematic analysis~\cite{olsonWaysKnowingHCI2014}, which was conducted after transcribing the interviews. The first author single-coded the snippets according to the research questions and merged them into overarching themes. The first author conducted multiple rounds of coding, and identified differences across conditions, which were refined and validated using a deductive coding process.

Quantitative findings are derived from a Bayesian approach, which enhances transparency by interpreting posterior distributions and moving beyond binary thresholds~\cite{kay2016researcher}. Bayesian methods suit various sample sizes, leveraging maximum entropy priors to ensure conservative and robust inferences~\cite{mcelreath2018statistical}.

\subsection{Overall Cognitive Load from NASA-TLX}
\label{sec:cog_overall}
Weighted NASA-TLX uses a continuous $0$ to $100$ score, with higher values denoting greater cognitive load. We use predefined mappings of NASA-TLX scores to cognitive levels: low, medium, somewhat high, high, and very high, as described by~\citet{hart1988development}. Results are shown in Figure~\ref{fig:nasatlx-final1}, with value interpretations presented in Figure~\ref{fig:nasatlx-final2}.

Given the sparsity of the data, we modeled the weighted NASA-TLX scores as ordinal outcomes based on value interpretations. We developed a hierarchical Bayesian ordinal regression model with length as an ordinal predictor and interface type as a categorical predictor, using hierarchical priors for partial pooling. Interaction effects between length and interface are captured via a non-centered parameterization with an LKJ prior to account for correlations~\cite{mcelreath2018statistical}. We applied the same model to the NASA-TLX subscales; since these subscales lack inherent cognitive level interpretations, we constructed weighted bins for the ordinal regression. In our model, a latent variable represents a continuous measure of cognitive load, discretized into ordinal outcomes via thresholds. Details of this model and additional subscale results are provided in Appendix~\ref{apdx:model_tlx}.

\begin{figure*}[ht]
    \centering
    \includegraphics[width=0.8\textwidth]{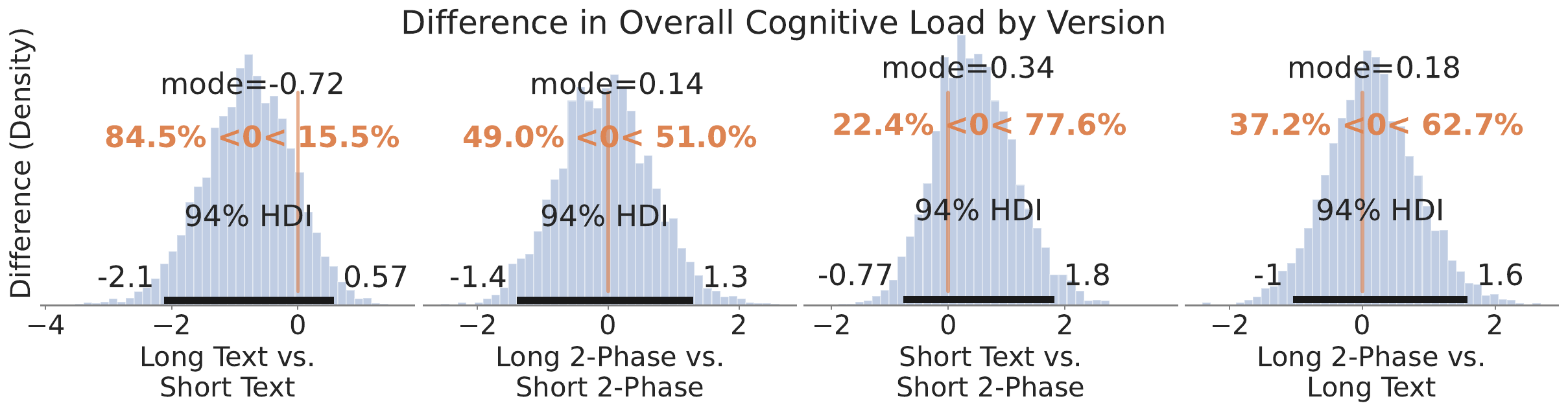}
    \caption{Posterior distributions of differences in latent cognitive load between experimental conditions. Values below 0 indicate reduced load.~\textbf{Main takeaway:} while the model does not indicate statistically significant differences, longer text interfaces are more likely to reduce cognitive load, and the two-phase interface has a higher probability of lowering cognitive load.}
    \Description{A grouped panel of four histograms titled "Difference in Overall Cognitive Load by Version," displaying posterior distributions of differences between experimental conditions. Each plot shows density (y-axis) against the difference (x-axis) with summarized statistics. None of the orange lines are outside of the interval. Each histogram features a density curve with credible intervals, a vertical reference line at zero, and key values in orange. These values emphasize the distribution characteristics and differences across versions.}
    \label{fig:weighted_cog_version}
\end{figure*}

In Bayesian analysis, the 94\% high-density interval (HDI) represents the range where the true parameter is most likely to lie. While the results (Figure~\ref{fig:weighted_cog_version}) in terms of differences in latent cognitive load are not statistically significant because 0 is within this range, the HDI quantifies probabilistic trends and accounts for uncertainty in a transparent manner.
\begin{itemize}[leftmargin=*]
    \item Increased option length with text interface trends to~\textit{reduced} cognitive load with a posterior probability of approximately $84.5\%$. This reflects a median cognitive load of $33.85$ (mean = $34.60$, SD = $17.69$) compared to a median of $39.00$ (mean = $43.23$, SD = $17.65$).
    \item Within short QSs, the two-phase interface trends to~\textit{reduced} cognitive load, with a posterior probability of $77.6\%$ supporting the reduction. Participants report a median cognitive load of $29.85$ (mean = $35.36$, SD = $18.17$) under the two-phase interface compared to a median of $39.00$ (mean = $43.23$, SD = $17.65$) under the text interface.
    \item For the long QSs, the two-phase interface trends an~\textit{increase} in cognitive load with a posterior probability of $62.7\%$. The median cognitive load is $42.70$ (mean = $42.02$, SD = $18.48$) under the two-phase interface compared to $33.85$ (mean = $34.60$, SD = $17.69$) in the text interface.
\end{itemize}

This result contradicts our hypothesis that more options would increase cognitive load and that interfaces can reduce it. Thus, we explore qualitative results to identify possible explanations. To understand the similarities and differences in sources of cognitive load (\textbf{RQ2b}), we analyze qualitative results across the six NASA-TLX subscales: mental demand, physical demand, temporal demand, effort, frustration, and performance. Detailed breakdown of each subscale are provided in Appendix~\ref{apdx:cog_qual}.

\subsection{Qualitative Analysis: Common Sources of Cognitive Load}
\label{sec:cog_common}
Our analysis reveals several themes across different cognitive load subscales. We focus on three themes common to all experimental conditions, omitting less related themes for clarity.

\textbf{Preference Construction} is cited by 97.5\% (N=39) of participants as a significant source of mental demand, consistent with prior literature suggesting that preferences are often constructed in context rather than fixed~\cite{lichtensteinConstructionPreference2006}. Specific tasks contributing to this demand include evaluating the relative importance between options (e.g.,\smallquote{S002}{Figuring out\bracketellipsis how much I prioritize option 1 over option 2}, 40\% (N=16)), making trade-offs due to limited resources (e.g.,\smallquote{S005}{\bracketellipsis very hard to take decisions~\ldots I felt that multiple options deserve equal amounts of credit~\ldots but you have given very limited credit.}, 42.5\% (N=17)), and deciding the exact number of votes (e.g.,\smallquote{S023}{\bracketellipsis having to pick how many upvotes would go to each one}, 70\% (N=30)).

\textbf{Budget Management} emerges as a source of both mental and temporal demand. 25\% (N=10) of participants describe the challenge of working with limited credits while trying to maximize their allocation (e.g.,~\smallquote{S032}{~\bracketellipsis for certain societal issues, you had to~\ldots take away from other issues you could support}). An equal percentage of participants find it mentally taxing to keep track of remaining credits (e.g.,~\smallquote{S006}{~\bracketellipsis looking at the remaining credits, I'm trying to mentally divide that up before I start allocating}).

When assessing themes across all subscales, we identified patterns that highlights the underlying nature of participants' cognitive efforts across different contexts. Thus, we also coded interview snippets as~\textbf{Operational} and~\textbf{Strategic} actions in addition to goal-oriented actions such as Budget Management and Preference Construction.

\textbf{Operational Actions} refer to reactive efforts addressing immediate, tactical needs, which emerged across all experimental conditions. These actions involve direct task execution, responding to constraints without reflection on broader, long-term implications. Examples include adjusting choices to stay within budget (e.g.,~\smallquote{S003}{I had to alter~\bracketellipsis I kept going under budget}), re-reading options (e.g.,~\smallquote{S010}{I just had to reread it again}), completing questions efficiently (e.g.,~\smallquote{S010}{I was trying to be efficient in responding to the question}), and interacting with the survey interface (e.g.,~\smallquote{S018}{Like (deciding) one upvote or two upvotes\bracketellipsis}). 40\% (N=16) of participants attribute Operational actions to temporal demand. Additionally, 37.5\% (N=15) attribute this cause to frustration, and 32.5\% (N=13) attribute it to performance. While commonly cited across conditions, its distribution varies.

\subsection{Qualitative Analysis: Different Sources of Cognitive Load}
\label{sec:cog_diff}
There are several notable differences between the text and two-phase interfaces. 

First, regardless of length, when analyzing performance, which refers to a person's perception of their success in completing a task, participants describe their performances differently. We categorize them into indications of satisficing behaviors(``good enough''), exhausting their effort (i.e., ``done their best,''), or feeling positive (i.e., ``feeling good.'') There are almost twice as many participants using the two-phase interface to report a positive feeling about their final submission~(55\% v.s 30\% (N=11 vs. 6)).

Second, 70\% (N=14) of text interface participants attribute operational actions as contributors to effort, double the percentage observed in the two-phase interface group (35\%, N=7). This partially echoes the finding that 90\% (N=18) of text interface participants report mental demand from deciding the exact number of votes, compared to 60\% (N=12) in the two-phase interface group.

The distinction between the text and two-phase interfaces becomes more pronounced in the context of the long survey. 80\% of the long text interface participants (N=8) attribute operational actions to effort, compared to only 20\% (N=2) in the long two-phase interfaces. Conversely, 90\% of long two-phase interface participants (N=8) attribute effort to strategic actions, compared to 50\% (N=5) in the text interface. 

We also found differences in how preference construction differs in contributing to their mental demand and sources of effort. Opposite to operational actions, \textbf{strategic considerations} refer to considering about long term goals, determining priorities, considering broader implications, and considering option's more holistically. Consider the following quotes:

\aptLtoX[graphic=no,type=html]{
    \begin{enumerate}
        \item[\,] {\color{darkgray}{\it Trying to figure out what upvotes I should give~\bracketellipsis went back and forth between those two.~\bracketellipsis it was very mentally tasking for me. \qquad\hspace{0.1em}\textnormal{~~~\faCommentsO}\hspace{-0.2em}\texttt{\kern-0.2em S015~(LT)}}}
    \end{enumerate}
}{
    \begin{displayquote}
        Trying to figure out what upvotes I should give~\bracketellipsis went back and forth between those two.~\bracketellipsis it was very mentally tasking for me. \hfill \quoteby{S015~(LT)}
    \end{displayquote}
}

\aptLtoX[graphic=no,type=html]{
    \begin{enumerate}
        \item[\,]{\color{darkgray}{\it \bracketellipsis especially with so many different societal issues. How do I personally prioritize them? And to what extent do I prioritize them? \qquad\hspace{0.1em}\textnormal{~~~\faCommentsO}\hspace{-0.2em}\texttt{\kern-0.2em S009~(L2P)}}}
    \end{enumerate}
}{ 
    \begin{displayquote}
        \bracketellipsis especially with so many different societal issues. How do I personally prioritize them? And to what extent do I prioritize them? \hfill \quoteby{S009~(L2P)}
    \end{displayquote}
}

\texttt{S015} describes the~\textbf{operation} of locating tasks to find the right vote, whereas \texttt{S009}~\textbf{strategically} aligns higher-order values holistically. Regarding mental demand, 80\% of participants in the long text interface focused on a narrower scope, comparing fewer options (N=8), while only 30\% did so in the two-phase interface (N=3). Conversely, 90\% of participants in the long two-phase interface considered broader societal impacts and evaluated more options simultaneously (N=9), compared to 30\% in the text interface (N=3). Similar distinctions were evident in effort-related sources.

These differences highlight variations in \textbf{levels of engagement} with the survey content. Participants using the two-phase interface expressed higher satisfaction with their performance. For the long survey, they engaged with broader aspects across different options and strategically allocated their credits.

\subsection{Qualitative Analysis: Instances of Satisficing}
\label{sec:satisficing}
When individuals cannot process all available information, prior research has found that people exhibit~\textit{satisficing behaviors}, which refers to settling for \textit{good enough} rather than \textit{optimal} decisions~\cite{gigerenzerReasoningFastFrugal1996}. While we did not explicitly ask participants if they 'satisficed,' nor did we measure it quantitatively, we identified satisficing behaviors based on participants' explanations of how they completed the survey. For example,

\aptLtoX[graphic=no,type=html]{ 
    \begin{enumerate}
        \item[\,]    \color{darkgray}{\it \bracketellipsis you thought of enough things, you know, and so it wasn't the most effort I could put in because again, that would have been diminishing returns. I tried to think of enough things~\bracketellipsis and then move on.~\bracketellipsis    \qquad\hspace{0.1em}\textnormal{~~~\faCommentsO}\hspace{-0.2em}\texttt{\kern-0.2em S032 (ST)}}
    \end{enumerate}
 }{ 
    \begin{displayquote}
        ~\bracketellipsis you thought of enough things, you know, and so it wasn't the most effort I could put in because again, that would have been diminishing returns. I tried to think of enough things~\bracketellipsis and then move on.~\bracketellipsis
    \hfill\quoteby{S032 (ST)}
\end{displayquote}
} 

\aptLtoX[graphic=no,type=html]{ 
    \begin{enumerate}
        \item[\,]     \color{darkgray}{\it I felt like that (the response) was satisfied, but not perfect. Cause perfect is not a reality. \qquad\hspace{0.1em}\textnormal{~~~\faCommentsO}\hspace{-0.2em}\texttt{\kern-0.2em S036 (ST)}}
    \end{enumerate}
}{
    \begin{displayquote}
        I felt like that (the response) was satisfied, but not perfect. Cause perfect is not a reality. \hfill\quoteby{S036 (ST)}
    \end{displayquote}
 } 

This quote illustrates satisficing decision-making, where participants chose to settle for suboptimal outcomes. Satisficing was observed primarily at the beginning and end of the survey, where participants allocated large amounts of credit initially and then managed the remaining credits to confirm their final vote allocations. For instance, 

\aptLtoX[graphic=no,type=html]{ 
\begin{enumerate}
\item[\,]    \color{darkgray}{\it ~\bracketellipsis Because that (the credit) was what was left. [Laughter] I probably wouldn't use that on <optionA> instead of <optionB>.~\bracketellipsis \qquad\hspace{0.1em}\textnormal{~~~\faCommentsO}\hspace{-0.2em}\texttt{\kern-0.2em S015 (LT)}}
\end{enumerate}
 }{ \begin{displayquote}
    ~\bracketellipsis Because that (the credit) was what was left. [Laughter] I probably wouldn't use that on <optionA> instead of <optionB>.~\bracketellipsis \hfill\quoteby{S015 (LT)}
\end{displayquote}
 } 

\aptLtoX[graphic=no,type=html]{ 
    \begin{enumerate}
        \item[\,]    \color{darkgray}{\it \bracketellipsis it went negative, and then I just settled for just \$6 remaining. ~\bracketellipsis I don't think it's perfect. But I think I'm satisfied. Yeah, I'm satisfied.  \qquad\hspace{0.1em}\textnormal{~~~\faCommentsO}\hspace{-0.2em}\texttt{\kern-0.2em S033 (LT)}}
    \end{enumerate}
}{
    \begin{displayquote}
        \bracketellipsis it went negative, and then I just settled for just \$6 remaining. ~\bracketellipsis I don't think it's perfect. But I think I'm satisfied. Yeah, I'm satisfied.  \hfill\quoteby{S033 (LT)}
    \end{displayquote}
 } 

\aptLtoX[graphic=no,type=html]{ 
    \begin{enumerate}
        \item[\,]    \color{darkgray}{\it    ~\bracketellipsis when I had first started like looking at the first few, I was just doing it kinda like willy nilly, I'm not really paying that much attention to necessarily how many credits I had, or how many categories there were. \qquad\hspace{0.1em}\textnormal{~~~\faCommentsO}\hspace{-0.2em}\texttt{\kern-0.2em S041 (LT)}}
    \end{enumerate}
 }{
    \begin{displayquote}
        ~\bracketellipsis when I had first started like looking at the first few, I was just doing it kinda like willy nilly, I'm not really paying that much attention to necessarily how many credits I had, or how many categories there were. \hfill\quoteby{S041 (LT)}
    \end{displayquote}
 } 

Participants also exhibited satisficing behaviors regarding~\textit{defaults}, particularly when constructing their preferences. For example, participant \texttt{S003}, described how default placements influenced their final decisions:

\aptLtoX[graphic=no,type=html]{ 
    \begin{enumerate}
        \item[\,]    \color{darkgray}{\it    Honestly, if medical research~\bracketellipsis was the first one I saw, I think it would automatically give it a lot more. \qquad\hspace{0.1em}\textnormal{~~~\faCommentsO}\hspace{-0.2em}\texttt{\kern-0.2em S003 (ST)}}
    \end{enumerate}
}{
    \begin{displayquote}
        Honestly, if medical research~\bracketellipsis was the first one I saw, I think it would automatically give it a lot more. \hfill\quoteby{S003 (ST)}
    \end{displayquote}
}

Our qualitative analysis found that 60\% of short-text participants (N=6) and 50\% of long-text participants (N=5) expressed instances of satisficing behaviors when describing how they completed the survey, compared to none of the short two-phase participants and 30\% of long-text participants (N=3). These qualitative results highlighted potential satisficing behaviors across conditions.

\section{Clickstream data: Interface reduces edit distance in long surveys}
\label{sec:dist}
Following our findings on cognitive load, we analyze voting behaviors to identify differences in how participants cope with survey lengths, how interfaces influence their behavior, and why the long text interface might exhibit lower cognitive load. All data are publicly available\footnote{https://github.com/CrowdDynamicsLab/Quadratic-Survey-Dataset-and-Analysis} to ensure transparency and support further research. This measure reveals how participants navigate and engage with survey options. We examine three dimensions of this measure: edit distance per option, edit distance per action, and cumulative edit distance throughout the survey.

\begin{figure}[h!]
    \centering
    \includegraphics[width=0.47\textwidth]{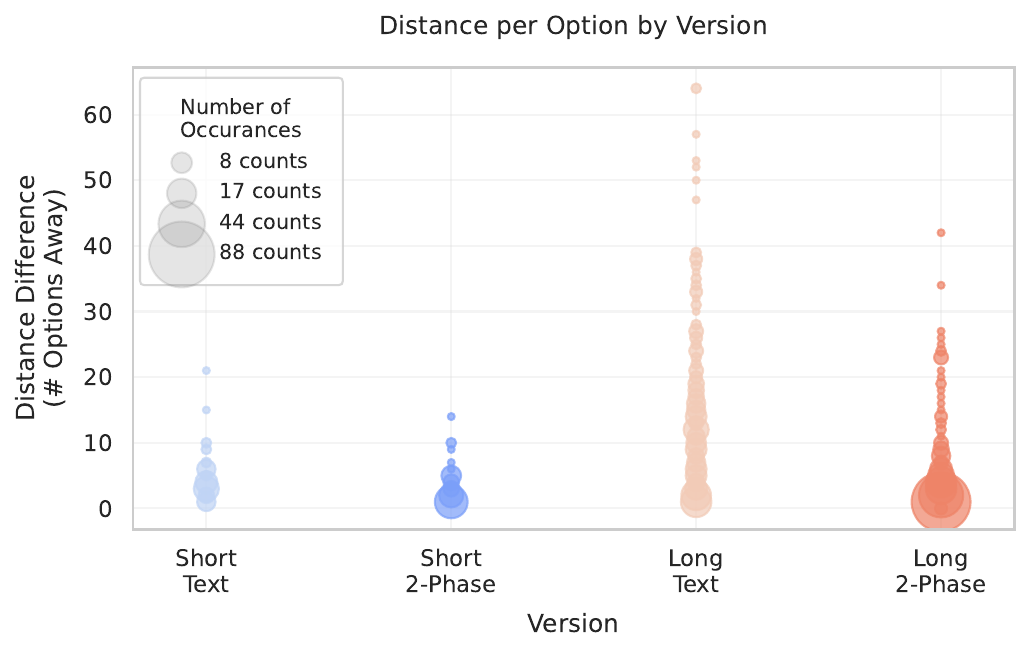}
    \caption{Edit Distance Per Option: We sum the total number of edit distances for each option, with the figure using the radius to indicate how often a specific edit distance occurred within an experimental condition.~\textbf{Main takeaway:} Participants in the two-phase interface completed their votes for more options with fewer edit distances, whereas the Long Text interface shows a long tail of options requiring a wider range of edit distances.}
    % \vspace{-10pt}
    \Description{A bubble plot titled "Distance per Option by Version" showing the distribution of distance differences (y-axis, measured as the number of options away) across four experimental versions (x-axis): Short Text, Short 2-Phase, Long Text, and Long 2-Phase. The size of the bubbles range from 8 counts (smallest bubble) to 88 counts (largest bubble). Trends: The largest bubbles (highest frequency) occur closer to zero distance difference, particularly in Short Text and Short 2-Phase. Higher distance differences appear more frequent in the Long Text and Long 2-Phase conditions. The visualization emphasizes the variability and frequency of distance differences for each version, with bubble size providing a visual cue for occurrences.}
    \label{fig:dist_per_option}
\end{figure}

\textbf{Edit distance per option:} We calculate the total number of options a participant traversed when adjusting votes for a single option. Figure~\ref{fig:dist_per_option} illustrates differences across experimental conditions, with the long text interface showing the largest variance in the distance traveled and the highest mean. We implement a hierarchical Bayesian framework to model edit distance differences across experimental conditions. The observed distance differences are modeled using an exponential distribution, where the scale parameter is linked to survey length (treated as an ordinal variable), interface type (treated as a categorical variable), interaction effects between length and interface, and controlling for individual user variability. The linear predictor includes a global intercept and slope for length, random effects for each interface condition with an LKJ prior that captures the correlations among interface categories, and user-specific random effects to account for individual heterogeneity. Appendix~\ref{sec:apdx:model_distance_option} includes the detailed model.

\begin{figure*}[h!]
    \centering
    \includegraphics[width=0.75\textwidth]{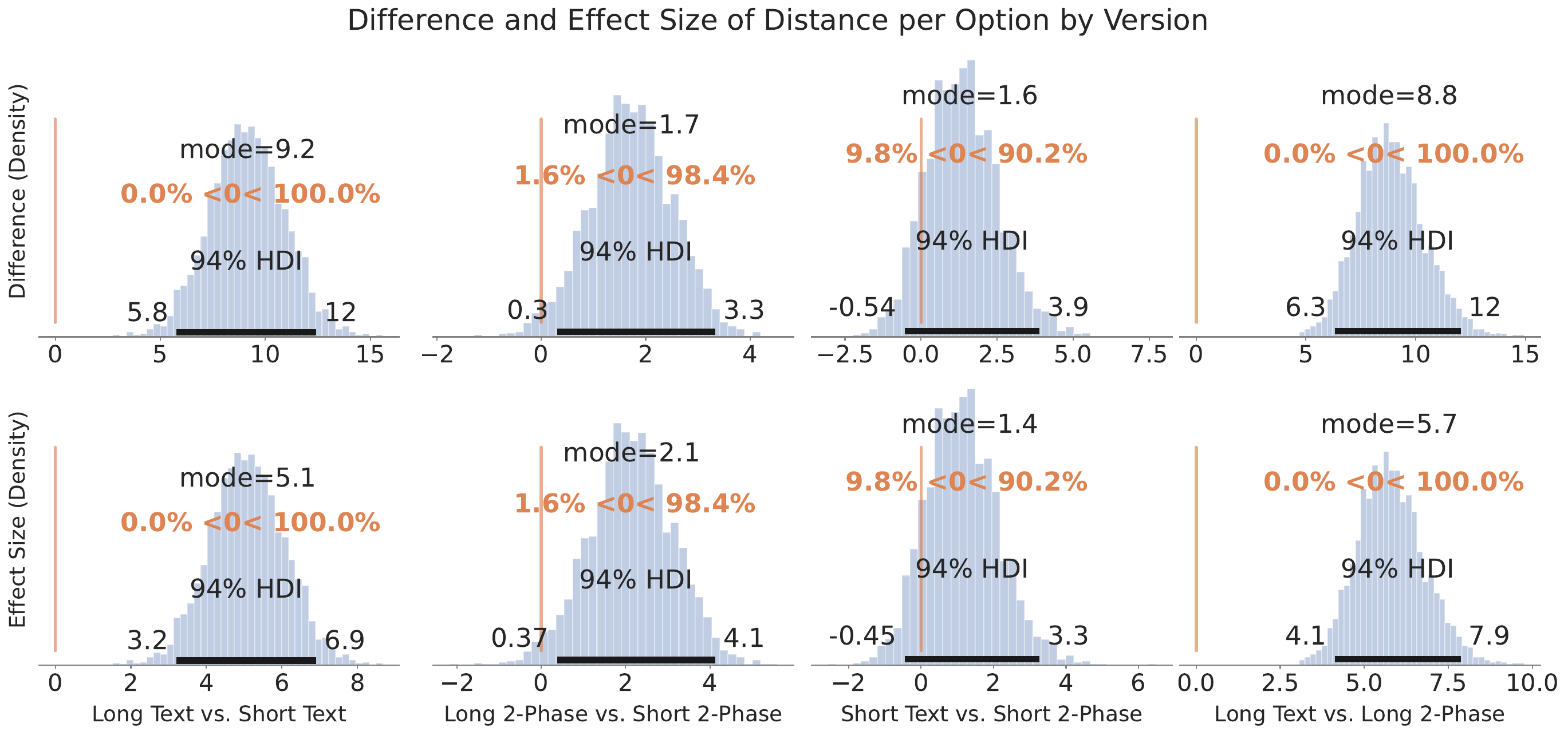}
    \caption{The figure shows the contrast distributions of the mean edit distance per option between pairwise experimental conditions, with the first row representing absolute differences and the second row depicting effect sizes.~\textbf{Main takeaway:} is that participants in the long text estimated more edit distance per option compared to those in the short text and the long two-phase condition. Notably, the long two-phase interface required estimated only slightly more edit distances despite the longer survey length.}
    \Description{A grid of histograms titled "Difference and Effect Size of Distance per Option by Version," showing posterior distributions for differences (top row) and effect sizes (bottom row) across four experimental comparisons. Each plot includes density (y-axis) against either difference or effect size (x-axis). Key values are annotated in orange above each histogram. The vertical line at zero serves as a reference point for interpreting the directionality of the differences and effect sizes. In this plot, the orange line lies outside of the ROPE for differences between long and short text; long 2-phase and short 2-phase; and long text and long two phase, signifying statistical differences among these comparisons.}
    \label{fig:dist_per_option_bayesian}
\end{figure*}

Figure~\ref{fig:dist_per_option_bayesian} illustrates the pairwise posterior distributions for differences in edit distances across experimental conditions. For example, the difference in edit distances between the short and long static interfaces has a mode of 9.1, with a 94\% highest density interval (HDI) of [6, 13]. This indicates that participants in the long text interface move approximately 9.1 steps more than those in the short text interface, with a high degree of confidence. The effect size is large (mode = 5.1, 94\% HDI = [3.3, 7.1]), suggesting a statistically significant difference, which is expected due to the greater number of options in the long text interface.

Similarly, two-phase interface participants make approximately 8.9 fewer steps per option (mode = 8.9, 94\% HDI = [6.4, 12]) than those in the long text interface, with a large effect size (mode = 5.7, 94\% HDI = [4.2, 7.9]). The increase in edit distances between the short and long two-phase interfaces is substantially smaller (mode = 1.7, 94\% HDI = [-0.01, 3.1]) compared to their static counterparts. Comparing the short text and short two-phase interfaces shows limited difference (mode = 1.3, 94\% HDI = [-0.78, 3.8]), though the posterior distribution favors fewer steps for the two-phase interface (89.3\% probability). The model suggests that the two-phase interface reduces edit distance per option, particularly for the long QS.

\begin{figure*}[h!]
    \centering
    \includegraphics[width=0.75\textwidth]{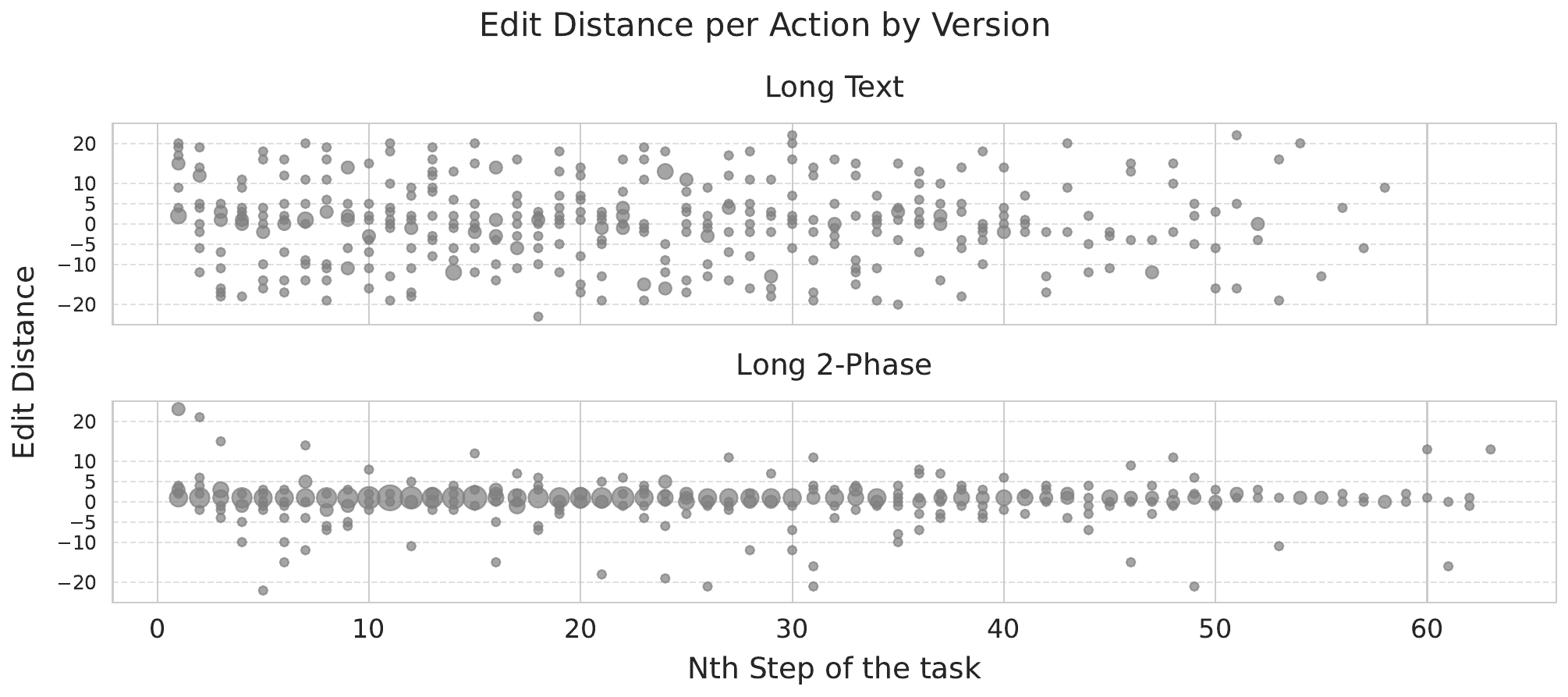}
    \caption{Edit Distance Per Action: This plot shows the frequency of specific edit distances at each step across the text interface and two-phase interface.~\textbf{Main takeaway:} Participants in the long two-phase interface tend to make adjustments closer to their previous actions, resulting in visually less variance in edit distances throughout the entire survey.}
    \Description{A two-panel scatter plot titled "Edit Distance per Action by Version," comparing edit distances across task steps for two conditions: Long Text (top panel) and Long 2-Phase (bottom panel). The Y-Axis represents the Edit Distance, ranging from -20 to 20, representing the magnitude of deviations in task execution. the X-Axis: Nth Step of the Task contains values increasing up to 60, indicating progression through task steps. Top Panel (Long Text): Shows a dispersed pattern of points with no clear trend, indicating variability in edit distances throughout the task steps. Bottom Panel (Long 2-Phase): Displays a more compressed spread of points closer to zero, suggesting reduced variability in edit distances compared to the Long Text condition. The scatter plot highlights differences in task execution consistency between the two versions, with denser clusters near zero indicating closer adherence to expected task paths.}
    \label{fig:step-over-distance}
\end{figure*}

\textbf{Edit distance per action:} Building on the statistical disparities observed in the previous analysis and the unique patterns exhibited by long text interface participants, we present analyses focusing on edit distance per action and cumulative edit distance throughout the survey between the long text and long two-phase interfaces. Edit distance per action measures how far participants move during each adjustment while completing the survey. Figure~\ref{fig:step-over-distance} illustrates how, at each step, the number of participants moving a given distance (represented by the size of the dots) varies across experimental conditions. Visually, participants move less on average per option within the two-phase interface, with lower variance at smaller scales. This indicates that participants are making local edits, meaning their adjustments tend to occur near their previous edits in terms of edit distance. This also highlights that the organization phase effectively adjusts option positions for easier access, despite participants still having the freedom to move across the interface as all options are presented to them.

In contrast to earlier analyses, we use a hierarchical Bayesian model (detailed in Appendix~\ref{sec:apdx:model_distance_variance}) to jointly estimate the mean and variance of edit distances across experimental conditions. The model assumes that edit distances are continuous and follow a normal distribution. This approach accounts for both central tendencies and variability, using separate predictors for the mean and variance. The model includes hierarchical effects for survey length, interface type, interactions between length and interface, and user-level random effects. Non-centered parametrization is used for survey length and interface type to improve convergence, while interaction effects are modeled with an LKJ prior to capture the correlations between factors. % User-level random effects reflected individual differences in behavior, incorporating variability into the model.

\begin{figure}[ht!]
    \centering
    \includegraphics[width=0.4\textwidth]{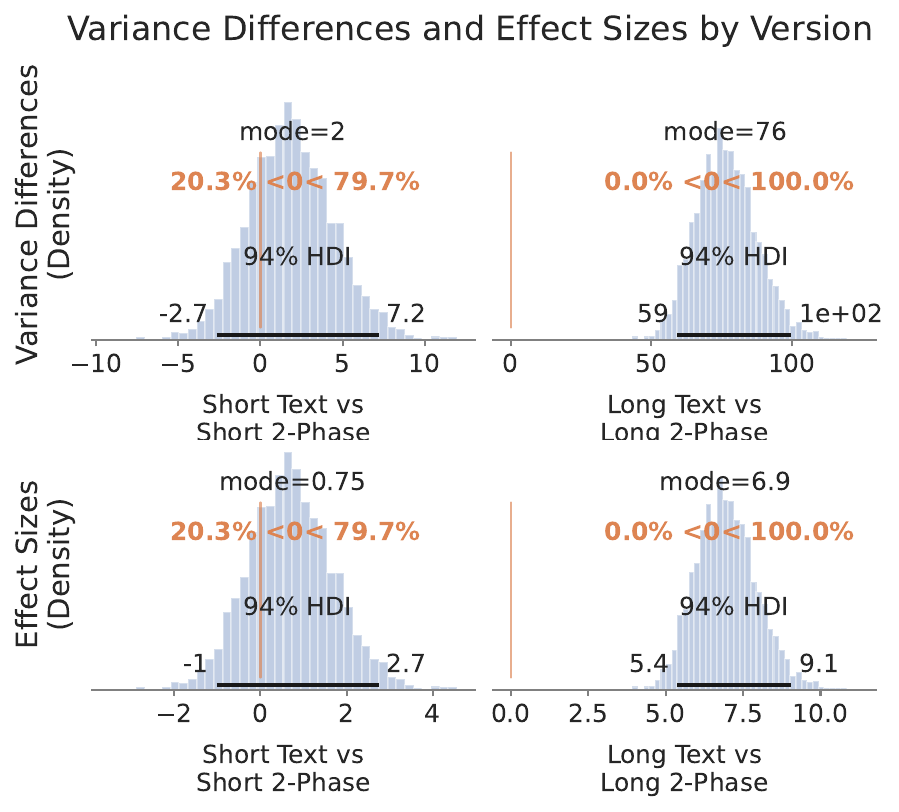}
    \caption{Posterior variance differences (left) and effect sizes (right) in mean edit distance per step between text and two-phase interfaces for different survey lengths.~\textbf{Main takeaway:} The long text interface had greater variance in edit distance per step, while differences in the short text condition were not statistically significant.}
    \label{fig:step-over-distance_bayesian}
    \Description{A grid of histograms titled "Variance Differences and Effect Size of Distance per Step by Version," displaying posterior distributions of variance differences (left) and effect sizes (right) for two experimental comparisons. Each plot includes density (y-axis) against variance differences or effect sizes (x-axis) with key statistics. Each histogram features annotated key values in orange and a vertical reference line at zero for interpretation. The distributions emphasize significant differences in variance and effect sizes across conditions, particularly for the Long Text vs. Long 2-Phase comparison, which shows consistently large and significant values.}
\end{figure}

Figure~\ref{fig:step-over-distance_bayesian} illustrates the posterior variance distributions, confirming our hypothesis. Participants in the long text interface exhibit greater variance in movement, frequently navigating across the interface, compared to those in the long two-phase interface. This is evidenced by a variance difference mode of 76 (95\% HDI = [59, 99]) and a large effect size (mode = 7.1, 95\% HDI = [5.5, 9.2]).

\begin{figure}[ht]
    \centering
    \includegraphics[width=0.47\textwidth]{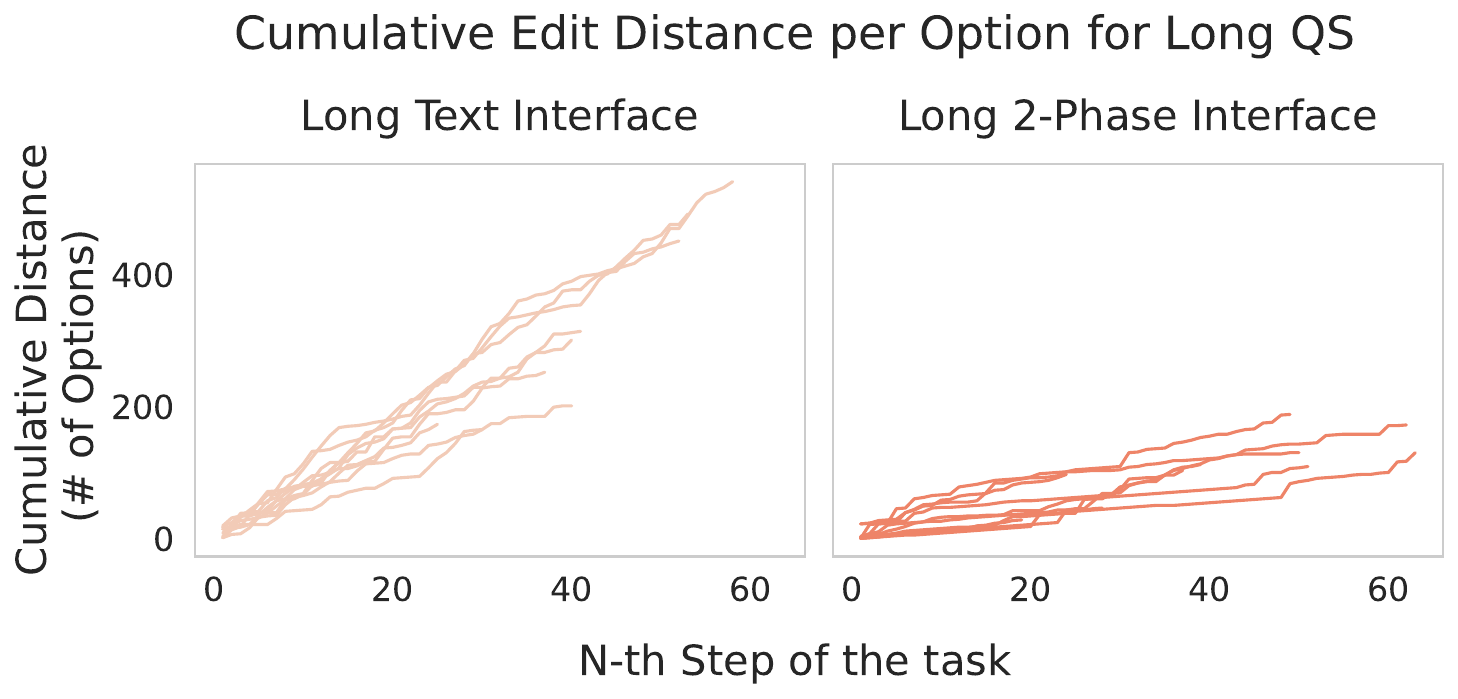}
    \caption{Cumulative edit distances over the survey for long text and long two-phase groups.~\textbf{Main takeaway:} The long two-phase interface encourages smaller, incremental adjustments, leading to a flatter slope than the text interface.}
    \label{fig:cumulative-distance}
    \Description{A two-panel line plot titled "Cumulative Edit Distance per Option for Long QS," comparing cumulative edit distances across task steps for two conditions: Long Text (left panel) and Long 2-Phase (right panel). Y-Axis: Cumulative Distance (# of Options), ranging from 0 to 500, indicating the accumulated edit distances over task steps. X-Axis: Nth Step of the Task, ranging from 0 to 60, representing task progression. Left Panel (Long Text): Shows multiple trajectories of cumulative edit distances with a wider spread, reaching up to 500, indicating higher variability and more significant cumulative distances across task steps. Right Panel (Long 2-Phase): Displays tighter trajectories with cumulative distances reaching approximately 300, suggesting more consistent and lower cumulative edit distances compared to the Long Text condition. The plot highlights differences in task completion consistency and edit distance accumulation between the two conditions, with the Long 2-Phase condition demonstrating more controlled and efficient task execution.}
\end{figure}

\textbf{Cumulative edit distance for a participant:} Figure~\ref{fig:cumulative-distance} illustrates how the two-phase interface reduces per-action distance, accumulating over time. Some long text participants traverse double the amount of distance to complete the task compared to the long two-phase participants. We model this growth rate using a hierarchical Bayesian regression model (Detailed in Appendix~\ref{sec:apdx:model_cum_distance}), with cumulative distance as the predictive variable. The experimental variables include interface type as a categorical variable, individual users modeled with random effects, and steps taken as a continuous variable. A truncated normal likelihood constrains cumulative distances to positive values and varies these distances across steps for each participant while masking incomplete data.

\begin{figure}[H]
    \centering
    \includegraphics[width=0.47\textwidth]{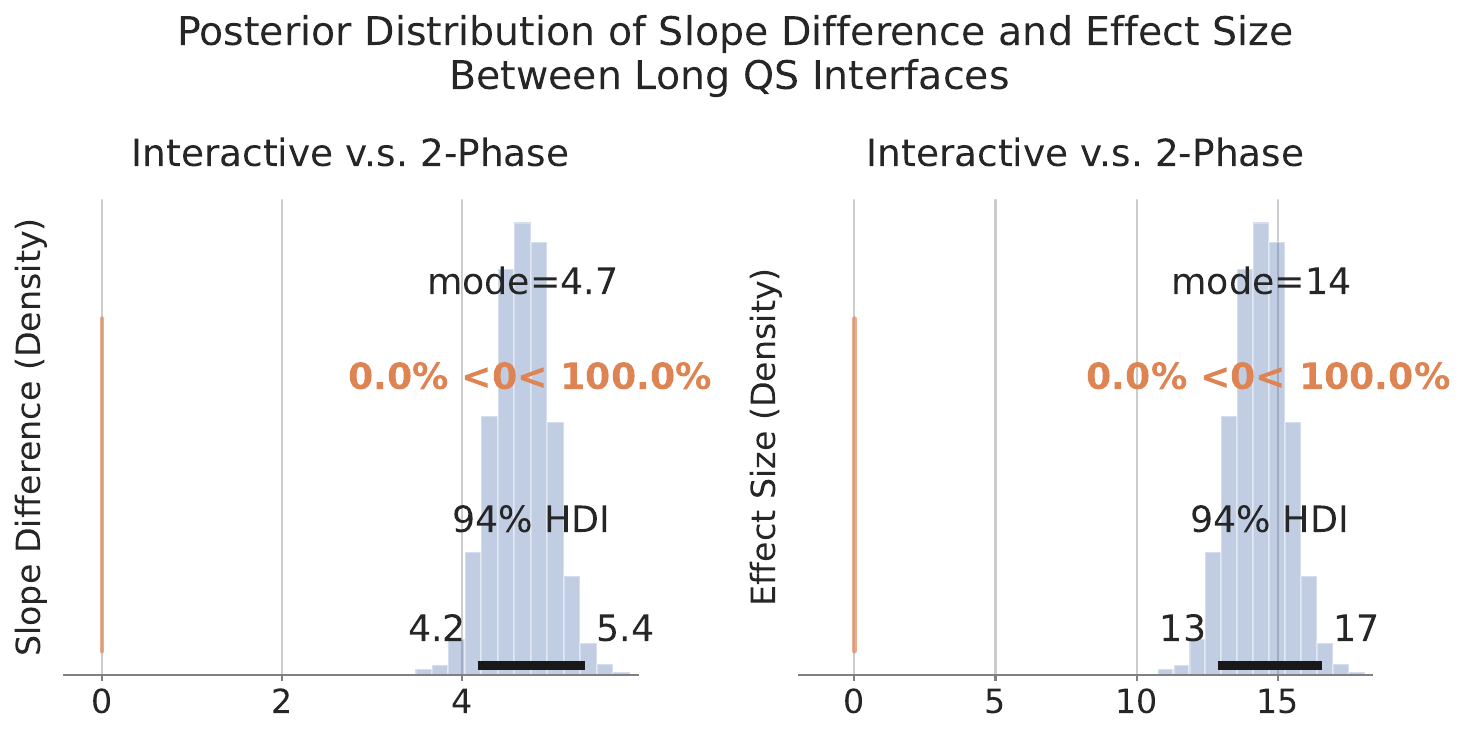}
    \caption{Posterior distribution of slope differences (left) and effect sizes (right) in cumulative edit distance between interactive and two-phase interfaces for long QSs.~\textbf{Main takeaway:} Participants in the interactive interface made larger adjustments compared to the two-phase interface.}
    \vspace{-10pt}
    \Description{A two-panel histogram titled "Posterior Distribution of Slope Difference and Effect Size Between Long QS Interfaces," showing the posterior distributions for slope differences (left) and effect sizes (right) between the Interactive and 2-Phase interfaces. Key values are annotated in orange above each distribution. Vertical reference lines at zero emphasize the directionality and significance of the differences. These results indicate that the Interactive interface has a higher slope than the 2-Phase interface and effect size measures.}
    \label{fig:slope-diff-effect}
\end{figure}

Figure~\ref{fig:slope-diff-effect} shows that the slope for the long text interface is approximately 4.7, meaning each step by the text interface would add 4.7 edit distance (94\% HDI = [4.2, 5.4]), compared to the long two-phase interface, which shows a statistically significant difference with a mode of 1.4 (94\% HDI = [1.3, 1.7]). These results explain that the variance in edit distance per action and the increase in per option edit distance are consistent across participants between the two groups, showing that the organization phase allows participants to focus on adjusting options within proximity without having to navigate the interface to locate and make adjustments throughout the voting phase.

\textbf{Evidence from qualitative analysis:} Recall the differences in sources of cognitive load between the two experimental conditions: while two-phase interface participants make localized adjustments with nearby options, they experience cognitive demand from preference construction due to broader considerations that involve more options and higher-order values. Similarly, the qualitative results highlight that long text interface participants construct narrower preferences, yet their edit distance indicates broader movements across options.

Fewer long two-phase interface participants (60\%, N=6) reported precise resource allocation as a source of demand compared to 90\% in the text interface (N=9). We interpret this as former participants construct preliminary preferences during the organization phase, easing them to concentrate vote decisions as they focus more on deliberate preference building rather than mere completion. Conveniently positioning options with similar preferences reduced the need to look for an option and traverse the interface, allowing participants remain engaged in vote adjustments.

\section{Clickstream data: Time participants spent}
\label{sec:timeAnalysis}

\begin{figure}[h]
    \centering
    \includegraphics[width=0.4\textwidth, trim=2 0 0 0, clip]{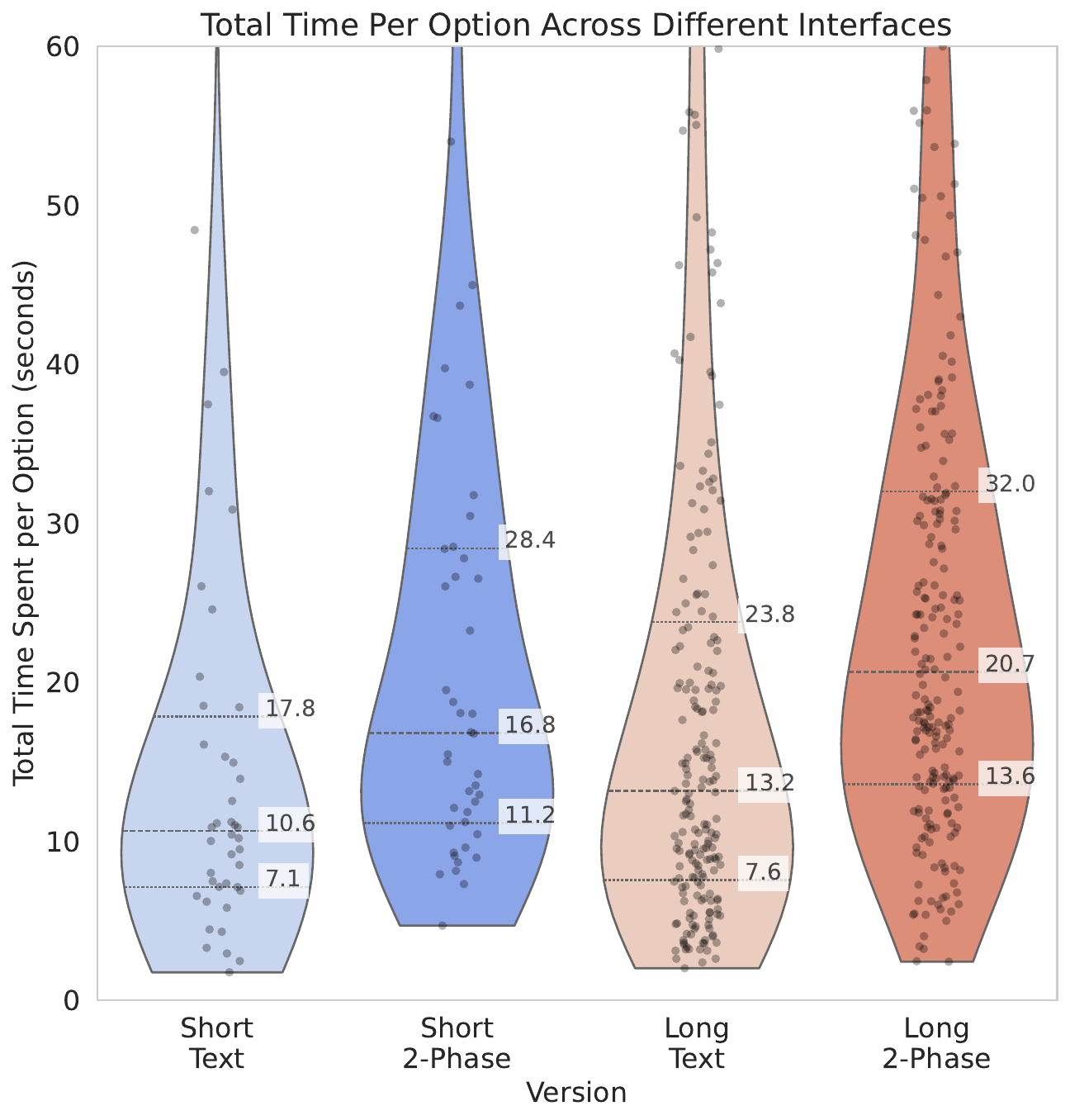}
    % \captionsetup{width=0.45\textwidth, justification=justified}
    \caption{Total Time per Option. Each dot represents the time a participant took to complete an option, with the plot's shape showing the distribution within each group. The wider it is, the more dots there are. The three horizontal lines indicate the 25th, 50th, and 75th percentile annotated with value. The two-phase interface skewed slightly higher than the text interface~\textbf{Main takeaway: } Two-phase interface participants spend longer time per option compared to its counterparts.}
    \vspace{-8pt}
    \Description{Violin plot showing total time spent per option in seconds across four interface versions: Short Text, Short 2-Phase, Long Text, and Long 2-Phase. The y-axis ranges from 0 to 60 seconds. Each violin plot has scattered dots representing individual data points. The shape of the Short Text plot is widest between 10 and 20 seconds, tapering at the top and bottom. The Short 2-Phase plot is the narrowest, with most dots concentrated between 10 and 20 seconds. The Long Text plot is narrow and widest near the bottom, between 5 and 15 seconds. The Long 2-Phase plot is widest near the top, between 20 and 40 seconds.}
    \label{fig:total_time}
\end{figure}

\begin{figure*}[ht!]
    \centering
    \includegraphics[width=0.85\textwidth]{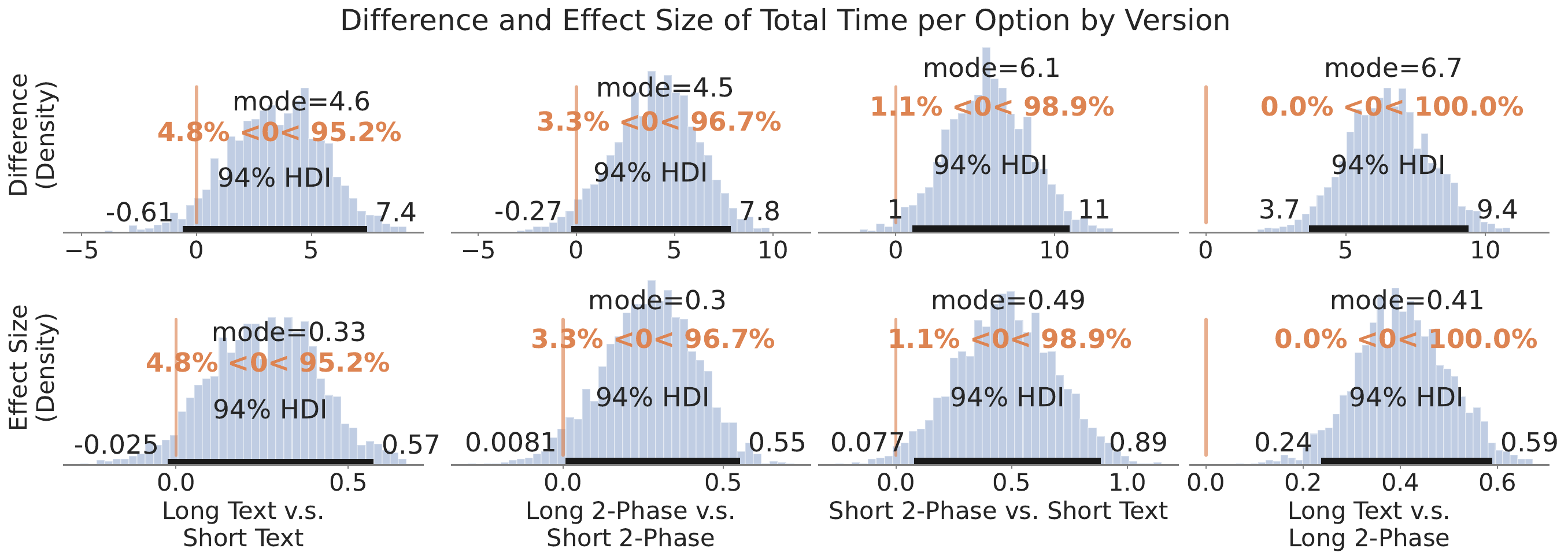}
    % \captionsetup{width=0.9\textwidth, justification=justified}
    \caption{The figure shows the contrast distributions of the mean time to complete per option between pairwise experimental conditions, with the first row representing absolute differences and the second row depicting effect sizes.~\textbf{Main takeaway: } is that participants in the long two-phase condition spent more time per option compared to those in the long text and short two-phase conditions. Additionally, short two-phase participants took longer per option than short text participants.}
    \Description{ A grid of histograms titled "Difference and Effect Size of Total Time per Option by Version," showing posterior distributions of differences (top row) and effect sizes (bottom row) across four comparisons. Key values are annotated in orange, with vertical reference lines at zero indicating significance. The orange line between the Long and short two phase, between short two phase and text, and between long text and long two-phase are outside of the ROPE, indicating statistical differences in Bayesian terms.}
    \label{fig:time_per_option_bayesian}
\end{figure*}

In addition to distance, participants in the short survey took an average of 2.7 minutes (short-text: $\mu$=2.3, $\sigma$=1.27; short two-phase: $\mu$=3, $\sigma$=1.02), while those in the long survey took 9.7 minutes (long-text: $\mu$=7.5, $\sigma$=3.45; long two-phase: $\mu$=11.95, $\sigma$=2.73). For a fairer comparison of interaction patterns, we analyze total ~\textbf{time-spend-per-option} using the QS system logs in this section. For participants in the two-phase interface conditions, this includes both organization and voting times for that option. The results are visualized in Figure~\ref{fig:total_time}.

Overall, participants spent slightly more time per option in the two-phase interface than in the text interface. To quantify these observations, we model the time data as predictive variables of separate Gamma distributions to characterize the continuous response times observed under distinct experimental conditions defined by survey length and interface type (Detailed in Appendix~\ref{sec:apdx:model_time}). Each of the four resulting subsets of data is modeled independently, with separate Gamma-distributed parameters governing the shape and rate of each group's time distributions.

We calculated the posterior differences between all pairwise comparisons of the four groups. The results in Figure~\ref{fig:time_per_option_bayesian} indicate that participants using the two-phase interface consistently spend more time per option than those using the text interface, regardless of survey length. For both the short and long QSs, participants most likely spend 6.1 seconds (94\% HDI = [1.0, 11.0]) and 6.7 seconds (94\% HDI = [3.7, 9.4]) more per option, respectively, with medium effect sizes of $d$=0.49 (94\% HDI = [0.077, 0.89]) and $d$=0.41 (94\% HDI = [0.24, 0.59]). In both cases, the intervals lie outside the ROPE of 0 ± 1, indicating statistical significance. These findings suggest that the two-phase interface encourages longer deliberation, particularly for long option surveys.

Some literature points out that increased time can lead to cognitive fatigue~\cite{kundingerReliableGroundTruth2020, karim2024examining}, which can impair decision-making. Other decision science literature suggests that longer decision times can indicate deeper cognitive processing~\cite{payneAdaptiveDecisionMaker1993, daniel2017thinking}. Our qualitative analysis points to the latter.

Descriptively, participants in the long two-phase condition remained actively engaged during the voting phase, editing their votes an average of 39.3 times per participant ($\sigma$=39.3, range=$19-63$) compared to 39.1 times ($\sigma$=13.29, range=$15-58$) in the long text condition. This suggests that the two-phase interface does not reduce engagement despite the additional organization step.

Quantitatively, other than the difference in operational thinking and strategic consideration discussed in Section~\ref{sec:cog_diff}, we find that 37.5\% of participants (N=15) who attribute time to \textit{Decision Making} as a source of temporal demand frame such demand differently. We label a participant as \textit{affirmative} if they describe the pressure to make decisions as a source of temporal demand. For example, \smallquote{S022}{So it didn't take too much time, but obviously there were a lot of things to consider, so there was some temporal demand.} is an affirmative statement. Conversely, we label a participant as \textit{negative} if they express concern about the time and effort they have already invested. For example, \smallquote{S024}{maybe I should just hurry up and make a decision.} is a negative statement.

50\% of participants (N=5) in the long two-phase group describe the pressure to make decisions affirmatively and none negatively. This suggests that their pressure stems from having too many remaining decisions to make, rather than from the time already invested. This is reflected in their higher average time spent per option and overall time spent ($\mu$=716.86 seconds, $\sigma$=164.04 seconds) completing the QS compared to the long text group ($\mu$=449.64 seconds, $\sigma$=206.97 seconds). We interpret these results that participants are thoughtfully engaged in constructing their preferences and choose to invest additional time, rather than being driven by decision-related pressures or experiencing urgency.

Conversely, in the short text group, 50\% of participants (N=5) express concern about the time and effort they have already invested~(\smallquote{S024}{maybe I should just hurry up and make a decision.}) and none frame it affirmatively. Descriptively, participants in the short text group spend comparatively less time than those in the long QS (short text: $\mu$=139.83 seconds, $\sigma$=76.43 seconds; short two-phase: $\mu$=178.78 seconds, $\sigma$=61.07 seconds). This suggests that participants in the short text group expect themselves to complete the task sooner than they actually do. 

\begin{figure}[h]
    \centering
    \includegraphics[width=0.4\textwidth, trim=0 0 0 0, clip]{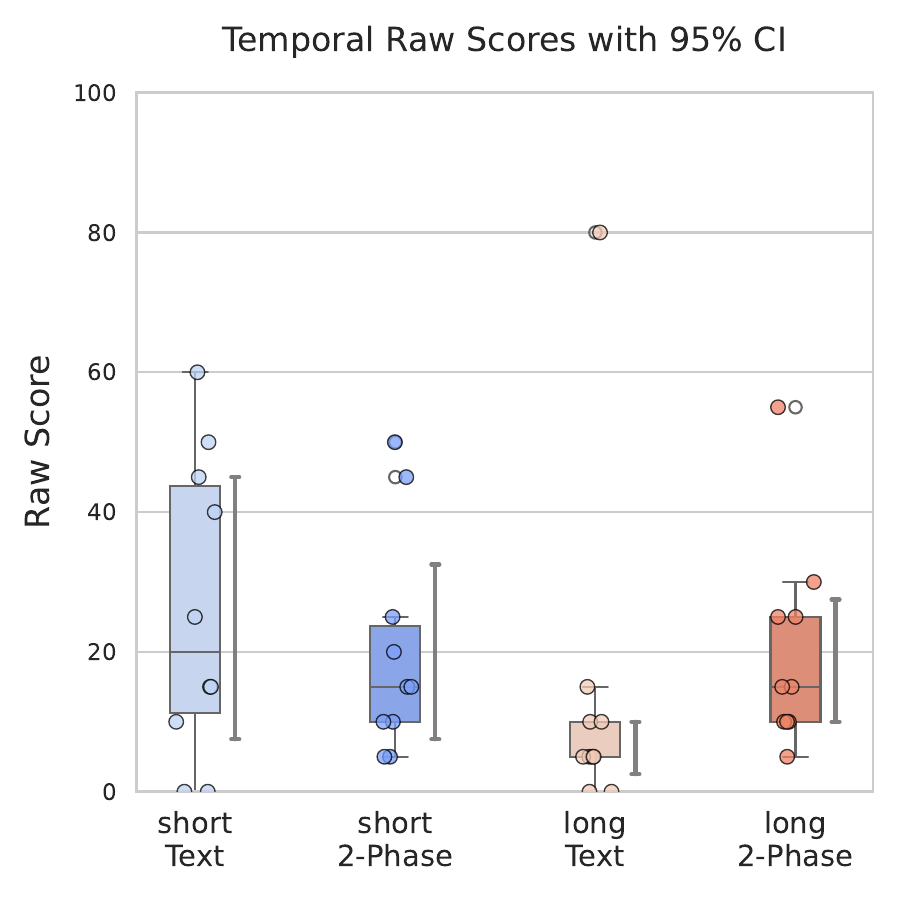}
%     \captionsetup{width=0.4\textwidth, justification=justified}
    \caption{Temporal Demand Raw Score: Each dot represents a participant's subscale response.~\textbf{Main takeaway:} Long text interface participants seem to express less temporal demand compared to the other experiment conditions.}
    \vspace{-12pt}
    \Description{A box plot titled "Temporal Raw Scores with 95\% CI," comparing raw temporal demand scores across four conditions: Short Text, Short 2-Phase, Long Text, and Long 2-Phase. Y-Axis represents 0 to 100 raw score, representing temporal demand. Short Text exhibited highest temporal demand with a wide spread of scores. Short 2-Phase showed moderate scores with reduced variability compared to Short Text. Long Text showed lowest temporal demand with minimal spread. Long 2-Phase showed moderate temporal demand, higher than Long Text but lower than Short Text. The plot shows boxplots with individual data points overlaid and whiskers indicating 95\% confidence intervals.}
    \label{fig:temporal_cog_score}
\end{figure}

Surprisingly, participants in the long text interface exhibit lower temporal demand compared to both the short text and long two-phase interfaces (Figure~\ref{fig:temporal_cog_score}). Bayesian analysis (Appendix~\ref{sec:temporal_subscale_bayesian}) supports this finding, with posterior probabilities of 86.1\% and 86.7\%, respectively. This result is notable considering participants spent more time per option compared to those in the short text interface and traversed the longest distance among all three groups (Section~\ref{sec:dist}). In addition, only 30\% of participants (N=3) mention the time spent making a decision as a source of temporal demand. One possible explanation is that some participants are satisficing, as we pointed out in Section~\ref{sec:satisficing}.  

In summary, we interpret the result that participants in the two-phase interface spend more time per option as a sign of deeper cognitive processing. This is further supported by examining participants' nuanced voting behaviors under budget constraint conditions for the long QS, which we omit here for brevity. Notably, two-phase interface participants make more small vote adjustments (i.e., adding or removing at most 2 votes on an option) when they have fewer remaining credits, further supporting our claim that they experience deeper engagement with preference construction, which we elaborate on further in Appendix~\ref{apdx:additional_results_behavior}.

\section{Discussion and Future Work}
\label{sec:discussion}

In this section, we interpret our findings on cognitive load and respondent behavior in a QS. We highlight the rationale and elements behind the two-phase interface for preference construction and its potential to mitigate satisficing behaviors. We also offer usage and design recommendations for practitioners and outline future directions for improving QS interfaces.

\subsection{Two-phase interface: a worthwhile trade-off}
Survey designers seek thoughtful responses from participants. This means the interface should balance survey usability, respondent satisfaction, and the effort individuals invest in their responses. Our results indicate that the two-phase interface encouraged deeper participant engagement with the options and reduced satisficing behaviors, despite its increased time per option and higher cognitive load for the long QSs.

\subsubsection{Analysis through the lens of cognitive load theory}
Cognitive load theory~\cite{swellerCognitiveLoadTheory2011}, when applied to QSs, identifies three components of cognitive load: intrinsic load (the cognitive demand required to understand questions and response options), germane load (associated with deeper processing and preference evaluation), and extraneous load (stemming from navigating and operating the survey interface).

Participants were randomly assigned to experimental conditions, with survey lengths containing options randomly drawn from a common pool to control for intrinsic load within the same group. 

When a QS is short, participants can engage with all options simultaneously. Participants using the two-phase interface traded a slightly longer survey response time for a potential reduction in cognitive load and edit distance. We interpret this as participants freeing up cognitive demand from extraneous load for germane load, prompting them to better construct and express their preferences.

When a QS is long, participants face more options, resulting in a higher intrinsic load at the start of the survey. We believe the two-phase interface traded longer survey response time and a potential increase in cognitive load for deeper engagement with the survey. This heightened cognitive load likely stemmed from making comparisons in both the organization and voting phases. Quantitatively, participants spent more time per option, suggesting deeper engagement while exerting limited extraneous load, as evidenced by shorter traversals during voting. Qualitatively, participants reported experiencing demand primarily from strategic considerations (germane load) rather than operational actions (extraneous load), which were common among text interface participants.

While some might argue that the additional organizing phase offers participants more opportunities to familiarize themselves with the options compared to text interface participants, the greater overall edit distance and high variance in edit distance per option suggest that text interface participants traversed the list frequently. This finding is further supported by qualitative data, where 70\% of long-text participants (N=7) reported scanning the list while voting. This behavior suggests that while long-text participants had opportunities to familiarize themselves with the options, the explicit organization phase encouraged deeper reflection on their preferences.

The effect of the two-phase interface shows nuanced differences influencing cognitive load outcomes; however, both analyses suggest that the two-phase interface \textit{shifted} participants' cognitive focus when completing QS.

\subsubsection{Potential in limiting Satisficing}
Qualitative findings (Section~\ref{sec:satisficing}) on potential satisficing behavior highlight the importance of careful consideration when deploying a long QS. However, the two-phase interface appeared to limit satisficing behaviors, as evidenced by fewer observations compared to the long text interface for the long QS and none for the short QS. We believe the potential reasons lie in the design of the two-phase interface, which scaffolds the preference construction process.

The deliberate one-option-at-a-time presentation during the voting task in the two-phase interface reduced reliance on defaults and encouraged deeper reflection using cognitive strategies such as \textit{\smash{problem decomposition}}~\cite{simonSciencesArtificial1996} and \textit{\smash{dimension reduction}}, both of which are known to reduce cognitive overload.

When asked about their experience with the interface, four participants highlighted how the organization phase supported their preference construction.~\texttt{S013} illustrated how the one-option-at-a-time approach reduced the dimensions of decision-making:

\aptLtoX[graphic=no,type=html]{ 

\begin{enumerate}  
\item[\,]\color{darkgray}{\it \bracketellipsis it (organization phase) gives you time to just focus on that single thing and rank it based on how you feel at that moment. \qquad\hspace{0.1em}\textnormal{~~~\faCommentsO}\hspace{-0.2em}\texttt{\kern-0.2em S013 (S2P)}  }
\end{enumerate}

 }{ \begin{displayquote}  
\bracketellipsis it (organization phase) gives you time to just focus on that single thing and rank it based on how you feel at that moment. \hfill\quoteby{S013 (S2P)}  
\end{displayquote}  
 } 

This focused mode enabled deeper reflection. When considering relative preferences among QS options,~\texttt{S009} described how it structurally decomposed the problem:

\aptLtoX[graphic=no,type=html]{ 

\begin{enumerate}  
\item[\,]\color{darkgray}{\it \bracketellipsis to have a preliminary categorization of all the topics ~\bracketellipsis (allowed me) to think about and process~\bracketellipsis digest all the information prior to actually allocating the budget~\bracketellipsis \qquad\hspace{0.1em}\textnormal{~~~\faCommentsO}\hspace{-0.2em}\texttt{\kern-0.2em S009 (L2P)}  }
\end{enumerate}

 }{ \begin{displayquote}  
\bracketellipsis to have a preliminary categorization of all the topics ~\bracketellipsis (allowed me) to think about and process~\bracketellipsis digest all the information prior to actually allocating the budget~\bracketellipsis \hfill\quoteby{S009 (L2P)}  
\end{displayquote}  
 } 

This quote highlighted how participants' deliberation occurred during the organization phase, enabling them to focus on constructing preferences without worrying about budget management---both of which are cited sources of cognitive load. Although direct measurement of satisficing behavior reduction is challenging, qualitative data and participant feedback suggest that the two-phase interface potentially limits such behaviors. Based on this evidence, we recommend that long QSs be implemented with a two-phase interface and sufficient time for participants to complete the process. We suggest future research investigate the mental processes underlying satisficing behaviors in long QSs. 

\textbf{In summary,} we argue that the trade-off of a longer completion time and potentially higher cognitive load in the two-phase interface is justified. Drawing on cognitive load theory, the interface fosters deeper engagement with the options. Additionally, our qualitative findings and participant feedback suggest that the interface may reduce satisficing, aligning with decision-makers' goals of obtaining thoughtful and deliberate responses from participants.

% ============================== %
\subsection{Preference Construction guided by Organize, Then Vote}
Completing a QS involves a series of in-situ, difficult decision tasks as participants construct their preference over unfamiliar options~\cite{lichtensteinConstructionPreference2006}, as one participant reflected:

\aptLtoX[graphic=no,type=html]{ 
\begin{enumerate}
\item[\,] \color{darkgray}{\it Oh, there are other aspects that I never care about.~\bracketellipsis Why (should) I spend money on that? \qquad\hspace{0.1em}\textnormal{~~~\faCommentsO}\hspace{-0.2em}\texttt{\kern-0.2em S037 (L2P)}}
\end{enumerate}

 }{ \begin{displayquote}
Oh, there are other aspects that I never care about.~\bracketellipsis Why (should) I spend money on that? \hfill\quoteby{S037 (L2P)}
\end{displayquote}
 } 

We believe the two-phase interface supported participants' preference construction process when faced with unfamiliar options.

First, 40\% of long-text participants (N=4) found it challenging to facilitate differentiation without organization tools that would allow grouping or drag-and-drop, as~\texttt{S025} said:

\aptLtoX[graphic=no,type=html]{ 

\begin{enumerate}
\item[\,] \color{darkgray}    {\it I would like to be able to like, click and drag the categories themselves so I could maybe reorder them to like my priorities.~\bracketellipsis make myself categories and subcategories out of this list~\ldots If I could organize it. \qquad\hspace{0.1em}\textnormal{~~~\faCommentsO}\hspace{-0.2em}\texttt{\kern-0.2em S025 (LT)}}
\end{enumerate}

 }{ \begin{displayquote}
    I would like to be able to like, click and drag the categories themselves so I could maybe reorder them to like my priorities.~\bracketellipsis make myself categories and subcategories out of this list~\ldots If I could organize it. \hfill\quoteby{S025 (LT)}
\end{displayquote} }

In contrast, 60\% (N=6) of long two-phase participants appreciated the upfront introduction of all options, which enabled them to organize and use drag-and-drop features to facilitate QS completion. Not only did participants use drag-and-drop options post-voting to reflect and ensure correct vote allocation, but drag-and-drop also enabled participants, like~\texttt{S039}, to make fine-grained comparisons between options:

\aptLtoX[graphic=no,type=html]{ 

\begin{enumerate}  
\item[\,] \color{darkgray}{\it    I think the system was actually really helpful because I could just drag them.~\bracketellipsis I can really compare them, I can drag this one up here, and then compare it to the top one~\bracketellipsis \qquad\hspace{0.1em}\textnormal{~~~\faCommentsO}\hspace{-0.2em}\texttt{\kern-0.2em S039 (S2P)}  }
\end{enumerate}

 }{ \begin{displayquote}  
    I think the system was actually really helpful because I could just drag them.~\bracketellipsis I can really compare them, I can drag this one up here, and then compare it to the top one~\bracketellipsis \hfill\quoteby{S039 (S2P)}  
\end{displayquote}  
 } 

This supports our intention of applying~\citet{svensonDifferentiationConsolidationTheory1992}'s differentiation and consolidation theory, in which participants attempt to identify differences and eliminate less favorable options. The organization phase and the drag-and-drop supported some degree of differentiation process.

\aptLtoX[graphic=no,type=html]{ 

\begin{enumerate}
\item[\,] \color{darkgray}{\it ~\bracketellipsis the hardest part deciding in which category of place (prefernce bin) each issue is. \qquad\hspace{0.1em}\textnormal{~~~\faCommentsO}\hspace{-0.2em}\texttt{\kern-0.2em S021 (L2P)}}
\end{enumerate}

 }{ \begin{displayquote}
    ~\bracketellipsis the hardest part deciding in which category of place (prefernce bin) each issue is. \hfill\quoteby{S021 (L2P)}
\end{displayquote}
 } 

This quote by~\texttt{S021} best represents the potential of the organization phase in separating part of the difficult decisions one needs to make when differentiating their preferences during preference construction. With the selected options, the shorter edit distance of long two-phase interface participants suggested that they were consolidating their identified preferences through votes.

% ========================= %

\subsection{What We Learned: Quadratic Survey Usage and Design Recommendations}
This study represents a crucial step toward developing better interfaces to support individuals responding to QSs by providing a deeper understanding of how survey respondents interact with QSs and the sources of cognitive load. In this subsection, we outline usage and design recommendations applicable to all applications of the quadratic mechanism.

\subsubsection{QS: Prioritizing Fewer Options or High-Stakes Evaluations}
We recommend deploying a QS with smaller sets of options or for critical evaluations, such as eliciting stakeholders' preferences before making investment decisions in hospital infrastructure. Our findings indicate that cognitive challenges and time requirements increase significantly as the number of options grows. For a long QS, while the two-phase interface helps mitigate some challenges, it does not eliminate them entirely, making adequate deliberation time essential. If a two-phase interface is unavailable, survey designers should present options in advance to allow participants to familiarize themselves and reflect before completing the QS.

\subsubsection{Facilitate Quadratic Mechanism Applications through Categorization, Not Ranking}
In a QS, the final ranking of preferences is typically a byproduct of vote allocation rather than a deliberate ranking effort. Participants did not explicitly rank options; instead, their preferences emerged dynamically through the voting process. To better support this preference construction, future quadratic mechanism interface designs should focus on helping participants categorize options effectively rather than ranking them directly. Facilitating differentiation among options is more critical than enabling precise manipulation for fine-tuning. We believe this approach extends beyond QSs to other ranking-based survey tools, such as ranked-choice voting and constant-sum surveys. Further research should examine how implementing such functionality influences survey respondents' mental models.

\subsection{Future work: Opportunities for Better Budget Management}
Budget management emerged as one of the participants' most prominent challenges, which the two-phase interface did not address. 35\% of participants (N=14) emphasized that current quadratic mechanism applications support automated calculations, but noted their insufficiency. We identified three challenges for future work:

First, participants struggled to decide on an initial vote allocation. Some distributed credits equally across options, while others used $1$, $2$, or $3$ votes as starting points. A few anchored their decisions to the tutorial's example of four upvotes. This suggests a need to better understand whether individuals have absolute value preferences among options. Second, 12.5\% of participants (N=5) expressed confusion about the relationship between budget, votes, and outcomes, despite understanding their definitions. They struggled to make trade-offs between votes and budget, leading to frustration and hampered decision-making. Third, determining the absolute amount of credits in a QS is highly demanding. Designing interfaces and interactions to address the cold start challenge and help participants decide on the absolute vote value, while also considering ways to limit direct influences, remains an open question.

We believe that, with a well-designed interface backed by real-time computing and a better understanding of how individuals calculate trade-offs, we can provide innovative solutions to help participants more easily express their preferences using QSs.

\section{Limitations}
\label{sec:limitations}
Evaluating the QS interface is challenging because of its novelty. We identified several limitations that warrant further research.

\paragraph{Individual differences in cognitive capacity}
Variations in individual cognitive capacity influenced participants' performance and cognitive scores. For example, participants with greater experience in decision-making may be better able to manage multiple options.  A within-subject study could clarify shifts in cognitive load, but deconstructing established preferences and altering options introduces additional complexity. Therefore, we opted for this in-depth, between-subject study, although the small sample size may introduce noise, potentially distorting the measurement of cognitive load. Future research should aim to quantify the impact of different QS interfaces on cognitive load at a larger scale. Furthermore, participants completed this study in a controlled laboratory environment, with options displayed on a large screen. Future work should also investigate how individuals respond to QSs on smaller devices and in less controlled environments.

\paragraph{Limited experience with QSs}
Participants lacked prior experience with the QS interface. After completing a tutorial and quiz, participants proceeded to perform tasks using the QS interface. While participants understood the mechanics of QSs, their familiarity with the interface likely influenced their strategies and cognitive load. As quadratic mechanisms become more prevalent, future research could compare the performance of novices and experts.

\paragraph{Limitations of Time and Distance as Proxies for Decision-Making Effort}
While time and distance are common metrics for quantifying the effort involved in decision-making, they do not capture without noise. Participants may have considered multiple options simultaneously. We acknowledge that these metrics are approximate indicators of decision-making effort. Despite these limitations, this approach provides valuable insights into decision-making within our experimental constraints.

\paragraph{Other Limitations}
Finally, although we observe meaningful trends in the Bayesian statistical results, the small sample size limits our ability to establish statistical significance in cognitive load differences. Additionally, despite our best efforts to ensure transparency in the qualitative analysis, potential biases may have been introduced by relying on a single coder. Future work should address these limitations by incorporating larger sample sizes and multiple coders to enhance the reliability and generalizability of findings related to cognitive load in QSs.

\section{Conclusion}
This study introduces and evaluates a two-phase ``Organize-then-Vote" interface to help QS respondents construct their preferences. We examined how the interface affected cognitive load and response behaviors across societal issues of varying lengths through an in-lab study, NASA-TLX, and interviews. The interface's organization and voting phases, designed to reduce cognitive overload by structuring the decision-making process, allowed respondents to differentiate between options before voting. Results revealed that the two-phase design reduced participants' edit distance between vote adjustments throughout the survey and they spent more time per option. Qualitative insights highlighted that the two-phase interface encouraged more iterative and reflective preference construction and its potential for reducing satisficing behaviors even though it did not clearly reduce the overall cognitive load for the longer QS. Nonetheless, this design shift promoted deeper engagement and strategic thinking compared to the text-based interface, by distributing cognitive effort more effectively. By integrating the organization and drag-and-drop functions, the interface facilitated both preference differentiation and consolidation, making it easier for respondents to refine their decisions. This two-phase interface design supports the development of future software tools that facilitate preference construction and promote the broader adoption of QSs. Future research should explore how to better support individuals' budget allocation and design interfaces for smaller devices.

\begin{acks}
    We thank the voluntary participants who participated in the pretest, pilot, and the study. We thank all members of the Social Spaces Group and the Crowd Dynamics Lab for their support and early feedback. Special thanks to Yi-Ting Kuo, Hsin-Ni Yu, Yun-Shan Sam Yang, Katherine Chou, Chieh-Yun Chen, Prof. Brian Bailey, James Eschrich, Aditya Karan, Pranay Midha and the anonymous reviewers who provided valuable feedback to this work. This work is partially funded by The Center for Just Infrastructures.
\end{acks}

% \printbibliography
\bibliography{tcheng}

% TC:ignore
\appendix

% % Interface Design Appendix
% \section{Interface Design}
\newpage
\section{Interface design process}\label{apdx:design}
In this section, we outline the design process leading to our final interface.% As mentioned in the paper, our design iteration began from existing QV applications in the wild.

\begin{figure}[H]
    \centering
    % First subfigure
    \begin{subfigure}[b]{\linewidth}
        \centering
        \includegraphics[width=0.9\linewidth]{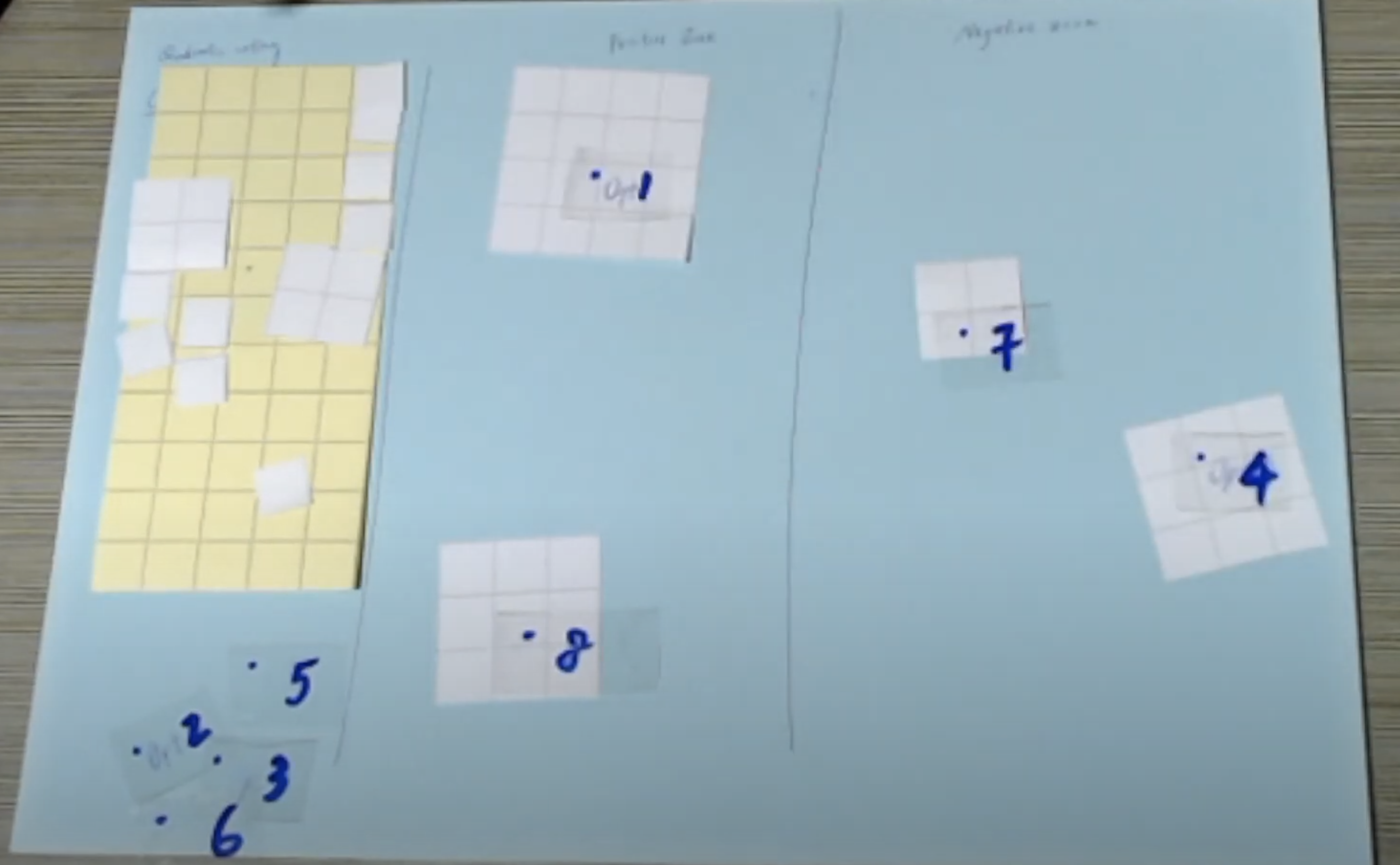}
        \caption{In this paper prototype, issues are denoted by different numbers that appear on mouseover. Pretest respondents can move options anywhere in the two sections of the interface, one denoting positive and one negative. The blocks represent the cost for each option, with no indication of the number of current votes. The credits are shown in the yellow box on the left.}
        \Description{An image of a paper prototype showing different sections for respondents to interact with. On the left side, a yellow grid contains small white squares, some of which are stacked and scattered outside the grid. The middle section labeled "Positive Zone" contains a large white square with a grid, labeled "Opt 1". The right section, labeled "Negative Zone," contains two white squares, each with grids, labeled with numbers 7 and 4. Additional small white squares labeled with numbers 2, 3, 5, 6, and 8 are positioned around the prototype. The yellow box on the left represents available credits.}
        \label{fig:horizontal_paper}
    \end{subfigure}

    \vspace{1em} % Adjusts vertical spacing between figures

    % Second subfigure
    \begin{subfigure}[b]{\linewidth}
        \centering
        \includegraphics[width=0.77\linewidth]{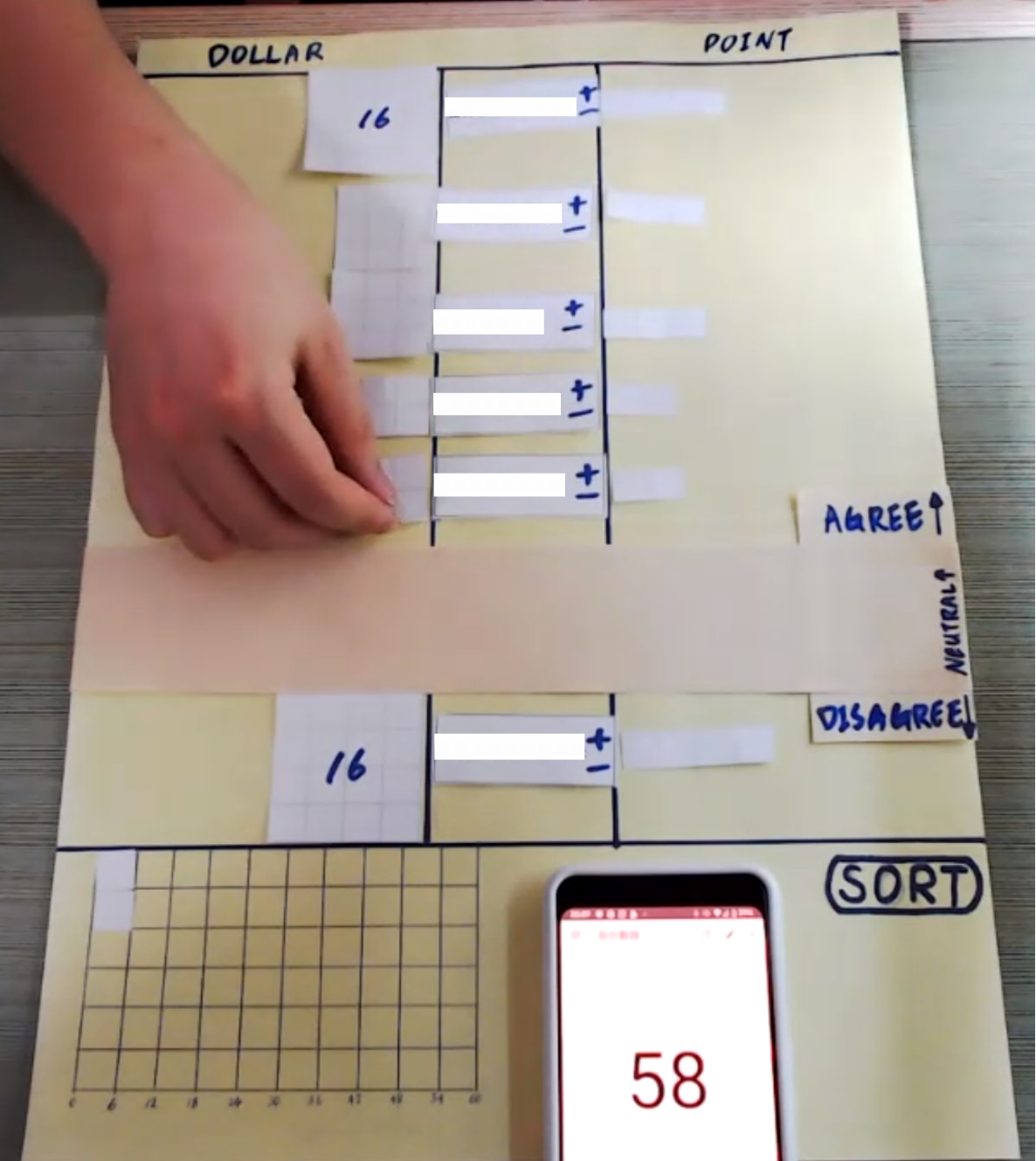}
        \caption{This paper prototype separates the positive and negative areas with a 'band' at the center. Undecided options are placed inside this band. The cost and the votes on both sides of the interface are denoted by small blocks. The budget is shown in the yellow box below the interface with a numerical counter.}
        \Description{A paper prototype interface where a person's hand is interacting with blocks. The prototype separates positive and negative areas using a wide horizontal band in the center, which holds undecided options. On the left side of the band, a column labeled "Dollar" shows a block marked with the number 16. On the right side, under a column labeled "Point," several rows have small blocks with plus and minus signs, indicating positive and negative areas. At the bottom left, a yellow box with a grid shows the available budget, marked with the number 16. A smartphone in the bottom right corner displays the number 58.}
        \label{fig:vertical_paper}
    \end{subfigure}

    \caption{Initial paper prototypes designed for QS interface.}
    \Description{This figure contains two subfigures showing two different paper prototypes.}
    \label{fig:qv_paper}
\end{figure}

\subsection{Prototype 1: Ranking-Vote}
Our first prototype emerged after various paper prototypes, such as those shown in~\Cref{fig:qv_paper}. Through pre-testing, we observed that participants engaging with QS needed interface support for organizing options and managing their credits. In this study, we decided to focus on the former.

\begin{figure}[H]
    \centering
    \includegraphics[width=0.43\textwidth, trim={7 7 7 8}, clip]{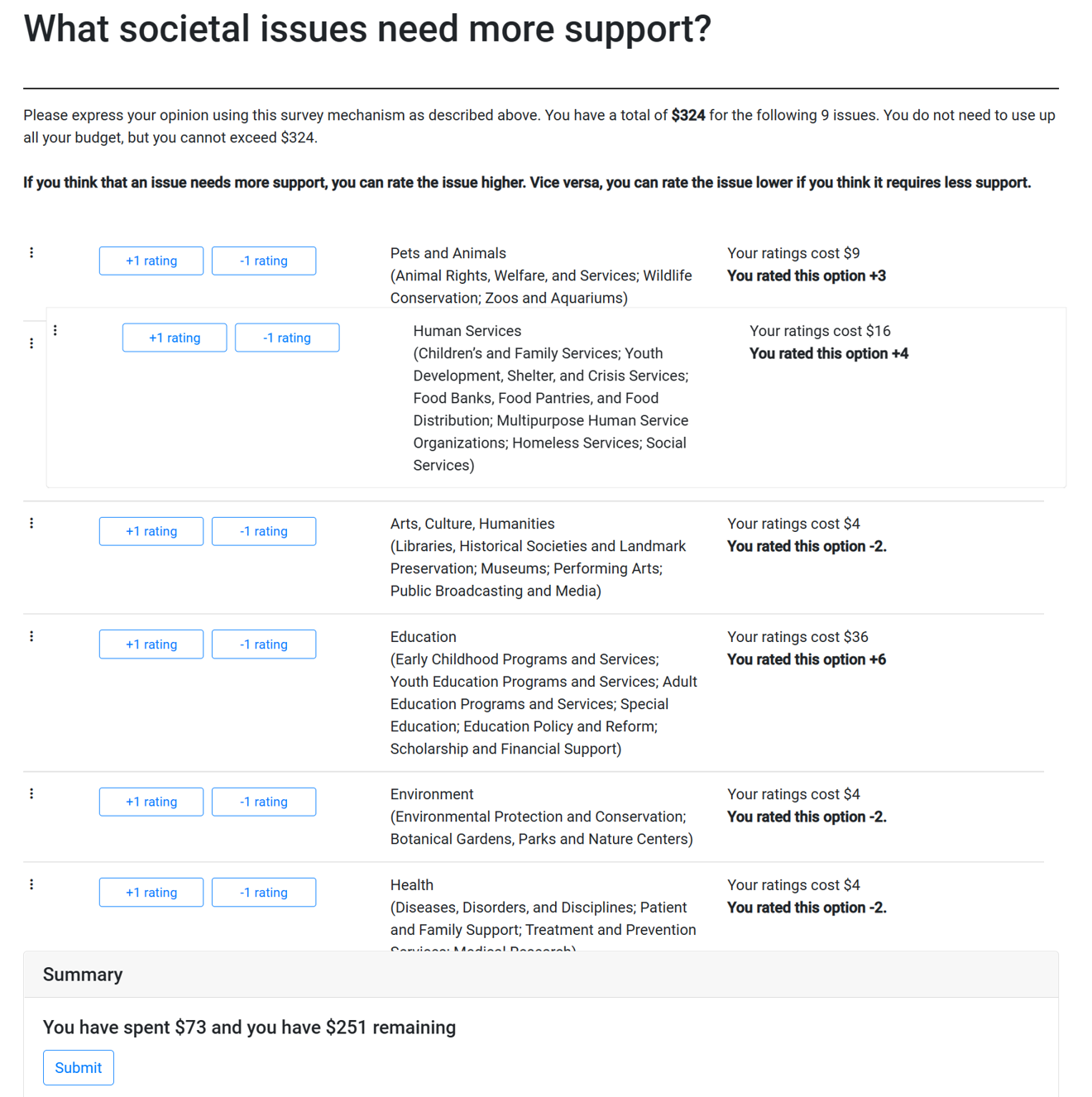}
    \caption{A Ranking-Vote Prototype: This prototype tests whether ranking options prior to voting helps establish an individual's relative preferences. Each option is draggable, allowing users to position it within the full list of options. Votes can be adjusted using the buttons on the left side of the interface, while the vote count and costs are displayed on the right. A summary box remains fixed at the bottom of the screen for easy reference.}
    \Description{A web interface showing a survey titled "What societal issues need more support?" The interface presents several societal issues in a list format, each with buttons labeled "+1 rating" and "-1 rating" to adjust the ratings. For example, the issue "Pets and Animals" shows a rating cost of \$9 with a +3 rating, and "Human Services" shows a rating cost of \$16 with a +4 rating. The remaining issues include Arts, Culture, Humanities, Education, Environment, and Health, each with their respective ratings and costs. At the bottom of the page is a summary box displaying the total spent (\$73) and the remaining balance (\$251). A "Submit" button is positioned below the summary.}
    \label{fig:qv_rank}
    \vspace{-3ex}
\end{figure}

Since participants needed to position options within the interface, and the end result formed a ranked list, we tested whether ranking options before voting would help establish an individual's relative preferences in Prototype 1 (~\Cref{fig:qv_rank}). This prototype allowed respondents to reposition options before voting. However, pre-test respondents rarely moved the options and questioned the necessity of a full ranking, as it did not influence their QS submission. Additionally, many were unaware that the options were draggable. These findings suggest that a full ranking is unnecessary for establishing relative preferences. Therefore, we decided to ask respondents to select a subset of options rather than requiring a full ranking of all options.

\begin{figure*}[p]
    \centering
    \begin{subfigure}[b]{0.47\textwidth}
        \centering
        \includegraphics[width=0.95\textwidth]{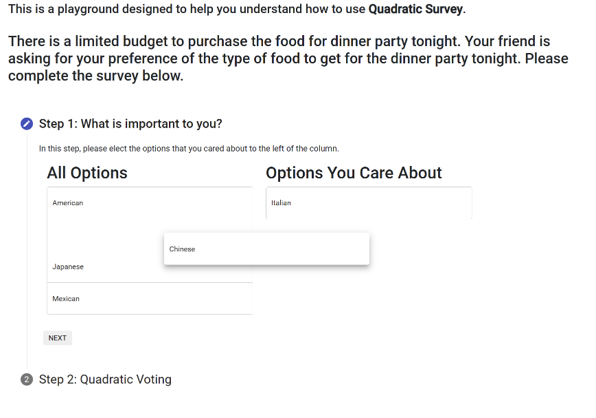}
        \caption{Options are dragged and dropped to the 'Option You Care About' box.}
        \Description{A web interface showing the first step of a quadratic voting prototype. The screen is titled "What is important to you?" On the left, under "All Options," a list of food types is displayed, including American, Japanese, and Mexican. One option, "Chinese," is being dragged to the right column labeled "Options You Care About," which already includes "Italian." A "Next" button appears at the bottom of the interface.}

        \label{fig:qv_select_selection}
    \end{subfigure}
    \hfill
    \begin{subfigure}[b]{0.47\textwidth}
        \centering
        \includegraphics[width=0.9\textwidth]{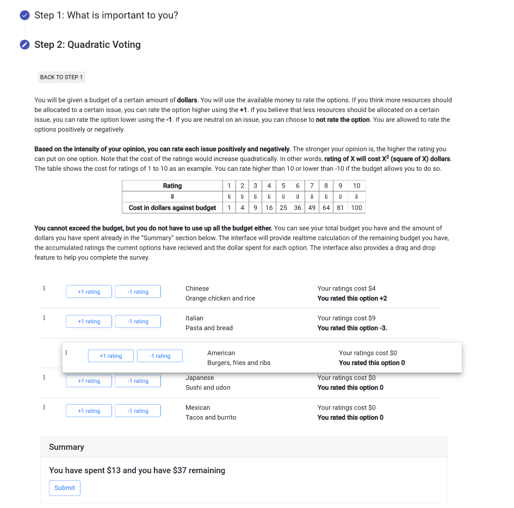}
        \caption{The previous step collapses showing all voting options.}
        \Description{The second step of a quadratic voting prototype showing a list of food options for voting. The voting interface displays food items such as Chinese, Pasta and bread, and American. Next to each item are "+1 rating" and "-1 rating" buttons for adjusting votes. Each option also shows the cost of votes, and a summary box at the bottom displays the amount spent (\$13) and the remaining balance (\$37).}

        \label{fig:qv_select_vote}
    \end{subfigure}
    \caption{A Select-then-Vote Prototype: The goal of this prototype is to nudge participants to focus on a subset of options to vote, rather than ranking all of them. This prototype introduces a two-step voting process. As shown in Fig.~\ref{fig:qv_select_selection}, the first step involves selecting options for further consideration. Important options are placed at the top of the list for voting shown in Fig.~\ref{fig:qv_select_vote}, but options can be placed anywhere on the list if desired. The rest of the controls follows the previous prototype.}
    \Description{A two-step quadratic voting prototype interface. In the first step (Subfigure 1), users drag and drop food options from a list on the left, such as American and Mexican, to the "Options You Care About" box on the right. In the second step (Subfigure 2), users vote on their selected options by adjusting the ratings using "+1 rating" and "-1 rating" buttons. A summary at the bottom shows the total amount spent and remaining balance.}

    \label{fig:qv_select}
\end{figure*}

\begin{figure*}[p]
    \centering
    \begin{subfigure}[b]{0.45\textwidth}
        \centering
        \includegraphics[width=\textwidth]{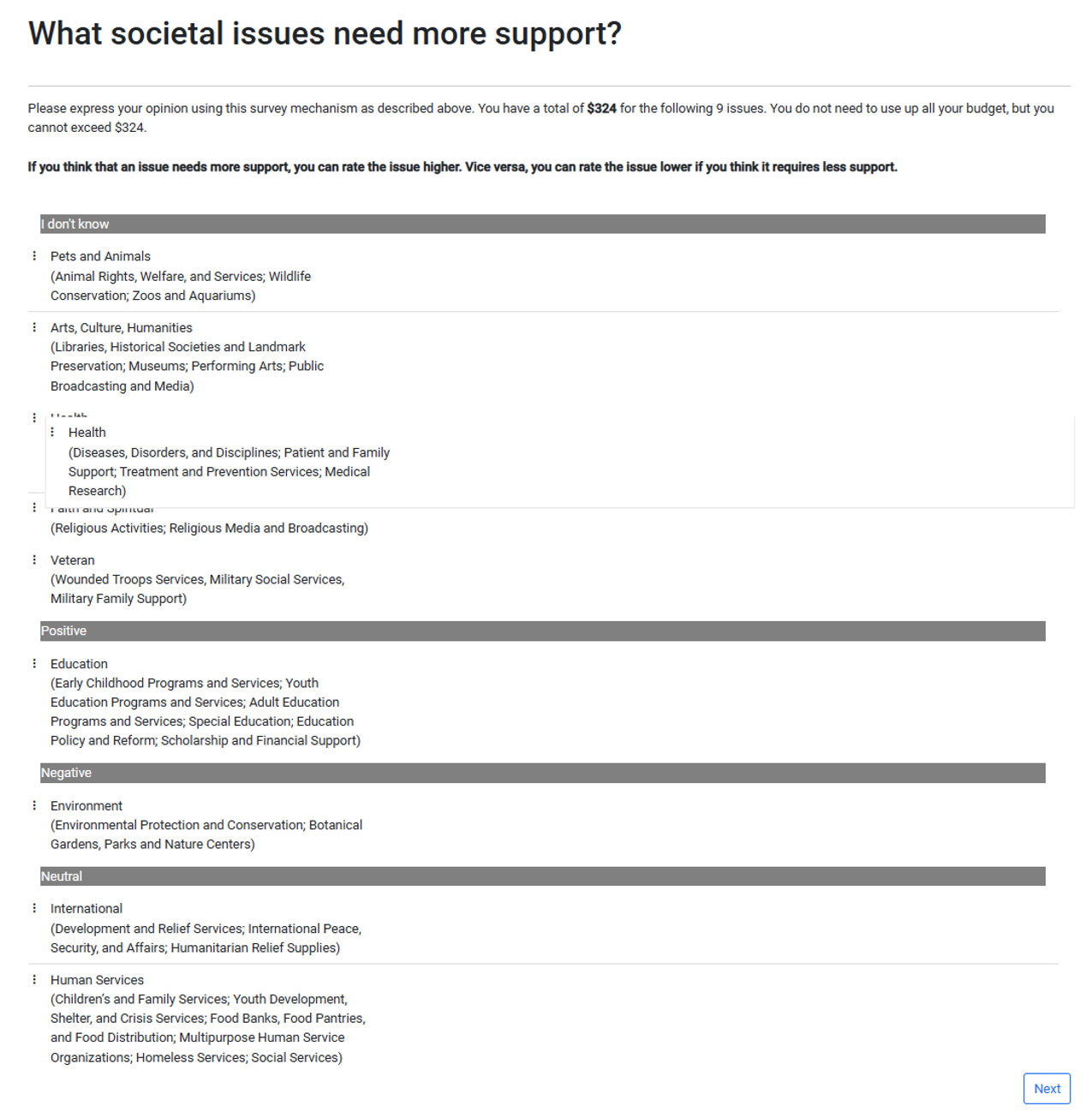}
        \caption{The Organization Interface: Options are shown initially in the first bin labeled as `I don't know.' Survey respondents can then drag and drop these options into the latter bins: Lean Positive, Lean Neutral, or Lean Negative. Only the details of each option are shown on this interface.}
        \Description{A web interface displaying a survey titled "What societal issues need more support?" Initially, all options are placed in the first bin labeled "I don't know." The listed options include Pets and Animals, Arts, Culture, Humanities, Health, Veterans, and others. Each option is accompanied by a brief description. Below the "I don't know" bin are three other bins: "Positive," "Negative," and "Neutral." Survey respondents can drag and drop options into these bins based on their preferences. A "Next" button is located at the bottom right corner of the interface.}
        \label{fig:qv_org_p1}
    \end{subfigure}
    \hfill
    \begin{subfigure}[b]{0.45\textwidth}
        \centering
        \includegraphics[width=\textwidth]{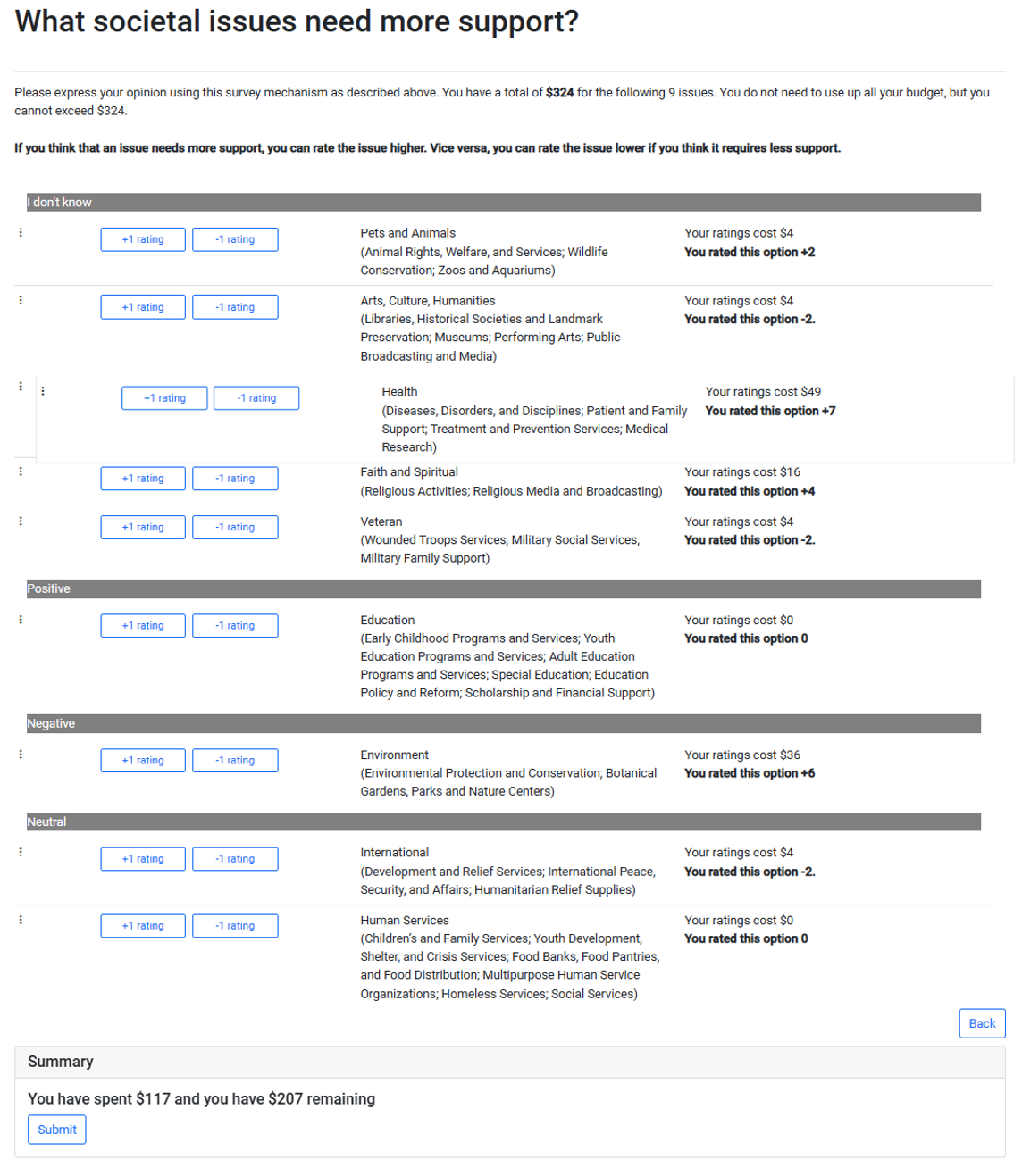}
        \caption{The Voting Interface: Voting controls appear on the left side of each option, showing the current votes and associated costs on the right. A budget summary sticks to the bottom of the screen.}
        \Description{A web interface displaying a survey titled "What societal issues need more support?" Each option in the list has voting controls, with "+1 rating" and "-1 rating" buttons appearing on the left side of each issue. The issues are grouped into categories such as "I don't know," "Positive," "Negative," and "Neutral." To the right of each issue, the cost of the rating is shown along with the number of votes cast. For example, "Pets and Animals" has a rating cost of \$4 with a +2 rating, while "Health" has a rating cost of \$49 with a +7 rating. At the bottom of the page, a summary box shows the total amount spent (\$117) and the remaining balance (\$207), with a "Submit" button below it.}

        \label{fig:qv_org_p2}
    \end{subfigure}
    \caption{Organize-then-Vote Prototype: The goal of this prototype is to encourage participants to derive finer grain categories among options before voting. Survey respondents first organize their thoughts into categories and then vote on the options.}
    \Description{The figure shows an Organize-then-Vote prototype with two main steps. In the first step, the left figure, users organize options by dragging and dropping them into different categories: "I don't know," "Positive," "Neutral," or "Negative." In the second step, the right figure, users vote on these organized options using "+1 rating" and "-1 rating" buttons, with voting costs and current ratings displayed. A summary section shows the total spent and remaining budget.}
    \label{fig:qv_org}
\end{figure*}

\subsection{Prototype 2: Select-then-Vote}
Based on feedback from Prototype 1, instead of \textit{allowing} individuals to rank options, Prototype 2 implemented a two-phase process that \textit{intentionally} asks respondents to select options to express opinions before voting. 

As shown in Figure~\ref{fig:qv_select}, survey respondents selected their preferred options (Figure~\ref{fig:qv_select_selection}), and the interface positioned these options at the top of the list for voting (Figure~\ref{fig:qv_select_vote}). We identified several issues during the prototype 2 pretest: many respondents marked most options as 'options they care about,' which undermined the design's purpose. Additionally, the lack of clear distinction between selected and unselected options confused respondents about the necessity of Step 1. Thus, we need a clearer distinction and connection between the two phases to effectively construct relative preferences.

\subsection{Prototype 3: Organize-then-Vote}
Figure~\ref{fig:qv_org} shows the final prototype, which builds on our previous takeaway by introducing finer-grained groupings and establishing a clearer connection between option organization and voting position. Specifically, we provided three categories: Lean Positive, Lean Negative, and Lean Neutral. Initially, respondents see all options listed under a section labeled 'I don't know,' which displays only the option descriptions and not the vote controls. They are then asked to move these options into one of the three categories. On the subsequent page, voting controls and additional information appear for each option, reinforcing the connection between option grouping, position, and voting controls.

Feedback indicated that survey respondents are comfortable with the two-phase organize-then-vote design, demonstrating it as a central strategy for our interface development. However, we identified several areas for enhancement: First, the dragging and dropping mechanism in the organization phase is cumbersome and may inadvertently suggest a ranking process, contrary to our intentions. Second, placing unorganized options at the top of the voting list is counterintuitive. Third, the voting controls are disconnected from the option summaries, dividing attention between the left and right sides of the screen. These insights guided refinements in the final two-phase interface, adhering to the organize-then-vote framework.
 % Done
\section{Voting Interface Breakdown}\label{apdx:relatedVoting}
In this section, we outline additional literature that informed this study. There are two sets of literature that we surveyed: Survey response format and voting interfaces.

\subsection{Survey response format}
Research in the marketing and research communities focusing on survey and questionnaire design, usability, and interactions examines the influence of presentation styles and `response format.'~\citet{weijtersExtremityHorizontalVertical2021} demonstrated that horizontal distances between options are more influential than vertical distances, with the latter recommended for reduced bias. Slider bars, which operate on a drag-and-drop principle, show lower mean scores and higher nonresponse rates compared to buttons, indicating they are more prone to bias and difficult to use. In contrast, visual analog scales that operate on a point-and-click principle perform better~\cite{toepoelSlidersVisualAnalogue2018}. These studies show how even small design changes can have a large impact on usability, highlighting the importance of designing interfaces that prioritize human-centered interaction rather than focusing solely on functionality.

\subsection{Voting Interfaces}
Compared to digital survey interfaces, voting interfaces are a specialized type of survey interface can significantly influence democratic processes~\cite{engstrom2020politics, chisnellDemocracyDesignProblem2016, civicdesignDesigningUsableBallots2015} and often have consequential impacts. We categorize these related works into three main categories detailed below:

\paragraph{Designs that shifted voter decisions: } For example, states without straight-party ticket voting~(where voters can select all candidates from one party through a single choice) exhibited higher rates of split-ticket voting~\cite{engstrom2020politics}. Another example from the Australian ballot showing incumbency advantages is where candidates are listed by the office they are running for, with no party labels or boxes.

\paragraph{Designs that influenced errors: } Butterfly ballots, an atypical design, may have influenced the outcome of the 2000 U.S. Presidential Election~\cite{wandButterflyDidIt2001}. It increased voter errors because voters could not correctly identify the punch hole on the ballot. Splitting contestants across columns increases the chance for voters to overvote~\cite{quesenberyOpinionGoodDesign2020}. On the other hand, \citet{everettElectronicVotingMachines2008} showed the use of incorporating physical voting behaviors, like lever voting, into graphical user interfaces increased satisfaction while maintaining efficiency and effectiveness.

\paragraph{Designs that incorporated technologies: } Other projects like the Caltech-MIT Voting Technology Project addresses accessibility challenges, resulting in innovations like EZ Ballot~\cite{leeUniversalDesignBallot2016}, Anywhere Ballot~\cite{summers2014making}, and Prime III~\cite{dawkinsPrimeIIIInnovative2009}. In addition,~\citet{gilbertAnomalyDetectionElectronic2013} investigated optimal touchpoints on voting interfaces, and~\citet{conradElectronicVotingEliminates2009} examined zoomable voting interfaces.

Response format literature and voting interfaces informed how interfaces significantly influence respondent behavior, decision accuracy, and cognitive load. These burdens are especially problematic for complex systems like QS, where high cognitive demands may deter researchers and users alike. Developing effective, human-centered interfaces for QS could enhance usability, reduce cognitive overload, and increase adoption in both research and practical applications.
 % Done
% \newpage

% % Experimental Setup Appendix
% \section{Experimental Setup}
\section{List of Options}
\label{sec:charityList}
We provide the full list of options presented on the survey.

\begin{itemize}[leftmargin=*]
    \item \textbf{Animal Rights, Welfare, and Services:} Protect animals from cruelty, exploitation and other abuses, provide veterinary services and train guide dogs.
    \item \textbf{Wildlife Conservation:} Protect wildlife habitats, including fish, wildlife, and bird refuges and sanctuaries.
    \item \textbf{Zoos and Aquariums:} Support and invest in zoos, aquariums and zoological societies in communities throughout the country.
    \item \textbf{Libraries, Historical Societies and Landmark Preservation:} Support and invest public and specialized libraries, historical societies, historical preservation programs, and historical estates.
    \item \textbf{Museums:} Support and invest in maintaining collections and provide training to practitioners in traditional arts, science, technology, and natural history.
    \item \textbf{Performing Arts:} Support symphonies, orchestras, and other musical groups; ballets and operas; theater groups; arts festivals; and performance halls and cultural centers.
    \item \textbf{Public Broadcasting and Media:} Support public television and radio stations and networks, as well as providing other independent media and communications services to the public.
    \item \textbf{Community Foundations:} Promote giving by managing long-term donor-advised charitable funds for individual givers and distributing those funds to community-based charities over time.
    \item \textbf{Housing and Neighborhood Development:} Lead and finance development projects that invest in and improve communities by providing utility assistance, small business support programs, and other revitalization projects.
    \item \textbf{Jewish Federations:} Focus on a specific geographic region and primarily support Jewish-oriented programs, organizations and activities through grantmaking efforts
    \item \textbf{United Ways:} Identify and resolve community issues through partnerships with schools, government agencies, businesses, and others, with a focus on education, income and health.
    \item \textbf{Adult Education Programs and Services:} Provide opportunities for adults to expand their knowledge in a particular field or discipline, learn English as a second language, or complete their high school education.
    \item \textbf{Early Childhood Programs and Services:} Provide foundation-level learning and literacy for children prior to entering the formal school setting.
    \item \textbf{Education Policy and Reform:} Promote and provide research, policy, and reform of the management of educational institutions, educational systems, and education policy.
    \item \textbf{Scholarship and Financial Support:} Support and enable students to obtain the financial assistance they require to meet their educational and living expenses while in school.
    \item \textbf{Special Education:} Provide services, including placement, programming, instruction, and support for gifted children and youth or those with disabilities requiring modified curricula, teaching methods, or materials.
    \item \textbf{Youth Education Programs and Services:} Provide programming, classroom instruction, and support for school-aged students in various disciplines such as art education, STEM, outward bound learning experiences, and other programs that enhance formal education.
    \item \textbf{Botanical Gardens, Parks, and Nature Centers:} Promote preservation and appreciation of the environment, as well as leading anti-litter, tree planting and other environmental beautification campaigns.
    \item \textbf{Environmental Protection and Conservation:} Develop strategies to combat pollution, promote conservation and sustainable management of land, water, and energy resources, protect land, and improve the efficiency of energy and waste material usage.
    \item \textbf{Diseases, Disorders, and Disciplines:} Seek cures for diseases and disorders or promote specific medical disciplines by providing direct services, advocating for public support and understanding, and supporting targeted medical research.
    \item \textbf{Medical Research:} Devote and invest in efforts on researching causes and cures of disease and developing new treatments.
    \item \textbf{Patient and Family Support:} Support programs and services for family members and patients that are diagnosed with a serious illness, including wish granting programs, camping programs, housing or travel assistance.
    \item \textbf{Treatment and Prevention Services:} Provide direct medical services and educate the public on ways to prevent diseases and reduce health risks.
    \item \textbf{Advocacy and Education:} Support social justice through legal advocacy, social action, and supporting laws and measures that promote reform and protect civil rights, including election reform and tolerance among diverse groups.
    \item \textbf{Development and Relief Services:} Provide medical care and other human services as well as economic, educational, and agricultural development services to people around the world.
    \item \textbf{Humanitarian Relief Supplies:} Specialize in collecting donated medical, food, agriculture, and other supplies and distributing them overseas to those in need.
    \item \textbf{International Peace, Security, and Affairs:} Promote peace and security, cultural and student exchange programs, improve relations between particular countries, provide foreign policy research and advocacy, and United Nations-related organizations.
    \item \textbf{Religious Activities:} Support and promote various faiths.
    \item \textbf{Religious Media and Broadcasting:} Support organizations of all faiths that produce and distribute religious programming, literature, and other communications.
    \item \textbf{Non-Medical Science \& Technology Research:} Support research and services in a variety of scientific disciplines, advancing knowledge and understanding of areas such as energy efficiency, environmental and trade policies, and agricultural sustainability.
    \item \textbf{Social and Public Policy Research:} Support economic and social issues impacting our country today, educate the public, and influence policy regarding healthcare, employment rights, taxation, and other civic ventures.
\end{itemize} % Done
\section{Demographic Breakdown}
\label{sec:apdx:demo}
\Cref{tab:age_gender_distribution} provides a detailed demographic breakdown per group.
\begin{table*}[h!]
\centering
\caption{Participant Age and Gender Distribution by Experimental Condition}
\label{tab:age_gender_distribution}
\begin{tabular}{lcccccccccc}
\hline
\textbf{Condition} & \textbf{Mean Age} & \textbf{SD} & \textbf{Range} & \textbf{25th} & \textbf{Median} & \textbf{75th} & \textbf{Male} & \textbf{Female} & \textbf{Non-binary} \\
\hline
Short Text      & 31.6  & 13.7 & 18--67 & 23.8 & 29.5 & 32.8 & 4 & 6 & 0 \\
Short Two-Phase   & 32.1  & 14.0 & 18--52 & 20.3 & 27.0 & 44.5 & 4 & 6 & 0 \\
Long Text       & 36.0  & 14.8 & 21--61 & 24.0 & 33.0 & 42.8 & 2 & 7 & 1 \\
Long Two-Phase    & 38.8  & 19.6 & 19--71 & 25.0 & 28.5 & 53.0 & 2 & 8 & 0 \\
\hline
\end{tabular}
\Description{ A table summarizing participant demographics and descriptive statistics for four conditions: Short Text, Short 2-Phase, Long Text, and Long 2-Phase.}
\end{table*}
 % Done

% % Results Appendix
% \section{Results}
\begin{figure*}[t!]
    \centering
    \begin{minipage}[t]{0.32\textwidth}
        \centering
        \includegraphics[width=\textwidth, trim=0 13 0 13, clip]{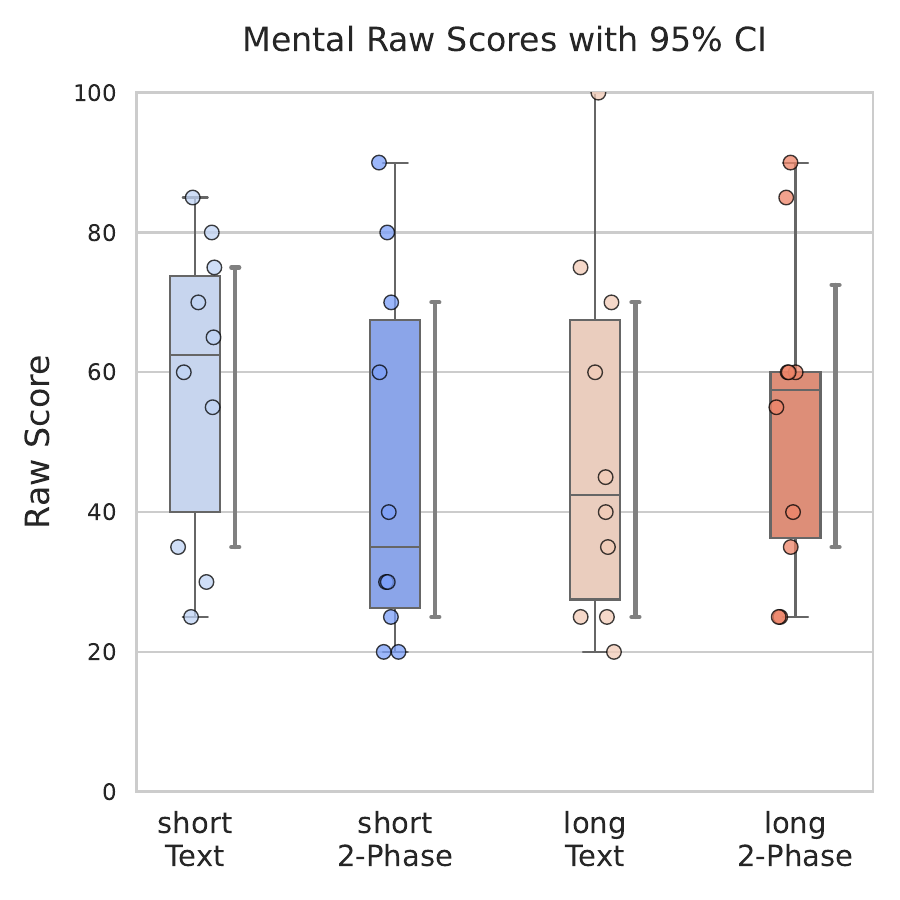}
        \caption{Mental Demand Raw Score: Across all four experiment groups, participants' reported mental demand is spread across a wide range with many participants experiencing high mental demand.}
        \Description{Box plot showing mental raw scores with 95\% confidence intervals across four interface versions: Short Text, Short 2-Phase, Long Text, and Long 2-Phase. The y-axis represents raw scores from 0 to 100. Each box plot includes individual data points. The Short Text and Short 2-Phase versions display wider score distributions, with medians around 60 and 40. The Long Text and Long 2-Phase versions have similar distributions but with slightly lower medians, 40 and 60, respectively. The plot shows a considerable spread in scores, with overlapping confidence intervals, indicating variability in mental demand across all groups.}
        \label{fig:mental_cog_score}
    \end{minipage}%
    \hfill
    \begin{minipage}[t]{0.32\textwidth}
        \centering
        \includegraphics[width=\textwidth, trim=0 13 0 13, clip]{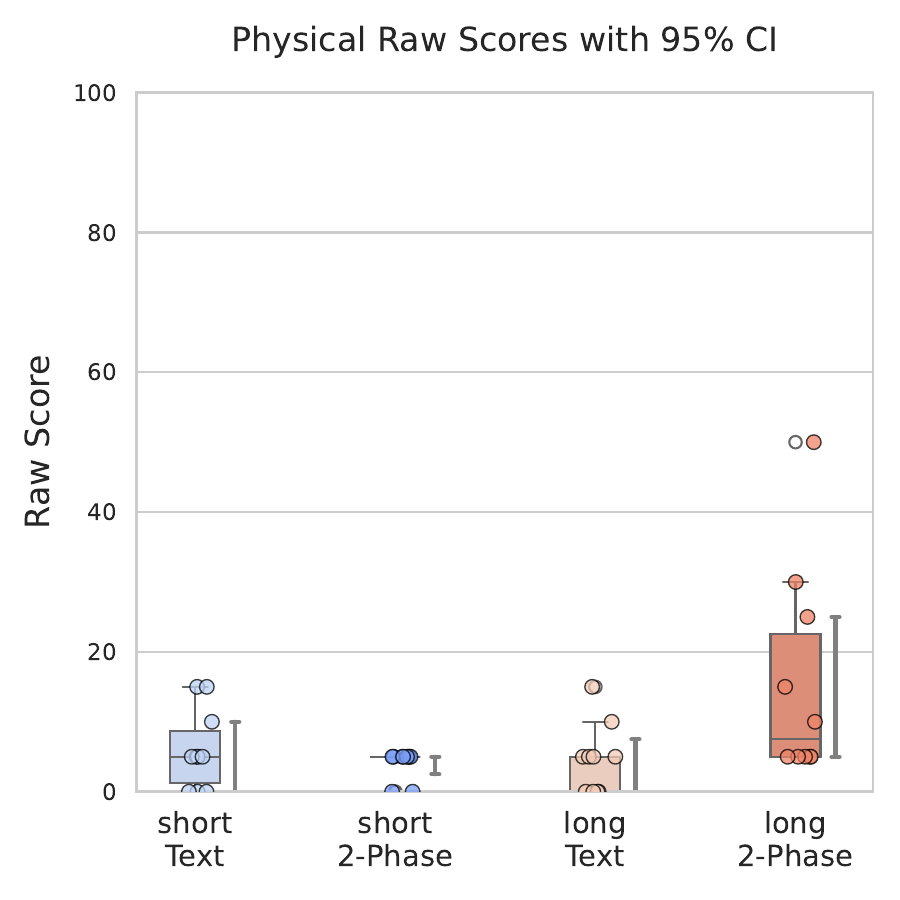}
        \caption{Physical Demand Raw Score: Participants other than the long two-phase interface reported minimal physical demand. The long two-phase interface had the highest physical demand, likely due to increased mouse clicks and extended time spent looking at the vertical screen.}
        \Description{A box plot showing the distribution of physical raw scores across four interface versions: Short Text, Short 2-Phase, Long Text, and Long 2-Phase. The y-axis represents the raw score ranging from 0 to 100. The box plots include individual data points, a central line for the median, and whiskers indicating the 95\% confidence interval. The Short Text, Short 2-Phase, and Long Text interfaces show minimal physical demand, with scores clustered below 20. The Long 2-Phase interface exhibits higher physical demand, with a few scores scattered up to around 60.}
        \label{apdxfig:physical_cog_score}
    \end{minipage}%
    \hfill
    \begin{minipage}[t]{0.32\textwidth}
        \centering
        \includegraphics[width=\textwidth, trim=0 13 0 13, clip]{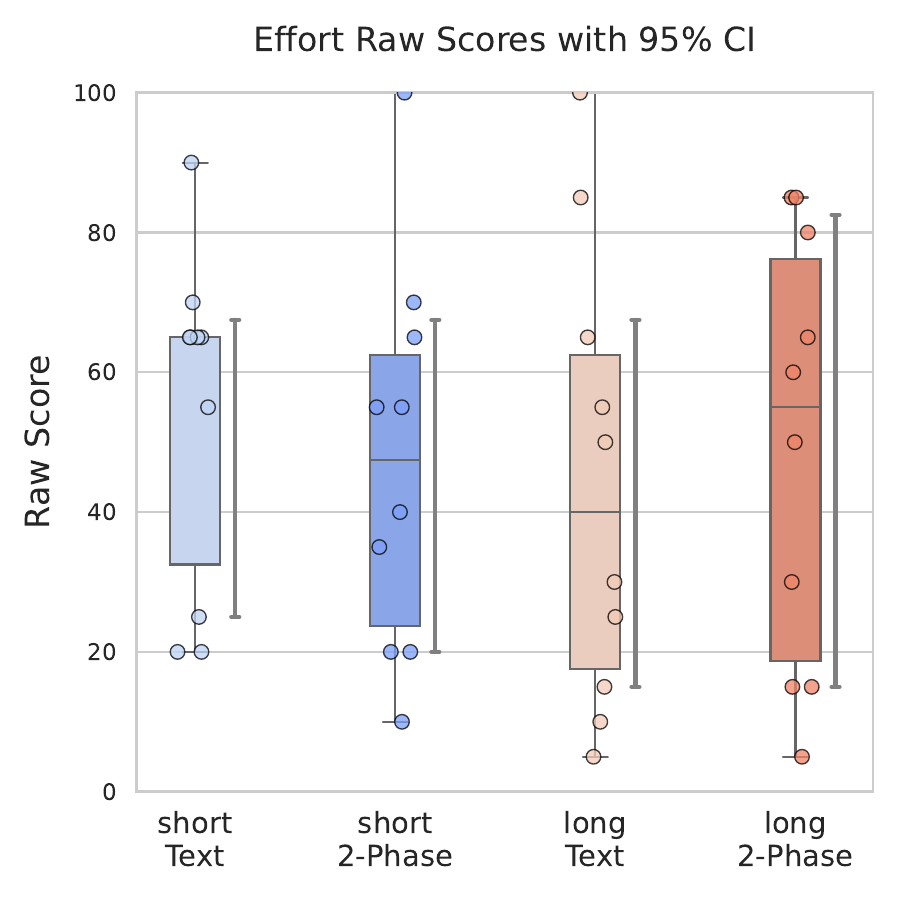}
        \caption{Effort Raw Score: Effort scores show indifference across groups. All groups had high variance of responses indicating some participants requires high amount of effort when completing QS regardless of length and interface}
        \Description{A box plot showing the distribution of effort raw scores across four interface versions: Short Text, Short 2-Phase, Long Text, and Long 2-Phase. The y-axis represents the raw score ranging from 0 to 100. Each box plot includes individual data points, a central line indicating the median, and whiskers representing the 95\% confidence interval. The Long 2-Phase interface shows the widest range of effort scores, while the other interfaces display more compact distributions. Data points are scattered within and outside the whiskers, reflecting variability in effort scores across the groups.}
        \label{apdxfig:effort_cog_score}
    \end{minipage}
\end{figure*}

\section{Detailed Qualitative Cognitive Load Breakdown}
\label{apdx:cog_qual}

We provide additional details on the six cognitive dimensions. Among all dimensions, we also provide the codes representing different types of demand in a table form. The shaded cells represent the percentage of participants citing each source of mental demand, allowing for comparison within columns. The abbreviations in the columns: ST (Short Text Interface), S2P (Short Two-phase Interface), LT (Long Text Interface), and L2P (Long Two-phase Interface). Short and Long refer to the sum across both interfaces; Text and 2P (Two-phase interface) refer to the sum across both survey lengths. We include Sparklines for comparisons across these experiment groups. Future studies can use these as initial codebooks to conduct interface studies on preference construction.

\begin{table*}[p]
   \caption{This table lists all the causes participants mentioned as contributing to their Mental Demand.}
   \Description{A table presenting mental demand across categories such as Budget Management (track credits, maximize usage), Preference Construction (prioritization, resource allocation), and Demand from Experiment Setup. Sparklines and percentage bars are used to visually represent the data across four interface versions (ST, S2P, LT, L2P) and experiment conditions (Short, Long, Text, Inter). The bars show higher mental demand in areas like Budget Management and Preference Construction, with trends highlighted by sparklines. Additional categories, such as External Factors and Demand due to Interface, are also represented with percentage bars.}
    \label{tbl:mental}
    \includegraphics[width=0.85\linewidth]{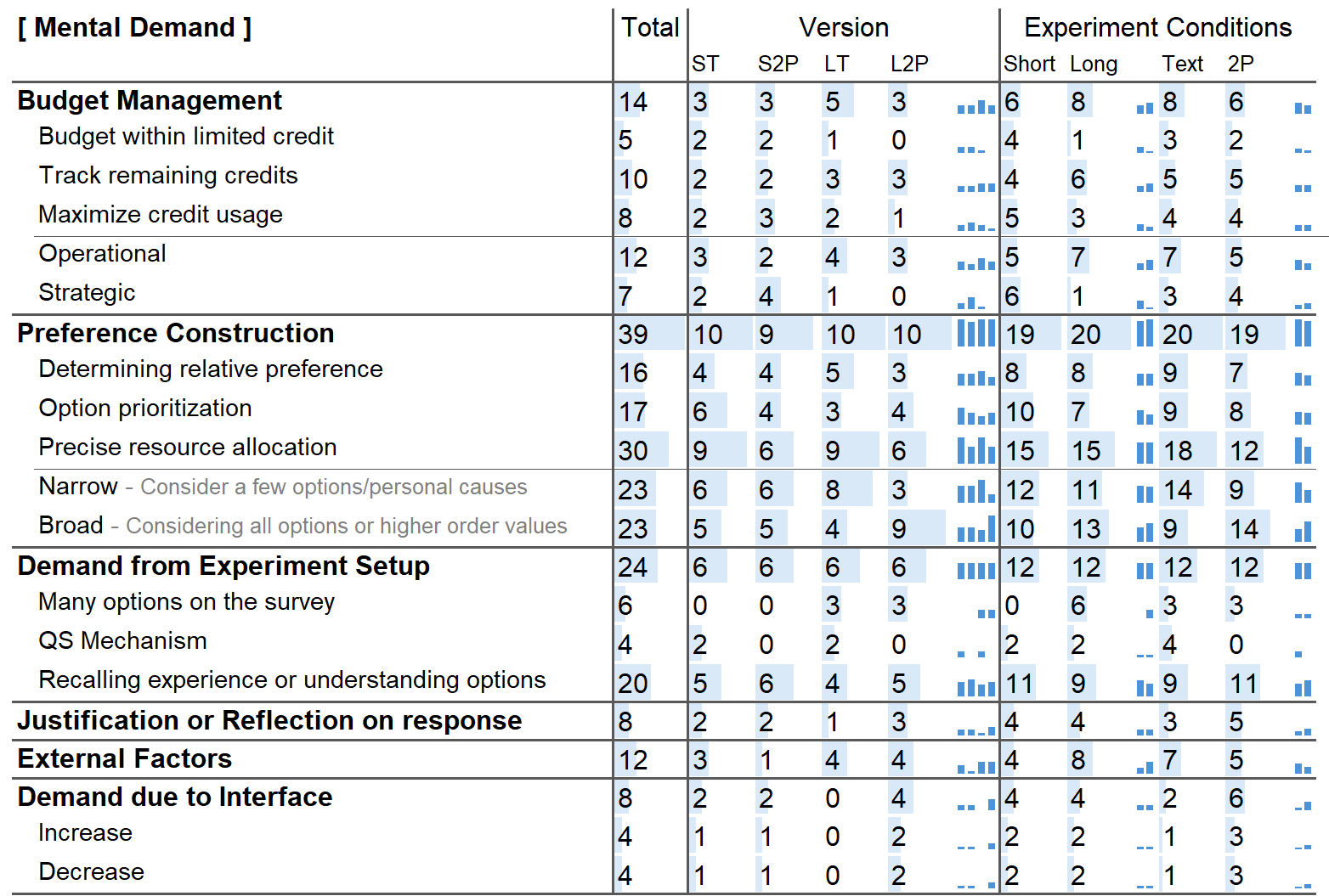}
\end{table*}

\begin{table*}[p]
    \caption{Physical Demand Causes: Most participants expressed little or no physical demand. Results reflected that participants in the long two-phase interface required more actions, hence the higher mention of mouse usage as a source.}
    \Description{A table showing physical challenges experienced by participants, including Reading, Mouse, Vertical Screen, and None/Little physical effort, with sparklines and percentage bars visualizing the data trends. Data is split across four interface versions (ST, S2P, LT, L2P) and experiment conditions (Short, Long, Text, Inter). The Mouse category has the highest counts, with trends clearly visible via sparklines and bars, while Reading and Vertical Screen challenges have lower values. The None/Little row shows participants reporting minimal physical effort with percentage bars illustrating the distribution.}
    \label{apdx:physical_table}
    \includegraphics[width=0.85\linewidth]{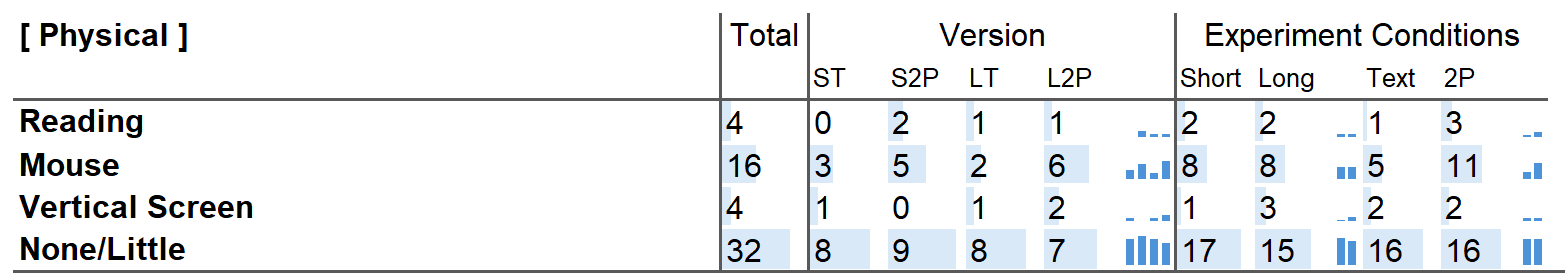}
\end{table*}

\begin{table*}[p]
    \caption{Effort Sources: Participants using the text interface focused more on operational tasks, while those using the two-phase interface focused more on strategic planning.}
    \Description{A table summarizing participant effort as Operational and Strategic (personal and global), with data visualized using sparklines and percentage bars. The table is split into four versions (ST, S2P, LT, L2P) and experiment conditions (Short, Long, Text, Inter), with counts and corresponding bars for each category. The operational effort shows participants managing tasks, while the strategic effort captures balancing personal preferences with societal concerns. Sparklines highlight trends across different conditions. A None/Little/A bit row shows participants exerting minimal effort, visualized with bars.}

    \label{apdx:effort_table}
    \includegraphics[width=0.85\linewidth]{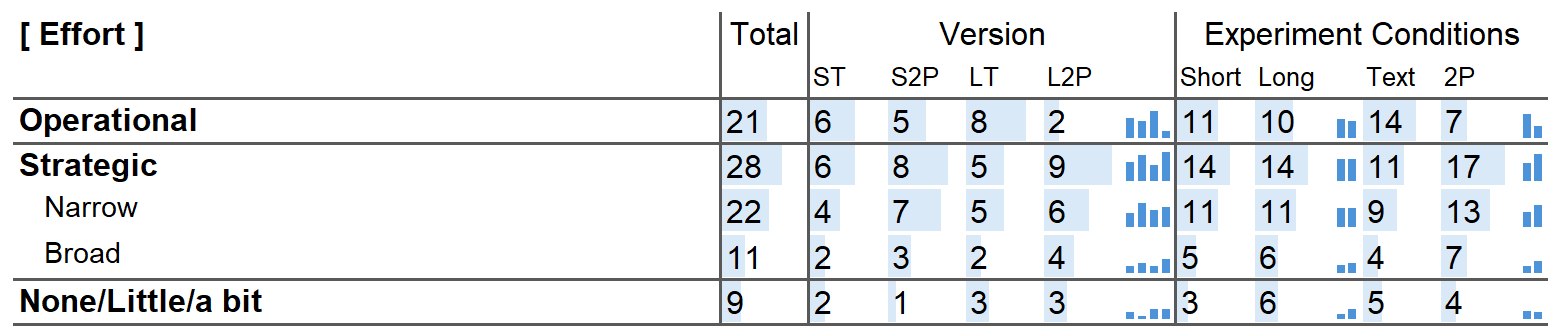}
\end{table*}

\subsection{Sources of Mental Demand}
\label{apdx:mental}
Mental demand refers to the amount of mental and perceptual activity required to complete a task.~\Cref{tbl:mental} lists all qualitative codes and Figure~\ref{fig:mental_cog_score} shows the boxplot of participant's subscale response. For thematic groups, we grouped them as source of demand (e.g., tracking remaining credits) and also of scope (e.g., Operational) as separated by the light gray line within each row.

\subsection{Sources of Physical Demand} 
\label{apdx:physical}
Physical demand refers to the physical effort required to complete a task, such as physical exertion or movement. Most participants reported minimal physical demand~($N=32$), reflected in the low NASA-TLX physical demand scores~(Figure~\ref{apdxfig:physical_cog_score}). Notably, $11$ out of $20$ participants who used the two-phase interface mentioned physical demand from using the mouse, reflecting interacting with two interfaces. This is further supported by the raw NASA-TLX physical demand scores~(Figure~\ref{apdxfig:physical_cog_score}), which show a significant visual difference between short and long two-phase interfaces as well as between text and two-phase interfaces in long surveys. Table~\ref{apdx:physical_table} presents all the relevant codes across experiment groups.

\begin{table*}[p]
    \caption{Performance Causes: Most causes are shared across experiment conditions. We provided qualitative interpretations of their own performance assessments.}
    \Description{A table detailing participant performance across Operational Action (budget control, preference reflection, limited resources), Social Responsibility (decision maker, outcome uncertainty), and Performance Assessment (did their best, feel good, good enough). The table includes sparklines and percentage bars to visualize the distribution of performance data across four interface versions (ST, S2P, LT, L2P) and experiment conditions (Short, Long, Text, Inter). Operational action and social responsibility are visually represented, with sparklines highlighting performance trends.}

    \label{tbl:perf}
    \includegraphics[width=0.85\linewidth]{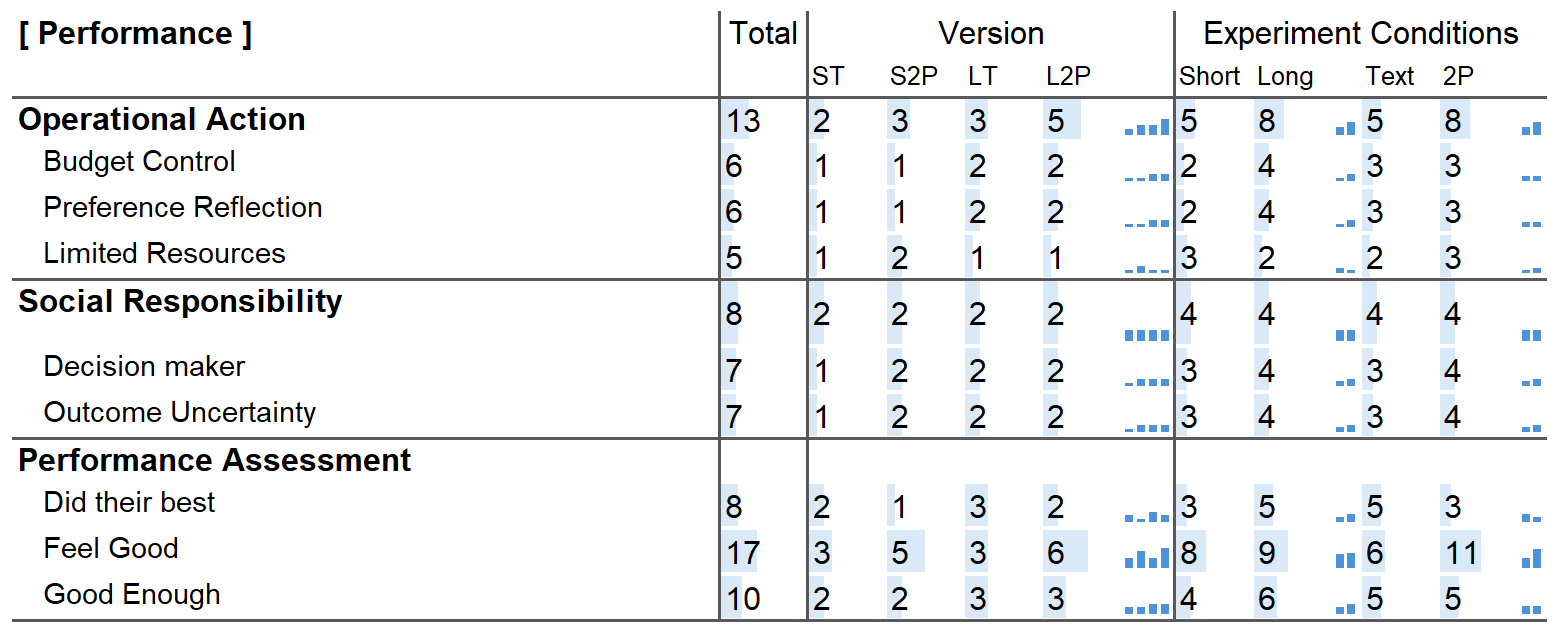}
\end{table*}
\begin{table*}[p]
    \caption{Temporal Demand Sources: Decision-making and Operational Tasks are the main causes. Participants framed their decision-making sources differently.}
    \Description{A table categorizing temporal challenges across Budget Management, Decision Making (affirmative and negative), and Operational (task completion, efficiency). The data includes sparklines and percentage bars to visualize patterns across four versions (ST, S2P, LT, L2P) and experiment conditions (Short, Long, Text, Inter). Affirmative decision-making shows higher values than negative decision-making, with percentage bars indicating the relative distribution. Temporal operational tasks show consistent effort across conditions, as reflected in the sparklines.}

    \label{tbl:temporal}
    \includegraphics[width=0.85\linewidth]{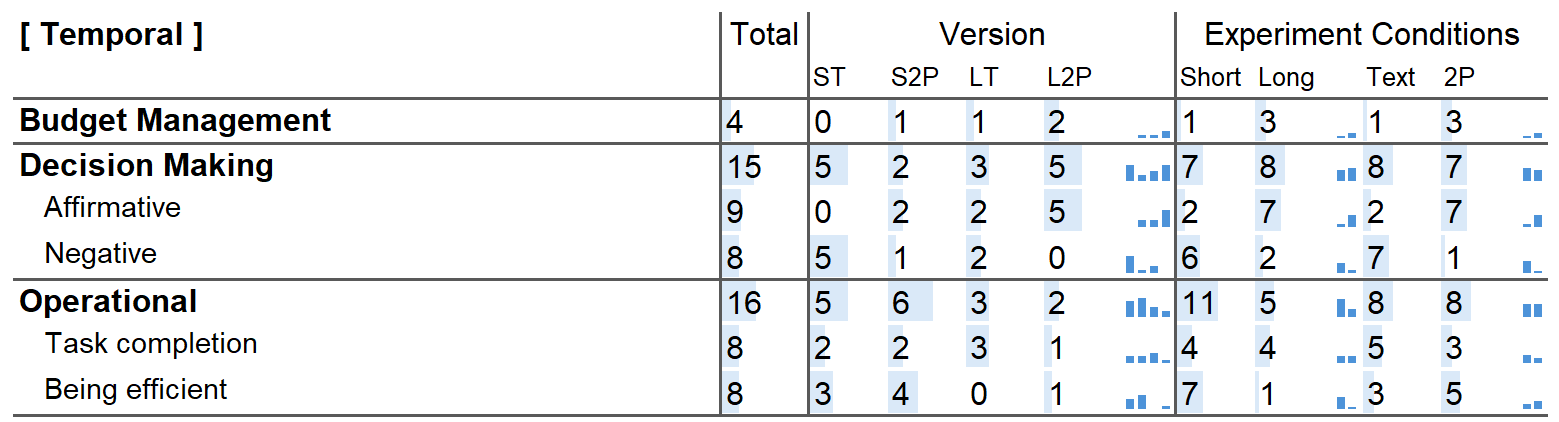}
\end{table*}

\subsection{Source of Effort}
\label{apdx:effort}
Effort refers to how hard participants felt they worked to achieve the level of performance they did. Since effort includes both mental and physical resource intensity, refer to \Cref{apdx:mental} and \Cref{apdx:physical} for definitions. Raw NASA-TLX effort scores~(Figure~\ref{apdxfig:effort_cog_score}) showed a similar spread across experiment groups, the qualitative analysis showed more distinction that participants using the two-phase interface considered options more comprehensively and felt less effort on completing operational tasks, similar to what we found on mental demands~(Section \ref{apdx:mental}). For this subscale, we grouped codes through the lens of scope. Table~\ref{apdx:effort_table} contains codes. 

$14$ of the $20$ participants using the text interface mentioned operational tasks as a source of effort, compared to $7$ participants using the two-phase interface, with the lowest mention in the long two-phase interface group~($N=2$).

\aptLtoX[graphic=no,type=html]{ 
\begin{enumerate}
\item[\,] \color{darkgray}{\it I wanted to bump up~(an option) maybe to 4 or <option> to 5 and realize I couldn't.~\bracketellipsis that would be effort came in of how do I want to really rearrange this to make it~(the budget spending) maximize? \qquad\hspace{0.1em}\textnormal{~~~\faCommentsO}\hspace{-0.2em}\texttt{\kern-0.2em S029 (ST)}  }
\end{enumerate}

 }{ \begin{displayquote}
I wanted to bump up~(an option) maybe to 4 or <option> to 5 and realize I couldn't.~\bracketellipsis that would be effort came in of how do I want to really rearrange this to make it~(the budget spending) maximize? \hfill\quoteby{S029 (ST)}
\end{displayquote}
 } 

In contrast, strategic planning was reported as an effort source by $11$ participants in the text interface, compared to $17$ participants in the two-phase interface, with nearly all participants in the long two-phase interface group~($N=9$) expressing effort related to it. In this subscale, we further categorize strategic planning into \textit{narrow} and \textit{broad} scopes as we did for mental demand~(\Cref{apdx:mental}). Participants using the two-phase interface~($N=7$) had nearly mentioned double~($N=4$) times regarding global strategies. For example:

\aptLtoX[graphic=no,type=html]{ 

\begin{enumerate}
\item[\,] ~\color{darkgray}{\it \bracketellipsis the effort was how to rank order these~(options) and allocate the resources behind the upvotes so that I can accurately depict what I want~\ldots say, a committee to focus on and allocate actual fungible resources, too. \qquad\hspace{0.1em}\textnormal{~~~\faCommentsO}\hspace{-0.2em}\texttt{\kern-0.2em S019 (L2P)}  }
\end{enumerate}

 }{ \begin{displayquote}
~\bracketellipsis the effort was how to rank order these~(options) and allocate the resources behind the upvotes so that I can accurately depict what I want~\ldots say, a committee to focus on and allocate actual fungible resources, too.  \hfill\quoteby{S019 (L2P)}  
\end{displayquote}
 }

\begin{table*}[p]
    \caption{Frustration Sources: Frustration comes from different levels of strategic operations or operational tasks.}
    \Description{A table with sparklines showing frustration categories: Strategic (higher-level and lower-level) and Operational, across four versions (ST, S2P, LT, L2P) and experiment conditions (Short, Long, Text, Inter). The table includes sparklines and percentage bars alongside the numerical counts to visually represent the data distribution across conditions. Strategic frustration is divided into higher-level and lower-level conflicts between personal preference and broader societal values. Operational frustration includes challenges such as credit management, adhering to the quadratic mechanism, making decisions, and understanding options. A final row captures participants reporting None/Little frustration, also visualized with percentage bars.}

    \label{tbl:fustration}
    \includegraphics[width=0.85\linewidth]{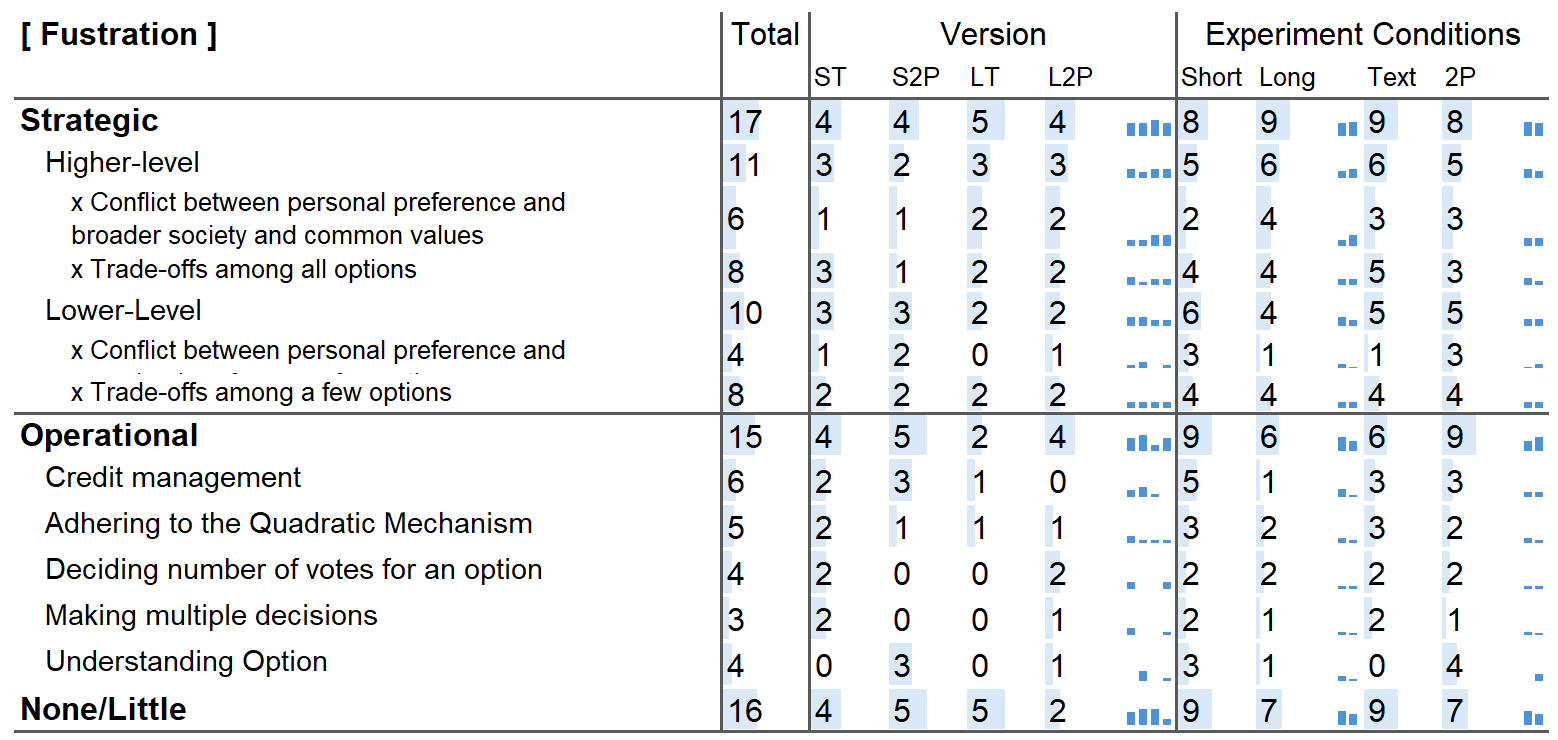}
\end{table*}

\subsection{Source from Performance}
\label{apdx:performance}

\begin{figure}[h]
    \centering
    \includegraphics[width=0.38\textwidth, trim=0 13 0 13, clip]{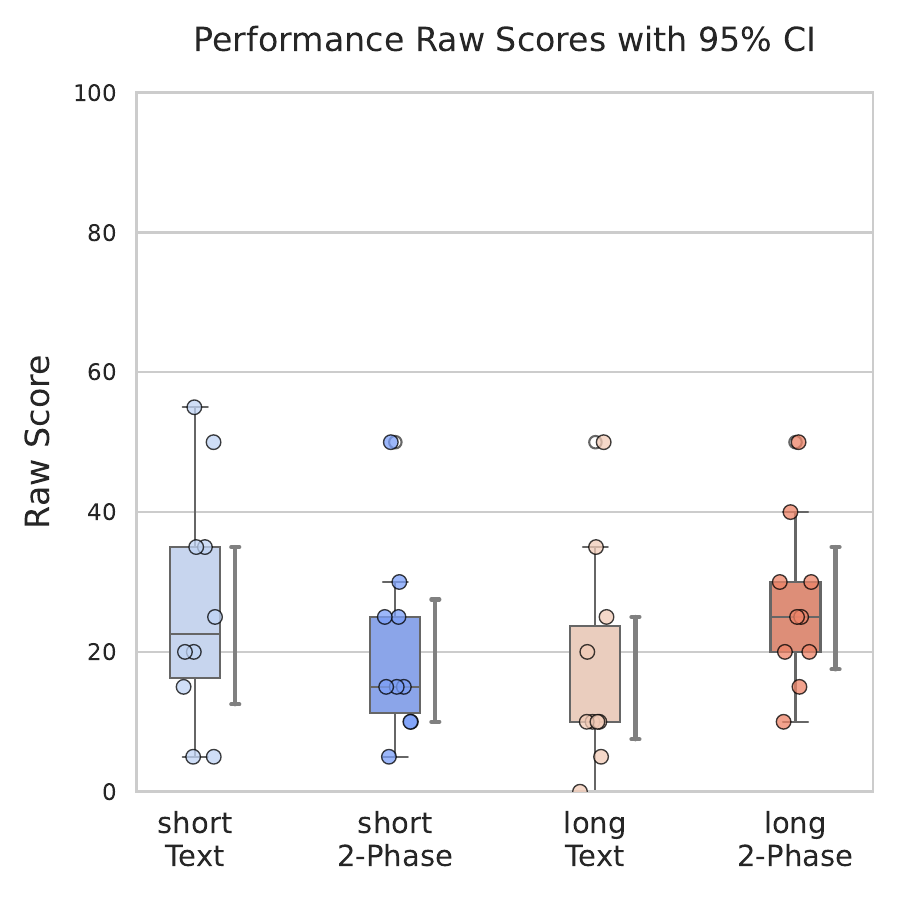}
    \captionsetup{width=0.9\linewidth, justification=justified}
    \caption{Performance Demand Raw Score: Participants showed indifferent performance raw scores across experiment conditions, all trending toward satisfactory.}
    \Description{A box plot displaying the distribution of performance raw scores across four interface versions: Short Text, Short 2-Phase, Long Text, and Long 2-Phase. The y-axis represents raw scores ranging from 0 to 100. Each box plot includes individual data points, a central line for the median, and whiskers representing the 95\% confidence interval. The performance scores appear relatively low across all conditions, with medians hovering between 20 and 40.}
    \label{fig:performance_cog_score}
\end{figure}

Performance refers to a person's perception of how successfully they have completed a task. Lower values indicate good perceived performance, while higher values suggest poor perceived performance. Raw NASA-TLX scores~(Figure~\ref{fig:performance_cog_score}) show that participants had similar performance scores, although we highlighted nuanced differences in the main text. In addition to the differences mentioned in the main text, an interesting theme that emerged across experimental conditions was that participants' identified that \textit{Social Responsibility} influenced their performance scores.~\Cref{tbl:perf} presents a detailed breakdown of our codes.

\paragraph{Social Responsibility.} This theme refers to concerns about performance when participants reflected on how their final vote counts would be perceived by others~(\smallquote{S041}{I don't want people to think that I just don't care about <ethnicity> people at all}) or how their votes might influence real-world decision-making~(\smallquote{S027}{Some of these things might \ldots have outcomes that I didn't foresee}).

\subsection{Temporal Demand}
\label{apdx:temporal}
Table~\ref{tbl:temporal} lists all the temporal demand codes.

\subsection{Frustration}
\label{apdx:frus}
Table~\ref{tbl:fustration} lists all codes related to participants' sources of frustration.

\begin{figure}[h]
    \centering
    \includegraphics[width=0.38\textwidth, trim=0 13 0 13, clip]{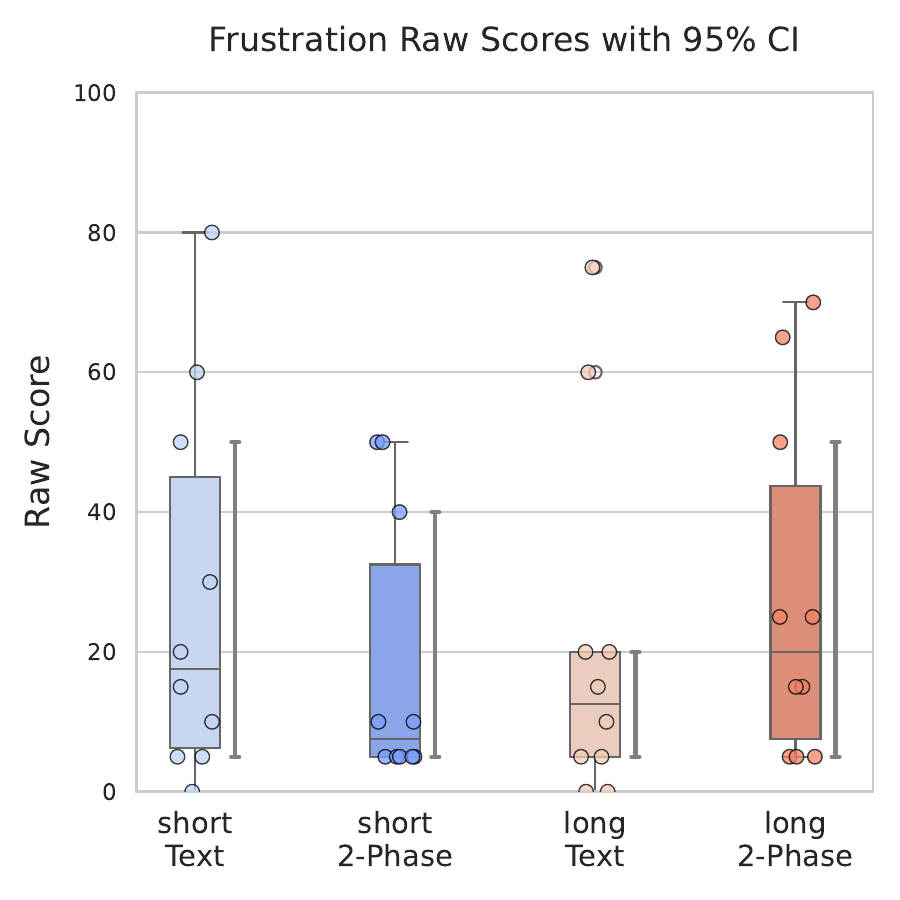}
    \captionsetup{width=0.9\linewidth, justification=justified}
    \caption{Frustration Raw Score: Participants other than the long text interface highlighted several operational tasks that led to frustration. All groups share causes from strategic planning.}
    \Description{Box plot showing frustration raw scores with 95\% confidence intervals across four interface versions: Short Text, Short 2-Phase, Long Text, and Long 2-Phase. The y-axis represents raw scores from 0 to 100. The Short Text interface has a wide range of scores, with several participants reporting high frustration. The Short 2-Phase interface shows lower median frustration with a narrower distribution. The Long Text interface displays the lowest frustration scores, while the Long 2-Phase interface shows a wide range, with some participants reporting high frustration. Individual data points are included, indicating variability across participants.}
    \label{fig:frustration_cog_score}
\end{figure}
\section{Additional voting behavior data}
\label{apdx:additional_results_behavior}
In this section, we describe additional voting behavior that we observed. The reason why we decided to focus on the percentage of remaining credits comes from prior literature `scarcity frames value'~\cite{Shah2015a}, a driver that makes researchers believe makes quadratic voting more accurate~\cite{chengCanShowWhat2021}. We did not follow~\citet{quarfoot2017quadratic} in counting accumulated votes over time due to varying total times across individuals.

\begin{figure*}[p]
    \centering
    \includegraphics[width=0.9\textwidth]{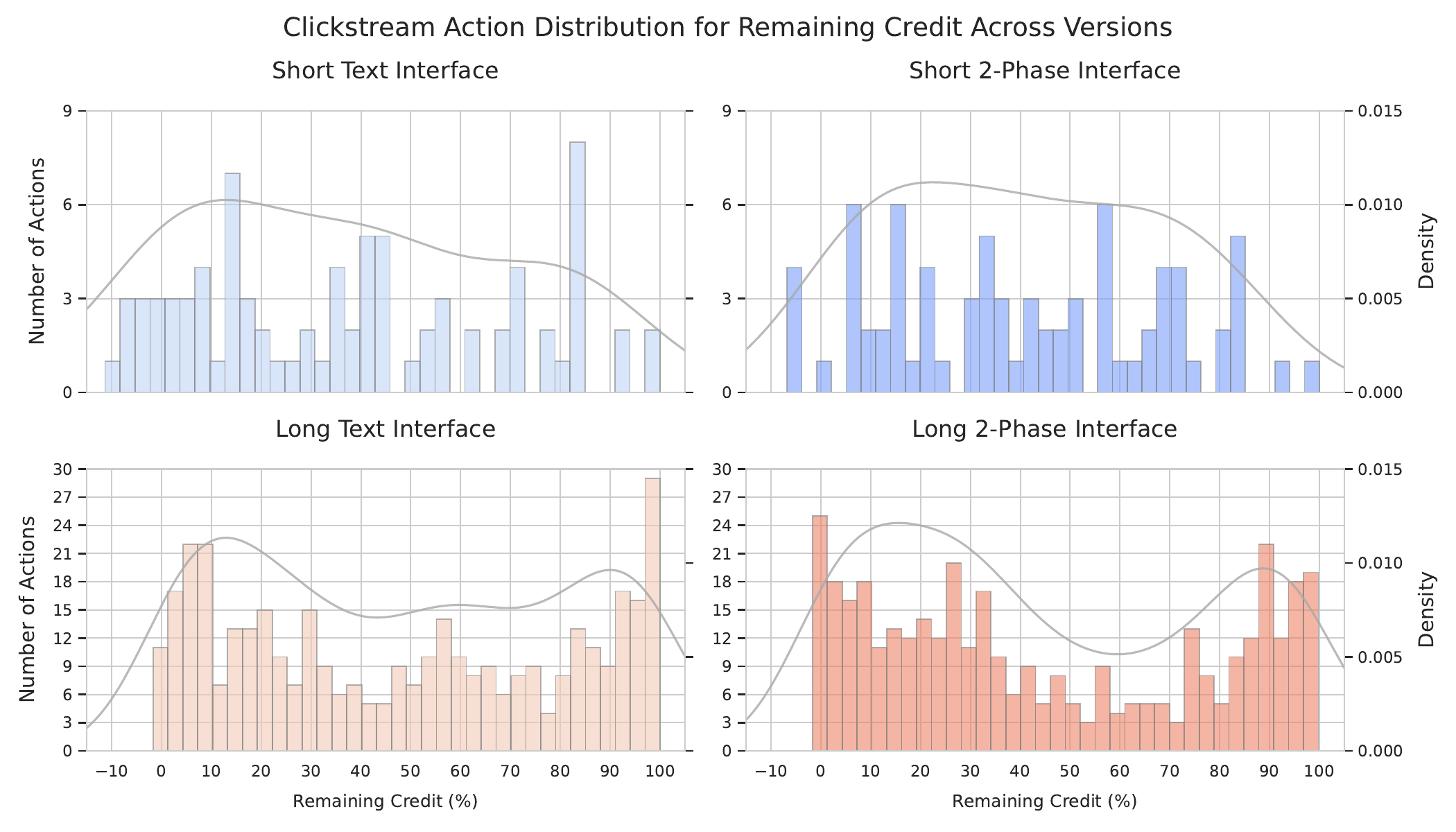}
    \caption{This plot counts the number of voting actions when there are $x$ percentages of credits remaining. A KDE plot is provided to help better understand the action distribution.}
    \Description{A four-panel histogram showing the number of voting actions across remaining credit percentages for four interface versions: Short Text, Short 2-Phase, Long Text, and Long 2-Phase. Each panel displays the number of actions on the y-axis and remaining credit on the x-axis, with an overlaid KDE curve representing density. In the top panels (Short Text and Short 2-Phase), actions are distributed relatively evenly, with small peaks around 10-20\% and 50\% remaining credit. The KDE curves show minor fluctuations. In the bottom panels (Long Text and Long 2-Phase), there are more pronounced peaks at 0-10\% and 100\% remaining credit, with broader distributions and smoother KDE curves indicating denser actions around these areas. The Long Text and Long 2-Phase interfaces exhibit more actions overall compared to the Short Text and Short 2-Phase interfaces.}
    \label{apdxfig:voting_all}
\end{figure*}

We observed the number of vote adjustments given a remaining vote credit percentage. Figure~\ref{apdxfig:voting_all} showed all the voting actions over the remaining credit for the four experiment conditions. Here we see two distinct patterns between the short survey and the long survey in terms of participant behaviors. In long surveys, participants exhibited more actions both when the budget was abundant and when it began to run out. This pattern was more pronounced with the long two-phase interface. This difference is why we further focused on the long QS group.

\begin{figure*}[p]
    \centering
    \includegraphics[width=0.9\textwidth]{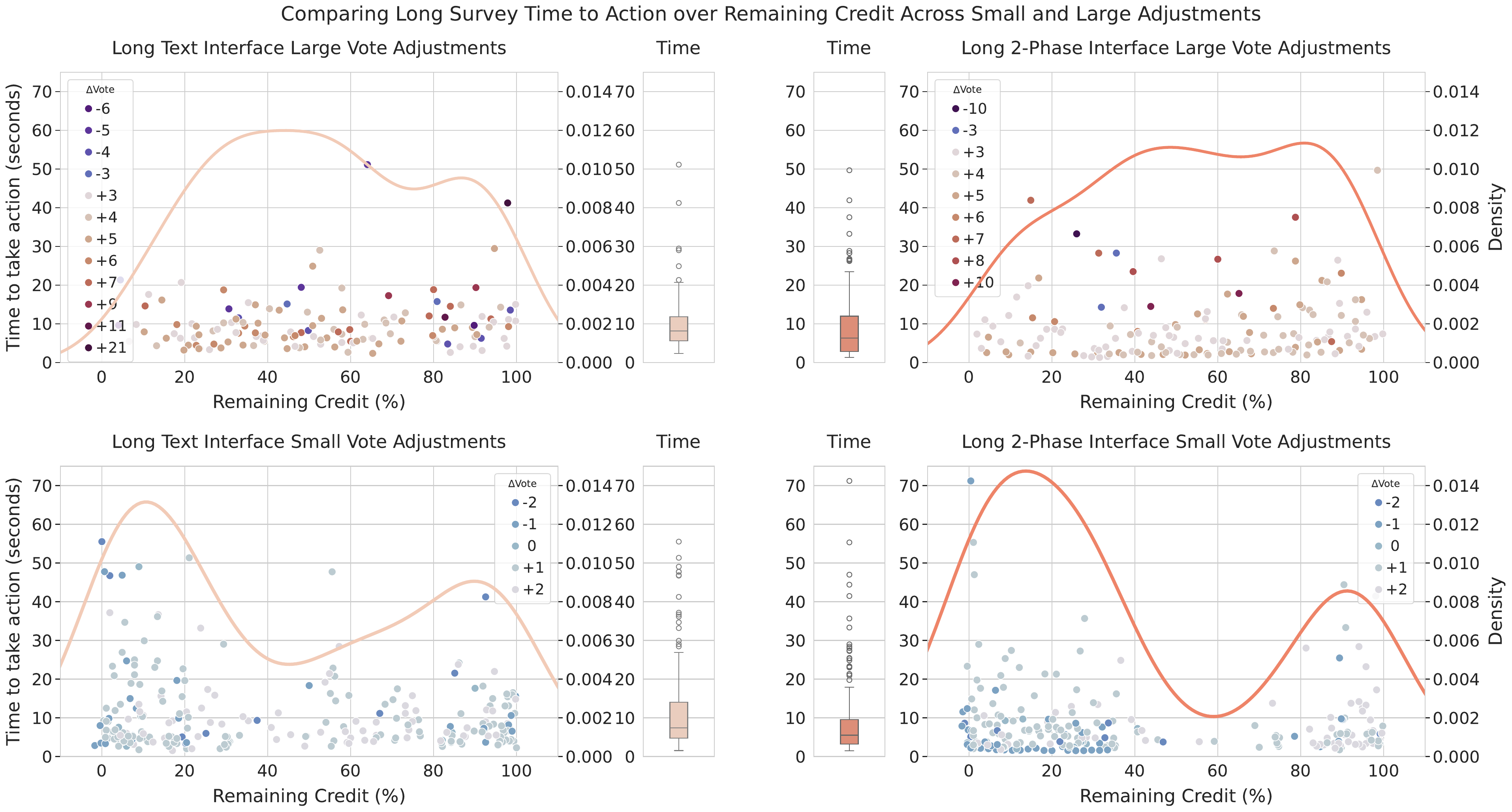}
    \caption{This plot further separates participants' interaction behavior based on the number of votes participants adjusted. We observed a bimodal interaction pattern across long QS when small vote adjustments are made.}
    \Description{A four-panel plot comparing the time to take action in seconds over remaining credit percentages for large and small vote adjustments in the Long Text and Long 2-Phase interfaces. Each panel includes scattered points and an overlaid KDE curve. The top left panel shows large vote adjustments for the Long Text interface, with peaks in the KDE curve around 20\% and 80\% remaining credit. The top right panel shows large vote adjustments for the Long 2-Phase interface, with two peaks in the KDE curve around 10\% and 100\%. The bottom left panel shows small vote adjustments for the Long Text interface, with scattered points and peaks in the KDE curve around 10\% and 90\%. The bottom right panel shows small vote adjustments for the Long 2-Phase interface, with a KDE curve peaking around 10\% and 100\% remaining credit. Box plots on the right side of each panel summarize the distribution of time taken to adjust votes for each interface.}    
    \label{apdxfig:voting_v3_v4}
\end{figure*}

Figure~\ref{apdxfig:voting_v3_v4} presents the comparison between when participants make small or large vote adjustments at different budget levels. Revisiting the KDE curve in the second row in Figure~\ref{apdxfig:voting_all} and the curve of the second row in Figure~\ref{apdxfig:voting_v3_v4} show a stronger bimodal distribution for small vote adjustments across interfaces. In fact, the bimodal distribution is more pronounced in the two-phase interface. This suggests that participants make small adjustments both at the beginning and toward the end of the QS. However, the two-phase interface shows more frequent and faster edits towards the end. In comparison, participants also made more large vote adjustments early on that spread more equally compared to the text interface. This indicates that participants had a clearer idea of how to distribute their credits across the options.

 % Done

% % Modeling Appendix
% % \section{Modeling}
\section{Modeling NASA-TLX Weighted Scores and Subscales}
\label{apdx:model_tlx}
This section first describes the hierarchical Bayesian ordinal regression model used for the NASA-TLX weighted scores and subscales. We then present the results for each subscale.

\subsection{Modeling Approach}

\subsubsection{Dependent variables}
\paragraph{NASA-TLX weighted scores} are transformed from a continuous $0$--$100$ scale to cognitive levels: low, medium, somewhat high, high, and very high, as described by~\citet{hart1988development}. This transformation helps the model adapt to sparse data. In our study, there were no participants who expressed "low" or "very high"; thus, we modeled the predictive variables as "medium," "somewhat high," and "high."

\paragraph{NASA-TLX subscale ratings} are transformed into ordinal groups using minimum frequency binning~\cite{frank2001simple}. Minimum frequency binning involves grouping adjacent response categories until each bin meets a predefined minimum number of observations. Since the subscale uses a 21-point Likert scale and we have 40 participants, the data are very sparse. Minimum frequency binning mitigates this by ensuring similar numbers of participants in each bin. We applied weighted bins across all participants within the same subscale, ensuring that each bin contained at least 10 participants.

\paragraph{Likelihood.} With these ordinal outcome variables, we designed $y_i$ as the observed ordinal category for participant $i$. Then:

\begin{equation}
    y_i \sim \text{OrderedLogistic}(\eta_i, \boldsymbol{\tau}),
    \label{eq:cog_main}
\end{equation}

where $\eta_i$ is the latent predictor, and $\boldsymbol{\tau}$ denotes the cutpoints demarcating the boundaries between the ordinal categories as in~\Cref{eq:cog_orderedTransfrom}. The cutpoints $\boldsymbol{\tau}$ ensure that $\tau_1 < \tau_2 < \cdots < \tau_{K-1}$ by construction.

\begin{equation}
    \boldsymbol{\tau} \sim \text{OrderedTransform}(\mathcal{N}(0, 1)^{K-1}),
    \label{eq:cog_orderedTransfrom}
\end{equation}

\subsubsection{Independent Variables and latent predictor}
For this model, we used three independent variables: length ($\gamma_i$, an ordinal variable), interface type ($\beta_I$, an categorical variable), and the interaction between the two ($\phi_{i,j}$) to construct the latent predictor $\eta_i$. Specifically, the latent predictor $\eta_i$ is constructed as:

\begin{equation}
    \eta_i = \alpha + \gamma_i + \beta_I[I_i] + \phi_{i,j},
    \label{eq:cog_regression}
\end{equation}

where: $\alpha$ is a global intercept drawn from $\mathcal{N}(0,1)$, $\gamma_i$ captures the (ordinal) effect of length, $\beta_I[I_i]$ is the effect for interface $I_i$, and $\phi_{i,j}$ is the interaction between length $i$ and interface $j$. 

Since length has two levels (short and long), we define the following equation to account for ordinality:
\begin{equation}
    \gamma_i = \mu_L + \beta_L \cdot L_i
    \label{eq:cog_ordinal}
\end{equation}

where $L_i \in \{0,1\}$, making $\gamma_i = \mu_L$ for the short condition and $\gamma_i = \mu_L + \beta_L$ for the long condition. We assign standard normal priors to these parameters: $\mu_L \sim \mathcal{N}(0,1)$ and $\beta_L \sim \mathcal{N}(0,1)$. 

\paragraph{Interface Effects.}
We model the interface effects using a non-centered parameterization to improve numerical stability and encourage partial pooling across the two interface levels. Specifically, we let $\mu_{\beta_I} \sim \mathcal{N}(0,1)$ and $\sigma_{\beta_I} \sim \mathrm{Exponential}(1)$ represent the shared mean and scale of the interface effects. We then sample a raw effect vector $\beta_{I_{\text{raw}}} \sim \mathcal{N}(0,1)^2.$ Combining these, we define:
\begin{equation}
    \beta_I = \mu_{\beta_I} + \sigma_{\beta_I} \cdot \beta_{I_{\text{raw}}}
    \label{eq:interface_reparam}
\end{equation}
where $\beta_I \in \mathbb{R}^2$ contains the effect for each of the two interface levels, 
and $\beta_I[I_i]$ indexes the effect for participant $i$'s interface. 

\paragraph{Interaction Effects} To capture potential interaction effects between length and interface types, we assign one interaction parameter, $\phi_{i,j}$, to each combination of length $i$ and interface $j$. Rather than sampling these $\phi_{i,j}$ directly, we employ a non-centered parameterization:
\[
  \boldsymbol{\phi} = L_{\Omega} \,\bigl(\sigma_{\phi} \odot z_{\phi}\bigr),
\]
where \(\boldsymbol{\phi}\) is a $2 \times 2$ matrix of interaction parameters (since we have 2 levels of length and 2 levels of interface), $z_{\phi} \sim \mathcal{N}(0,1)^{2\times2}$, $\sigma_{\phi} \sim \text{Exponential}(1)^{2\times2}$, and $L_{\Omega}$ is the Cholesky factor of a correlation matrix drawn from an $\text{LKJ}(2)$ prior. We then define
\[
    \phi_{i,j} 
    = 
    \bigl[\boldsymbol{\phi}\bigr]_{i,j},
\]
making $\phi_{i,j}$ a \emph{single scalar} drawn from the correlated matrix $\boldsymbol{\phi}$.

\subsubsection{Posterior predictive plots}
\begin{figure*}[h!]
    \centering
    \includegraphics[width=\textwidth]{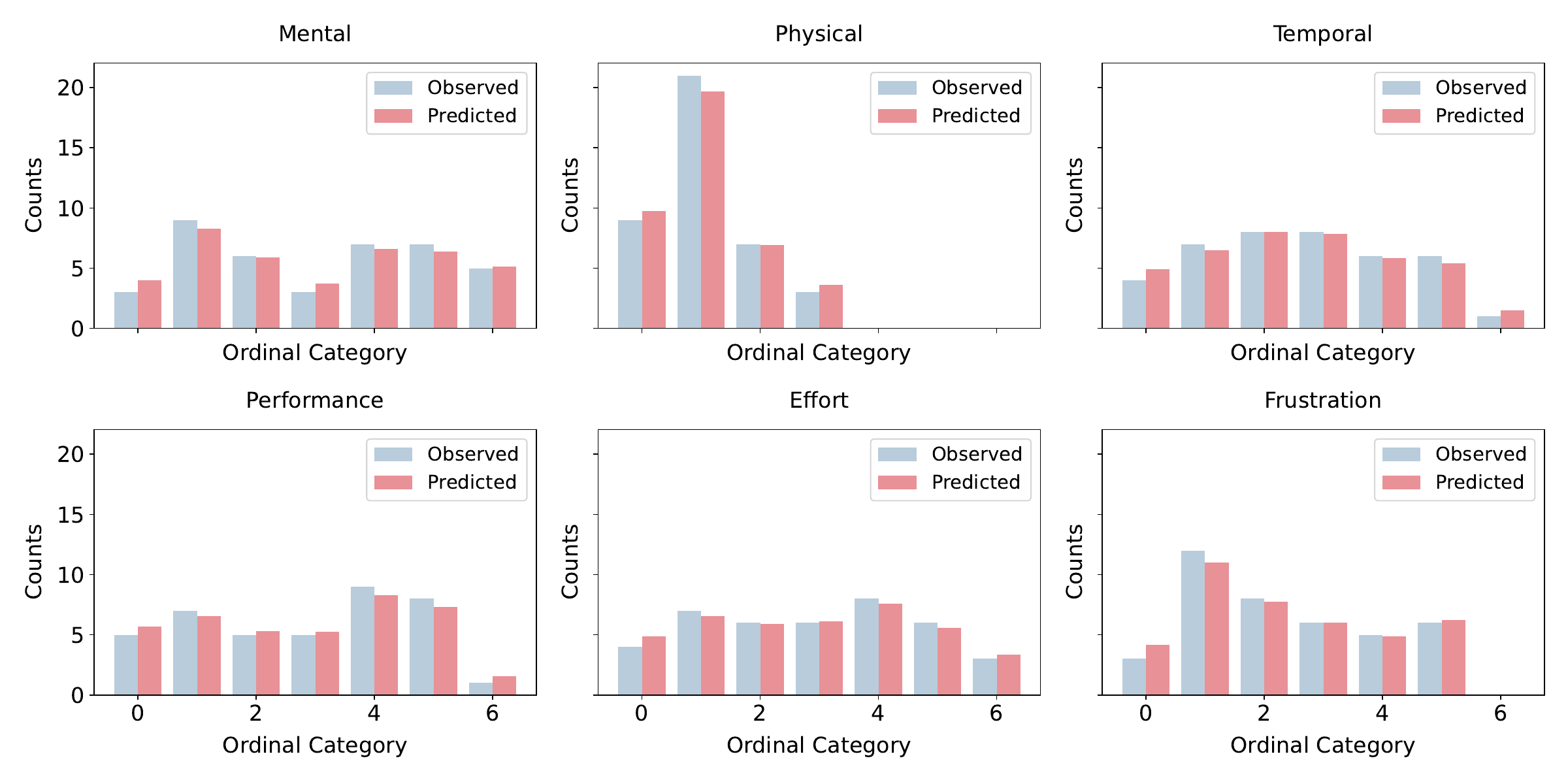}
    \caption{Posterior Predictions vs. observed data for NASA-TLX subscales. The plot shows the observed number of participants in each bin compared to the posterior predictions from the model. \textbf{Takeaway of the plot}: We believe that the model is reasonable at capturing the distribution of the subscales given the sparsity of the data.}
    \Description{ A collection of bar charts comparing observed versus predicted counts across six subscales: Mental, Physical, Temporal, Performance, Effort, and Frustration. Each chart compares counts (y-axis) across ordinal categories (x-axis), highlighting discrepancies between observed and predicted values for each subscale. Bars are color-coded to distinguish observed and predicted values.}
    \label{fig:observed_vs_predicted_all_subscale}
\end{figure*}

We conducted the Bayesian analysis using NumPyro, a widely used framework for Bayesian inference. We used Markov Chain Monte Carlo (MCMC) sampling, a method commonly applied in Bayesian inference. The model converged successfully, as evidenced by an $\hat{R}$ value of 1 for each subscale and the overall weighted TLX scores, indicating that multiple sampling chains converged. We plotted the posterior predictive distribution of the model to compare the observed data with the model's predictions. Figure~\ref{fig:observed_vs_predicted_all_subscale} shows the posterior predictions vs. observed data for the six subscales.

\subsection{Model Results}
\begin{figure*}[h!]
    \centering
    \includegraphics[width=0.75\textwidth]{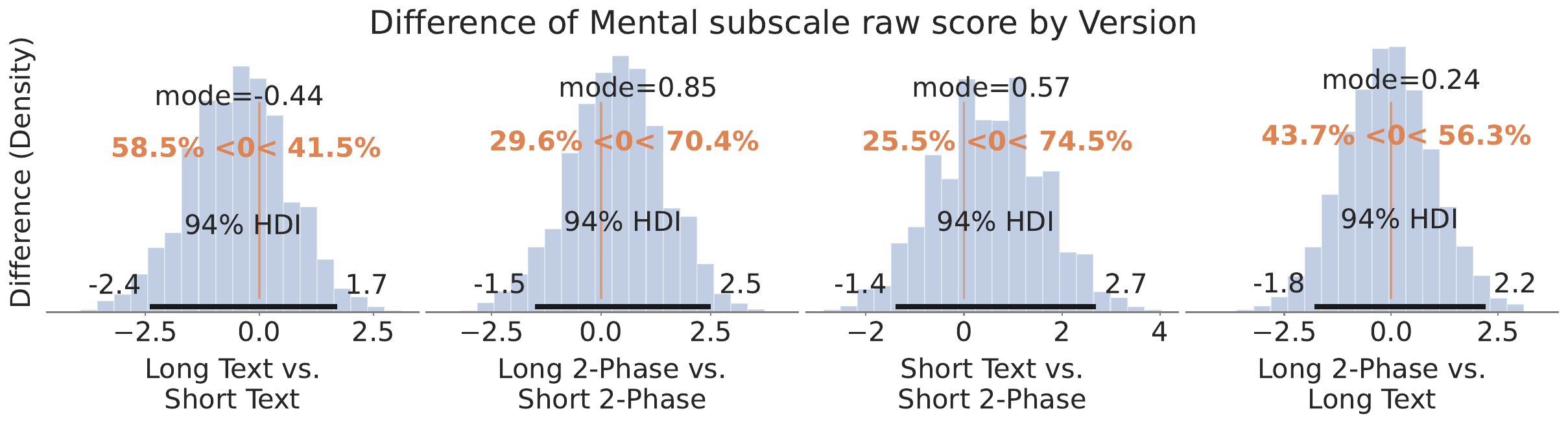}
    \caption{Differences in the mental subscale scores by version.~\textbf{Main Takeaway:} Participants in the long two-phase condition show trends to increase mental demand compared to the short two-phase. Within the short text condition, participants in the short two-phase condition show a trend to reduce mental demand.}
    \Description{A grouped panel of four histograms titled "Difference of Mental subscale raw score by Version," displaying posterior distributions of differences between various experimental conditions. Each plot shows a histogram of density (y-axis) versus difference (x-axis), with key summary statistics. Each histogram includes credible intervals, density curves, and a vertical line at zero for reference. Summary values are highlighted in orange and positioned at the top of each plot.}
    \label{fig:bayesian_mental_subscale}
\end{figure*}

\begin{figure*}[h!]
    \centering
    \includegraphics[width=0.75\textwidth]{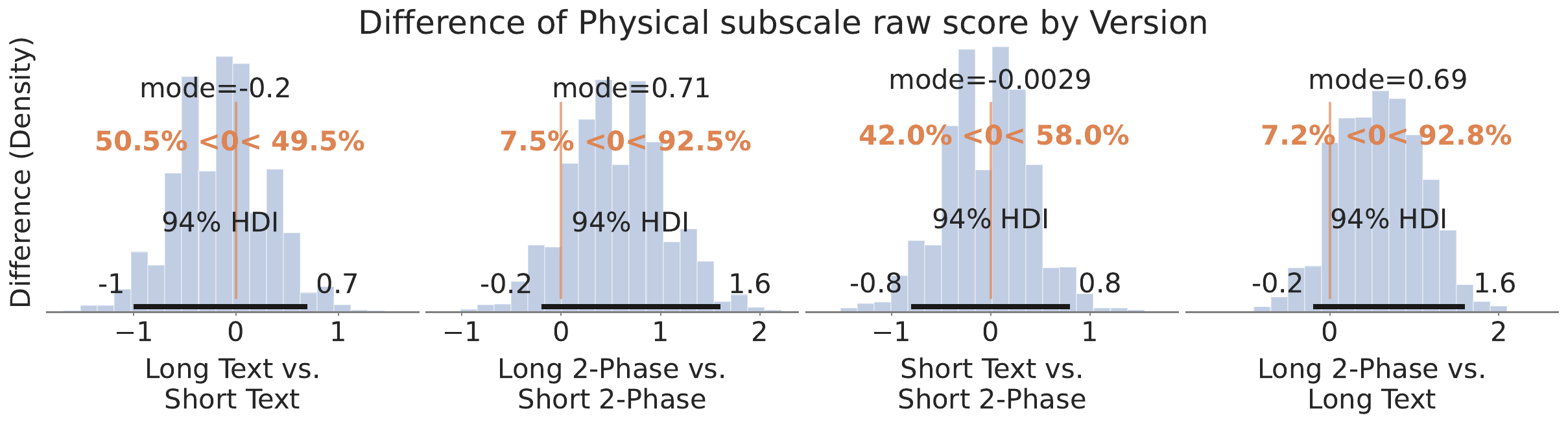}
    \caption{Differences in the physical subscale scores by version.~\textbf{Main Takeaway:} Participants in the long two-phase condition show trends to increase physical demand compared to short two-phase and long text despite the long text participants traversing higher edit distances.}
    \Description{A grouped panel of four histograms titled "Difference of Physical subscale raw score by Version," displaying posterior distributions of differences between various experimental conditions. Each plot shows a histogram of density (y-axis) versus difference (x-axis), with key summary statistics. Each histogram includes credible intervals, density curves, and a vertical line at zero for reference. Summary values are highlighted in orange and positioned at the top of each plot.}
    \label{fig:bayesian_physical_subscale}
\end{figure*}

\begin{figure*}[h!]
    \centering
    \includegraphics[width=0.75\textwidth]{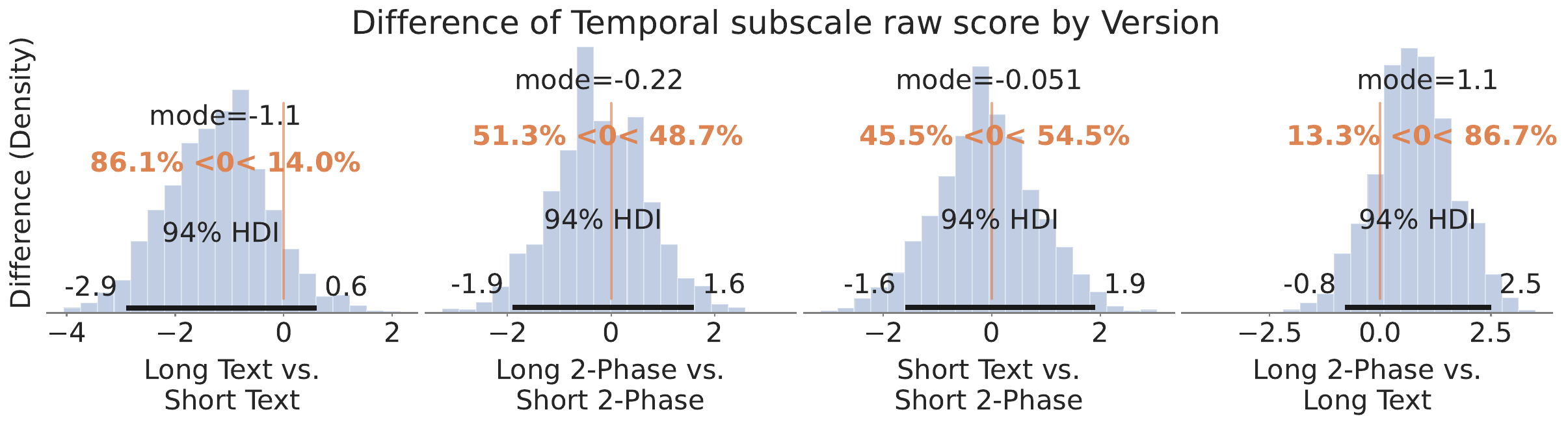}
    \caption{Differences in the temporal subscale scores by version.~\textbf{Main Takeaway:} Participants in the long text condition show a trend that it reduces temporal demand compared to the short text condition and the long two-phase condition.}
    \Description{A grouped panel of four histograms titled "Difference of Temporal subscale raw score by Version," displaying posterior distributions of differences between various experimental conditions. Each plot shows a histogram of density (y-axis) versus difference (x-axis), with key summary statistics. Each histogram includes credible intervals, density curves, and a vertical line at zero for reference. Summary values are highlighted in orange and positioned at the top of each plot.}
    \label{fig:bayesian_temporal_subscale}
\end{figure*}

\begin{figure*}[h!]
    \centering
    \includegraphics[width=0.75\textwidth]{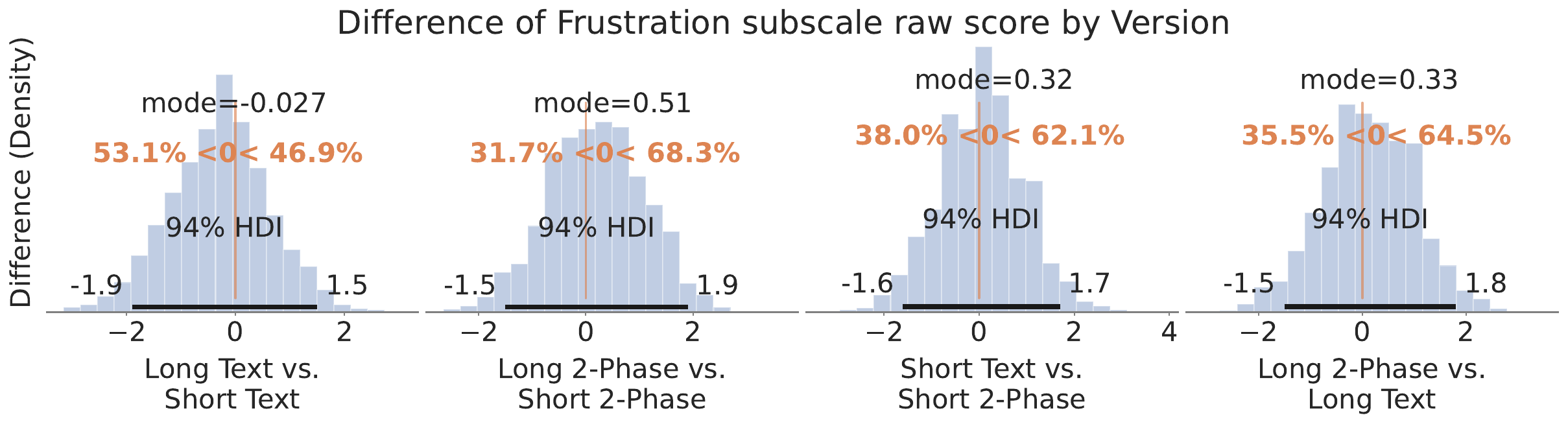}
    \caption{Differences in the frustration subscale scores by version.~\textbf{Main Takeaway:} The model does not see a significant difference in the frustration subscale between experiment groups other than a trend for participants in the long two-phase condition to have higher frustration than the short two-phase participants.}
    \Description{A grouped panel of four histograms titled "Difference of Frustration subscale raw score by Version," displaying posterior distributions of differences between various experimental conditions. Each plot shows a histogram of density (y-axis) versus difference (x-axis), along with key summary statistics. The plots highlight the density and credible intervals for the differences, with key values marked in orange (percentages) and labeled at the top of each distribution. The vertical line at zero serves as a reference point.}
    \label{fig:bayesian_frustration_subscale}
\end{figure*}

% Section for Mental subscale
\subsubsection{Mental Subscale}
Figure~\ref{fig:bayesian_mental_subscale} shows pairwise Bayesian results from mental demand highlighted 70.4\% of posterior probability that participants in the long two-phase condition had a higher mental demand compared to the short two-phase condition. On the other hand, the short text condition had a 74.5\% posterior probability of having a higher mental demand compared to the short two-phase condition. This is additional evidence that prompted us to believe that the participants in the short two-phase participants benefited from the organization phase. The sheer number of added options in the long two-phase condition may have added additional demand to participants, leading to higher mental demand.

% Section for Physical subscale
\subsubsection{Physical Subscale}
Figure~\ref{fig:bayesian_physical_subscale} shows the pairwise comparison of the physical subscale. Notable results show that there is a 86.1\% posterior probability that the long text condition had a lesser physical demand compared to the short text condition. This is counter intuitive as the long text participants actually traversed much higher edit distances. We are not clear what prompted their self reported value and requires future research. 

% Section for Temporal subscale
\subsubsection{Temporal Subscale}
\label{sec:temporal_subscale_bayesian}
Figure~\ref{fig:bayesian_temporal_subscale} shows the pairwise comparison of the temporal subscale. The results show that the long two-phase condition once again had a 74.6\% posterior probability of having a lower temporal demand compared to the short text condition. Conversely, participants in the long two-phase condition had a 71.1\% posterior probability of having a higher temporal demand compared to the short two-phase condition, reflecting the longer time they took to complete the survey questions. We believe that the lower temporal demand in the long two-phase condition is potential indicator of the participants' satisficing behavior.

% Section for Performance subscale
\subsubsection{Performance Subscale}
We omit the pairwise comparison of the performance subscale due to the mixed signals. We focused on the qualitative results analyzed in the main text.

% Section for Effort subscale
\subsubsection{Effort Subscale}
We omit the pairwise comparison of the effort subscale due to its similarity to the mental demand subscale. 

% Section for Frustration subscale
\subsubsection{Frustration Subscale}
Figure~\ref{fig:bayesian_frustration_subscale} shows the pairwise comparison of the frustration subscale. The results show that the long two-phase condition had a 68.3\% posterior probability of having a higher frustration compared to the short two-phase condition, likely due to the added number of options to assess.

\section{Modeling Total Time} \label{sec:apdx:model_time}

\subsection{Dependent Variables} The dependent variable is the total time $T_i$ spent on option $i$ measured in seconds. This measure captures both the duration participants took to vote and, where applicable, the time they spent organizing or reordering their options beforehand. We categorize the data into four experimental conditions: Short Text, Short Two-Phase, Long Text, and Long Two-Phase. These conditions are indexed by $k$, fit using separate submodels.

\subsection{Modeling Approach} We modeled the total time for each experimental condition using separate Gamma likelihood models. The Gamma distribution is well-suited for modeling positive continuous data, such as time measurements, which are often skewed and strictly positive. Equation~\ref{eq:time_main} shows the model for the total time. The shape parameter $\alpha_k$ and rate parameter $\beta_k$ were each assigned priors drawn from their own Gamma distributions, as described in Equations~\ref{eq:alpha_prior} and \ref{eq:beta_prior}.

\begin{align}
    T_i &\sim \text{Gamma}(\alpha_k, \beta_k) \label{eq:time_main} \\
    \alpha_k &\sim \text{Gamma}(2.0, 0.5) \label{eq:alpha_prior} \\
    \beta_k &\sim \text{Gamma}(1.0, 1.0) \label{eq:beta_prior}
\end{align}
\section{Modeling Edit Distance}\label{sec:apdx:model_distance}
This section presents our hierarchical Bayesian approaches for analyzing the edit distance data. We first describe a model for edit distance per option (\Cref{sec:apdx:model_distance_option}), followed by analysis for edit distance per action (\Cref{sec:apdx:model_distance_variance}). Finally, we detail a model for cumulative edit distances (\Cref{sec:apdx:model_cum_distance}).

\subsection{Model 1: Edit Distance per Option} \label{sec:apdx:model_distance_option}

\subsubsection{Likelihood}
The dependent variable in this model is the edit distance accumulated for each option, denoted by $D_i$, where $i$ refers to the $i$-th observation. Since $D_i$ must be positive, we model it using an exponential likelihood:

\begin{equation}\label{eq:distance_model_1_likelihood}
D_i \sim \text{Exponential}\bigl(\text{scale} = \lambda_i\bigr).
\end{equation}

\subsubsection{Independent variables and regression model}
We designed $\eta_i$ as the linear predictor that informs $D_i$ through the following transformation:
\begin{equation}\label{eq:transformation_model_1}
\lambda_i = \exp(\eta_i),
\end{equation}
where $\lambda_i$ is the scale (i.e., mean) parameter of the Exponential distribution, and thus must be positive.

This linear predictor:
\begin{equation}\label{eq:distance_model_1_eta}
    \eta_i = \gamma_i + \beta_I[I_i] + \phi_{ij} + U_i
\end{equation}
consists of four components: the length of the option $L_i$, interface type $I_i$, and interaction effect between both length and interface $\phi_{ij}$, and user effect $U_i$ which we describe in the following paragraphs.

\paragraph{Length.} Since length has two levels (short and long), we define:
\begin{equation}
    \gamma_i = \mu_L + \beta_L \cdot L_i 
    \label{eq:distance_model_1_eta_ordinal}
\end{equation}
where $L_i \in \{0,1\}$, making $\gamma_i = \mu_L$ for the short condition and $\gamma_i = \mu_L + \beta_L$ for the long condition. We assign standard normal priors to these parameters: $\mu_L \sim \mathcal{N}(0,1)$ and $\beta_L \sim \mathcal{N}(0,1)$. 

\paragraph{Interface.}
We model the interface effects using a non-centered parameterization to improve numerical stability and encourage partial pooling across the two interface levels. Specifically we let $\mu_{\beta_I} \sim \mathcal{N}(0,1)$ and $\sigma_{\beta_I} \sim \mathrm{HalfNormal}(0.5)$ represent the shared mean and scale of the interface effects. We then sample a raw effect vector $\beta_{I_{\text{raw}}} \sim \mathcal{N}(0,1)^2.$ Combining these, we define:
\begin{equation}
    \beta_I = \mu_{\beta_I} + \sigma_{\beta_I} \cdot \beta_{I_{\text{raw}}}
    \label{eq:distance_interface_reparam}
\end{equation}
where $\beta_I \in \mathbb{R}^2$ contains the effect for each of the two interface levels, 
and $\beta_I[I_i]$ indexes the effect for participant $i$'s interface. 

\paragraph{Interaction Effects} To capture potential interaction effects between length and interface types, we assign one interaction parameter, $\phi_{i,j}$, to each combination of length $i$ ($i \in \{0,1\}$) for short and long surveys and interface $j$ ($j \in \{0,1\}$) for the two interface types. Rather than sampling these $\phi_{i,j}$ directly, we employ a non-centered parameterization:
\[
  \boldsymbol{\phi} = L_{\Omega} \,\bigl(\sigma_{\phi} \odot z_{\phi}\bigr),
\]
where \(\boldsymbol{\phi}\) is a $2 \times 2$ matrix of interaction parameters (since we have 2 levels of length and 2 levels of interface), $z_{\phi} \sim \mathcal{N}(0,1)^{2\times2}$, $\sigma_{\phi} \sim \text{HalfNormal}(0.5)^{2\times2}$, and $L_{\Omega}$ is the Cholesky factor of a $2\times2$ correlation matrix drawn from an $\text{LKJ}(2)$ prior with shape parameter $\eta=3$. We then define
\begin{equation}
    \phi_{ij} = \bigl[\boldsymbol{\phi}\bigr]_{i,j}
\end{equation}
making $\phi_{ij}$ a \emph{single scalar} drawn from the correlated matrix $\boldsymbol{\phi}$.

\paragraph{Individual user effects.} 
Similar to the interface, we also applied a non-centered parameterization to user effects using the same approach:
\begin{equation}\label{eq:distance_model_1_user_effect}
    U_i = \mu_U + \sigma_U \cdot z_U 
\end{equation}

We assign weakly informative priors for the user effects: $\mu_U \sim \mathcal{N}(0,1)$ and $\sigma_U \sim \mathrm{Exponential}(0.5)$, which represent the shared mean and scale of the user effects. We use $z_U \sim \mathcal{N}(0,1)^{40}.$ to denote the $40$ participant's raw user effect vector. This approach allow us to capture user variations across all users.

\subsubsection{Posterior predictive plots}
\begin{figure}[h!]
    \centering
    \includegraphics[width=0.45\textwidth]{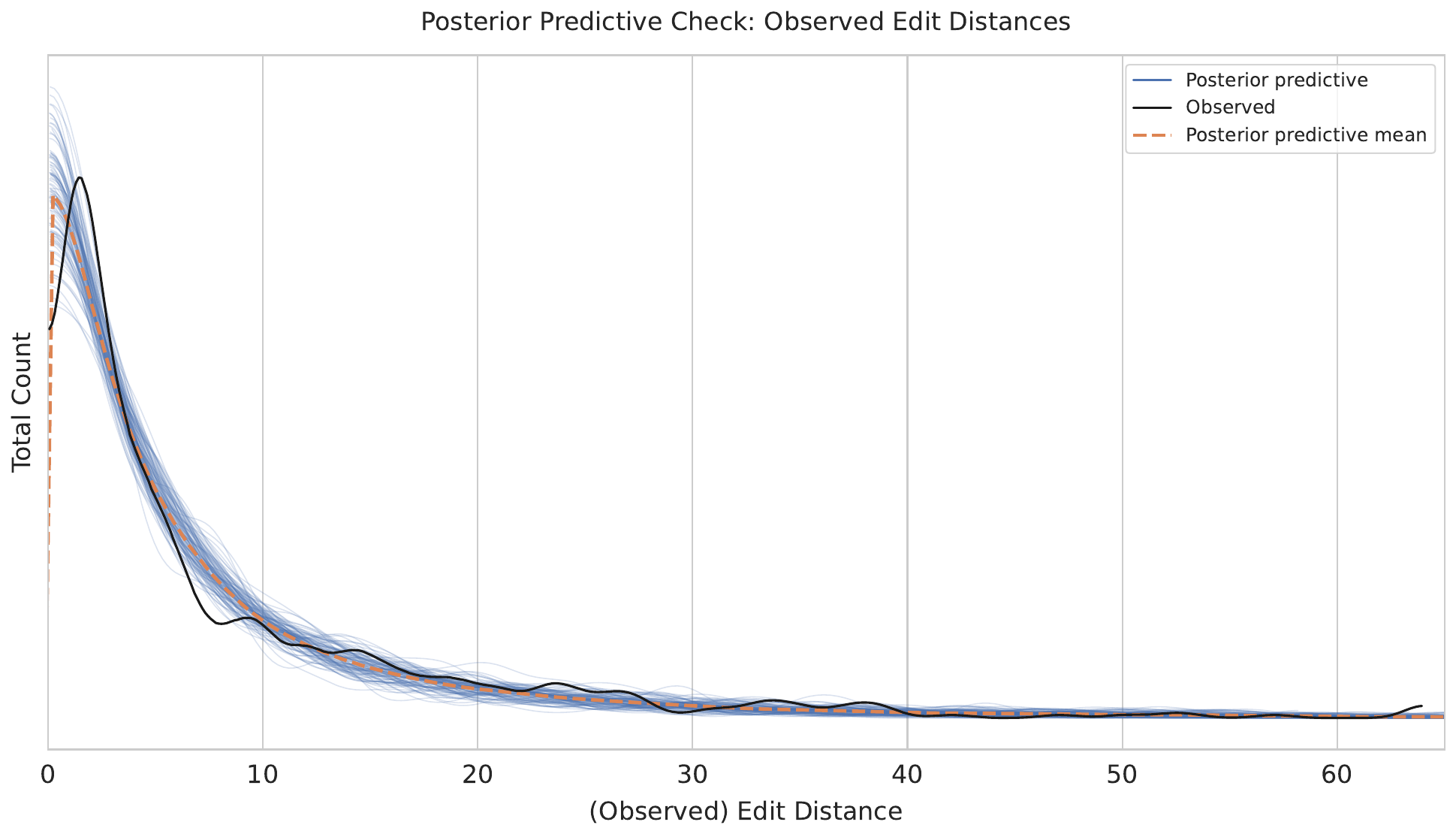}
    \caption{Posterior Predictions vs. observed data for edit distance per option. Each blue line represents a draw from the posterior distribution, while the black line represents the observed data. Dotted line represents the mean of the posterior data. \textbf{Takeaway of the plot}: We believe that the model is reasonable at capturing the distribution.}
    \Description{ A line plot titled "Posterior Predictive Check: Observed Edit Distances," comparing observed edit distances with model predictions. Multiple thin blue lines indicating individual model simulations. A solid black line representing actual observed data. A dashed orange line showing the average of posterior predictions. The plot illustrates strong alignment between observed data and the posterior predictive mean, with deviations primarily at higher edit distances. This indicates a well-calibrated model fit for most of the distribution.}
    \label{fig:ppc_distance_m1}
\end{figure}

Our Bayesian model converged successfully, as evidenced by an $\hat{R}$ value of 1 in the model summary. We plotted the posterior predictive distribution for the edit distance per option in Figure~\ref{fig:ppc_distance_m1}. This figure compares the models posterior predictive distribution with the observed data. 

\begin{figure*}[h!]
    \centering
    \includegraphics[width=0.75\textwidth]{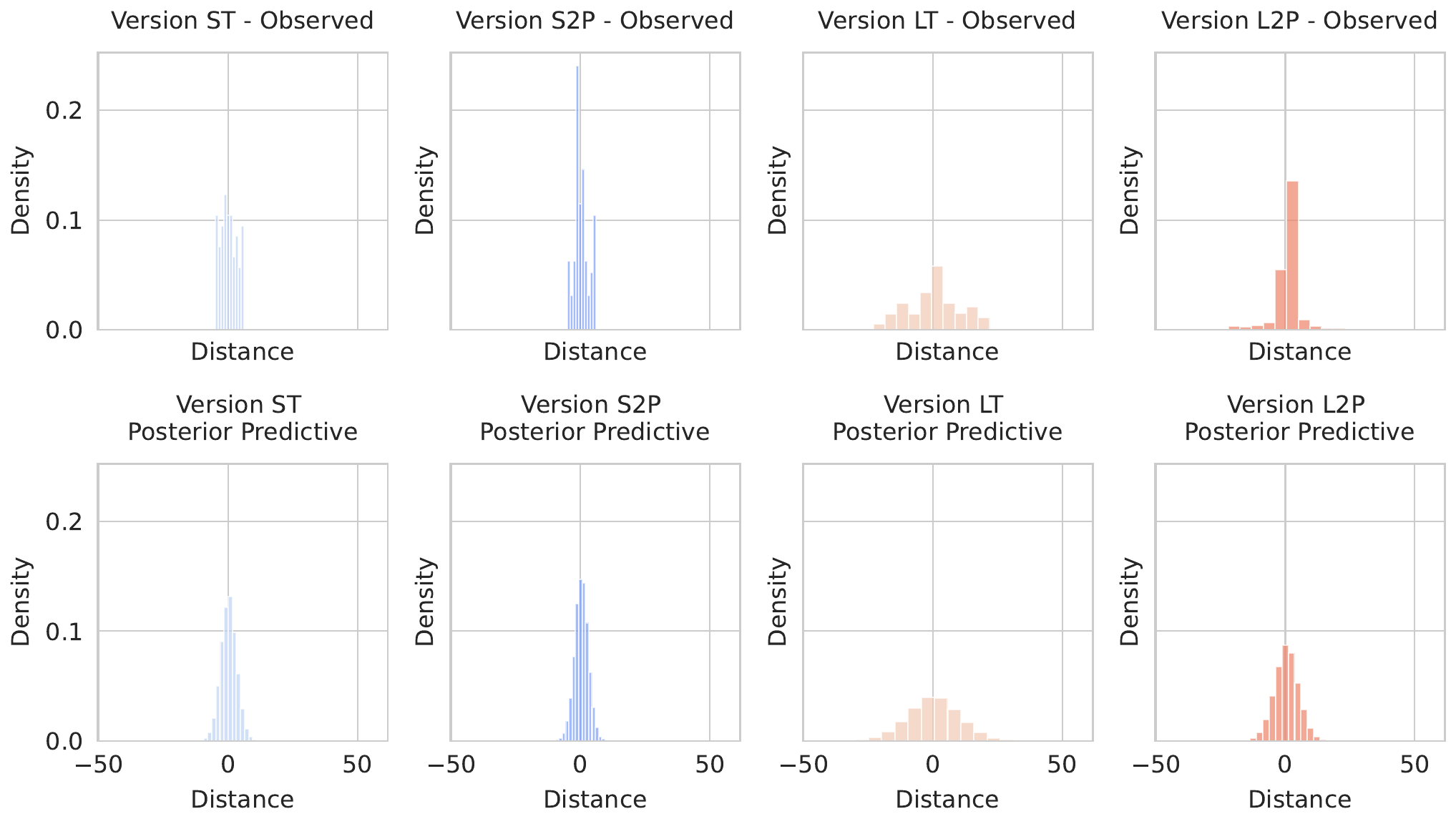}
    \caption{Posterior Predictions vs. Observed Data for Edit Distance per Option. The first row represents the distribution of edit distance per version. The second row shows the posterior predictions after multiple draws \textbf{Takeaway of the plot}: We believe that the model is reasonable at capturing the shape of the distributions though being slightly conservative for extreme values at the center. Future model enhancements could re-model them with a student-t distribution.}
    \Description{A grid of eight density plots comparing observed and posterior predictive distributions across four model versions (ST, S2P, LT, L2P). The top row displays observed distributions, and the bottom row displays posterior predictive distributions. Each plot has "Distance" on the x-axis ranging from -50 to 50 and "Density" on the y-axis. Observed distributions are more sparse, while posterior predictive distributions are tightly centered around zero for all model versions. Differences in spread and shape are noticeable, particularly between the LT and L2P models, which show wider and more peaked distributions compared to ST and S2P.}

    \label{fig:observed_vs_posterior_predictive_histogram_m2}
\end{figure*}

\subsection{Model 2: Edit Distance with Separate Mean and Variance Predictors} \label{sec:apdx:model_distance_variance}

\subsubsection{Likelihood}
The dependent variable for this model is the edit distance $D_i$, where positive values indicate a downward movement and negative values indicate an upward movement. To allow for different effects on both the mean and variance, we model $D_i$ using a Normal distribution:

\begin{equation}\label{eq:distance_model_2_likelihood}
D_i \sim \mathcal{N}\bigl(\mu_i, \sigma_{\text{obs},i}\bigr)
\end{equation}

\noindent Because our aim is to capture potential differences in variability (e.g., hypothesizing that a two-phase interface might yield lower oscillation than a text-based interface), we separately model both the mean $\mu_i$ and the standard deviation $\sigma_{\text{obs},i}$.

\subsubsection{Independent variables and regression model}

We specify two linear predictors: one for the mean $\mu_i$ (Equation~\ref{eq:distance_model_2_mu}) and one for the (logged) standard deviation $\log(\sigma_{\text{obs},i})$ (Equation~\ref{eq:distance_model_2_sigma}). Both linear predictors incorporate the following factors: the length of the option $L_i$, the interface type $I_i$, an interaction term $\phi_{ij}$, and a user-specific term $U_i$.

\begin{align} 
\mu_i &= \gamma_{\mu,i} + \beta_{I,\mu}[I_i] + \phi_{\mu,ij} + U_{\mu,i}, \label{eq:distance_model_2_mu}\\
log(\sigma_{\text{obs},i}) &= \gamma_{\sigma,i} + \beta_{I,\sigma}[I_i] + \phi_{\sigma,ij} + U_{\sigma,i}. \label{eq:distance_model_2_sigma} 
\end{align}

\paragraph{Length ($L_i$).} Similar to the previous model, we continue to define length as an ordinal value. In this model, the effect for mean and variance are modeled separately. We write: 
\begin{align} 
\gamma_{\mu,i} &= \mu_{L,\mu} + \beta_{L,\mu} \cdot L_i, \label{eq:distance_model_2_gamma_mu}\\
\gamma_{\sigma,i} &= \mu_{L,\sigma} + \beta_{L,\sigma} \cdot L_i. \label{eq:distance_model_2_gamma_sigma} 
\end{align}
For both the mean and variance parts, $\mu_{L,\mu}, \beta_{L,\mu}$ and $\mu_{L,\sigma}, \beta_{L,\sigma}$ capture how option length shifts the location and scale of the distribution, respectively. We assign weakly informative normal priors:
\begin{equation}
    \mu_{L,\mu}, \beta_{L,\mu}, \mu_{L,\sigma}, \beta_{L,\sigma} \sim \mathcal{N}(0, 1).
\end{equation}

\paragraph{Interface ($I_i$).} We treat the interface type as a categorical variable with two levels. As in Model 1, we use a non-centered parameterization for numerical stability and partial pooling. For the mean part, we define: \begin{equation}\label{eq:distance_model_2_beta_I_mu} 
\beta_{I,\mu}[I_i] = \mu_{I,\mu} + \sigma_{I,\mu} \cdot z_{I,\mu}[I_i].
\end{equation} and similarly for the variance part: 
\begin{equation}\label{eq:distance_model_2_beta_I_sigma} 
\beta{I,\sigma}[I_i] = \mu_{I,\sigma} + \sigma_{I,\sigma} \cdot z_{I,\sigma}[I_i].
\end{equation} 

\noindent We place weakly informative priors on the intercepts:
\begin{align}
\mu_{I,\mu}, \beta_{I,\mu}, z_{I,\mu}, \mu_{I,\sigma}, \beta_{I,\sigma}, z_{I,\sigma} \sim \mathcal{N}(0, 1),\\
\sigma_{I,\mu}, \sigma_{I,\sigma} \sim \text{HalfNormal}(0.5).
\end{align}.

\paragraph{Interaction Effects ($\phi_{ij}$).} We hypothesize that the effect of length might vary by interface. Similar to Model 1’s approach, we employ a non-centered parameterization with an LKJ correlation prior. Specifically, for both the mean and variance parts, we define: 
\begin{align} 
\phi_{\mu,ij} &= \bigl[L_{\Omega,\mu},\bigl(\sigma_{\phi,\mu} \odot z_{\phi,\mu}\bigr)\bigr]{i,j}, \label{eq:distance_model_2_phi_mu} \\
phi{\sigma,ij} &= \bigl[L_{\Omega,\sigma},\bigl(\sigma_{\phi,\sigma} \odot z_{\phi,\sigma}\bigr)\bigr]{i,j}, \label{eq:distance_model_2_phi_sigma} 
\end{align} 
where $i \in {0,1}$ (short or long) and $j \in {0,1}$ (two interface types). We continue the use of weakly informed priors:
\begin{align}
z_{\phi,\mu}, z_{\phi,\sigma} \sim \mathcal{N}(0, 1), \sigma_{\phi,\mu}, \sigma_{\phi,\sigma} \sim \text{HalfNormal}(0.5),\\
L_{\Omega,\mu}, L_{\Omega,\sigma} \sim \text{LKJ}(3).
\end{align}

\paragraph{Individual user effects ($U_i$).} To account for participant-level variability, we follow model 1 and adopt a non-centered parameterization but allow each user to have a distinct shift on both $\mu_i$ and $\log(\sigma_{\text{obs},i})$:
\begin{align} 
U_{\mu,i} &= \mu_{U,\mu} + \sigma_{U,\mu} \cdot z_{U,\mu,i}, \label{eq:distance_model_2_user_mu}\\
U_{\sigma,i} &= \mu_{U,\sigma} + \sigma_{U,\sigma} \cdot z_{U,\sigma,i}, \label{eq:distance_model_2_user_sigma}
\end{align} 
with priors:
\begin{align}
    \mu_{U,\mu}, \beta_{U,\mu}, z_{U,\mu,i}, \mu_{U,\sigma}, \beta_{U,\sigma}, z_{U,\sigma,i} \sim \mathcal{N}(0, 1),\\
    \sigma_{U,\mu}, \sigma_{U,\sigma} \sim \text{HalfNormal}(0.5).
\end{align}

\begin{figure*}[h!]
    \centering
    \includegraphics[width=0.8\textwidth]{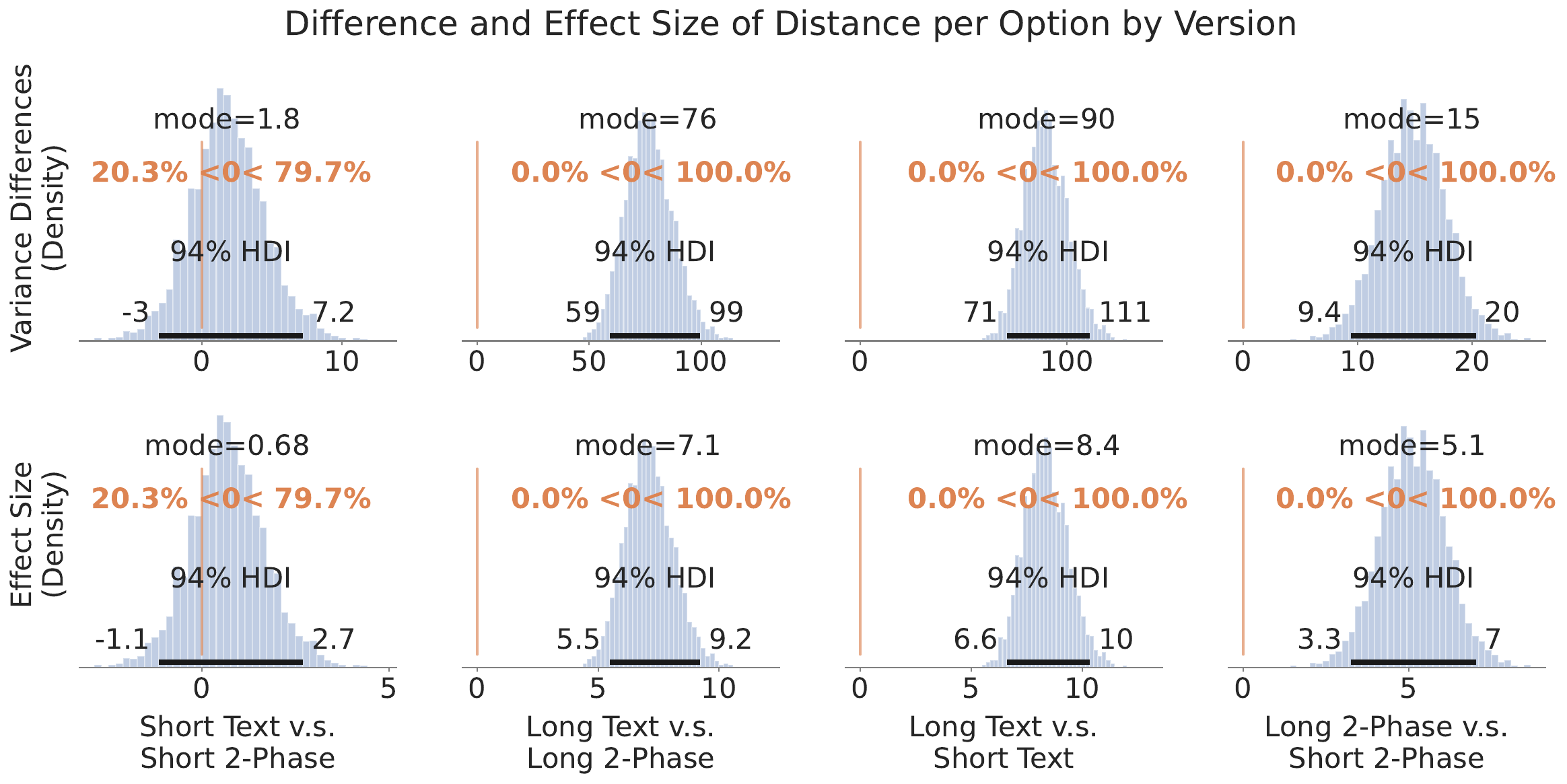}
    \caption{Differences in the variance of edit distance by version.~\textbf{Main takeaway: } This plot shows that with two-phase interface, there is a reduction in edit distance variance when the number of option grows.}
    \Description{A grid of eight density plots showing the differences and effect sizes of distances per option by version. The top row depicts variance differences, and the bottom row depicts effect sizes, both represented as density distributions. Each column corresponds to a comparison: Short Text vs. Short 2-Phase, Long Text vs. Long 2-Phase, Long Text vs. Short Text, and Long 2-Phase vs. Short 2-Phase. Each plot includes a mode value, a 94\% Highest Density Interval (HDI), and the proportion of the distribution below and above zero. Variance difference distributions on the top row show modes ranging from 1.8 to 90, with varying HDI ranges. Effect size distributions on the bottom row have modes between 0.68 and 8.4, with all 94\% HDIs above zero except for Short Text vs. Short 2-Phase.}

    \label{fig:bayesian_distance_variance}
\end{figure*}

\subsubsection{Posterior predictive plots}
Our Bayesian model converged successfully, as evidenced by an $\hat{R}$ value of 1 in the model summary. We plotted the posterior predictive distribution for the edit distance per option in Figure~\ref{fig:observed_vs_posterior_predictive_histogram_m2}. This figure compares the models posterior predictive distribution with the observed data.

\subsubsection{Model Results}
Figure~\ref{fig:bayesian_distance_variance} shows the pairwise comparison of the variance of edit distance in the first row followed by the effect size in the second row. In addition to the comparison within the same survey length, we provide all pairwise comparisons. A notable result that we omit from the main text is that if we compare the variance between the long and short text, and the variance between the long and short two-phase, we see that the text group had three times the standard deviation compared to the two-phase group. This indicates that the organization phase minimize the edit distance despite the increase in survey length.

\subsection{Model 3: Long QS Cumulative Edit Distance} \label{sec:apdx:model_cum_distance}

The dependent variable for this model is the cumulative edit distance $D_i$, a positive continuous variable measured at each step within a version for each user. We modeled this to test our hypothesis that for each participant, the growth rate of the edit distance is consistent. To accommodate its positive nature, we model $D_i$ using a Truncated Normal distribution:

\begin{equation}\label{eq:model3_likelihood}
D_i \sim \text{TruncatedNormal}(\mu_i, \sigma_{\text{obs},i}, \text{lower}=0),
\end{equation}
where the observation-specific standard deviation prior is:
\begin{equation}\label{eq:model3_prior_sigma}
\sigma_{\text{obs},i} \sim \text{HalfNormal}(0.3).
\end{equation}

\subsubsection{Independent Variables and Regression Model}
We incorporate the following independent variables: the step number when completing QS ($S_i$), the interface version ($V_i$), and user-specific effects ($U_i$). The interface version and user-specific effects are modeled using hyperpriors to capture variability across groups.

The linear predictor for $D_i$ is given by:
\begin{equation}\label{eq:model3_mu}
    \mu_i = \alpha_{\text{shared}} + \beta_v[V_i] \cdot S_i + U_i \cdot S_i,
\end{equation}
where $\alpha_{\text{shared}}$ represents the global intercept, $\beta_v[V_i]$ models the interface version effects, and $U_i$ captures individual user-specific effects. The intercept is assigned the following prior:
\begin{equation}\label{eq:model3_prior_shared}
    \alpha_{\text{shared}} \sim \mathcal{N}(2.0, 0.5).
\end{equation}

\paragraph{Interface Version ($V_i$).} Interface effects are modeled as:
\begin{equation}\label{eq:model3_prior_beta}
    \beta_v[V_i] \sim \mathcal{N}(\mu_{\beta}, \sigma_{\beta}),
\end{equation}
where the hyperparameters for the interface effect distribution are:
\begin{equation}
    \mu_{\beta} \sim \mathcal{N}(0.05, 0.05), \quad \sigma_{\beta} \sim \text{HalfNormal}(0.1).
\end{equation}

\paragraph{User Effects ($U_i$).} Instead of directly sampling $U_i$, we follow the reparameterization approach:
\begin{equation}\label{eq:model3_user_mu}
    U_i = \mu_U + \sigma_U \cdot z_{U,i},
\end{equation}
where we assign weakly informative priors $\mu_U \sim \mathcal{N}(0,1)$ and $\sigma_U \sim \text{HalfNormal}(0.1)$ to represent the shared mean and scale of the user effects. The term $z_{U,i} \sim \mathcal{N}(0,1)$ captures individual user variability, allowing us to model deviations across users while maintaining a structured prior.

\subsubsection{Posterior Predictive Plots}

\begin{figure}[h!]
    \centering
    \includegraphics[width=0.45\textwidth]{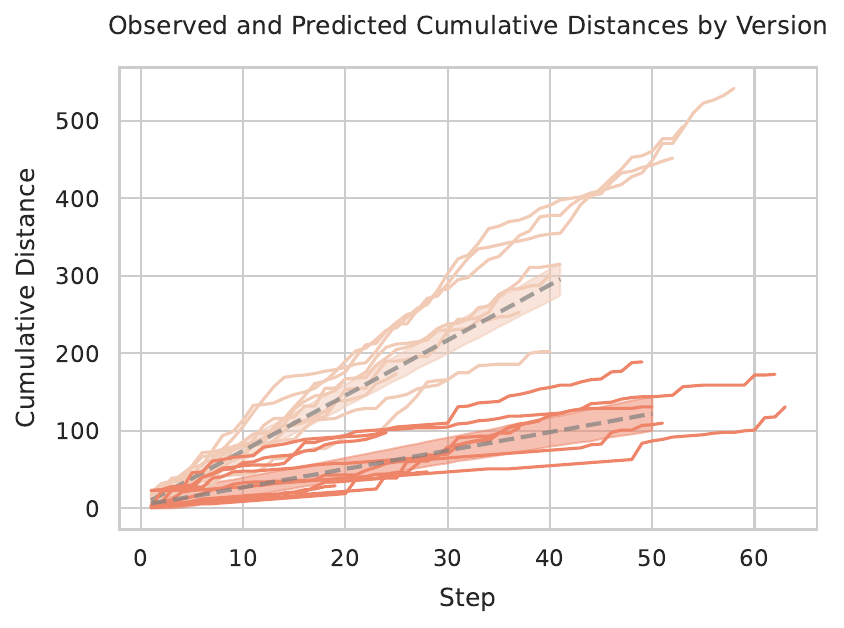}
    \caption{Posterior Predictions vs. observed data for cumulative edit distance. The plot showed each observed user's cumulative edit distance in different shades for the two groups of participants. Dotted line represent the posterior predictive mean. \textbf{Takeaway of the plot}: We believe that the model is reasonable at capturing slop of the cumulative trends.}
    \Description{ A line plot titled "Observed and Predicted Cumulative Distances by Version," showing cumulative distances across task steps for two conditions. Solid lines in varying shades of red and orange, representing actual cumulative distances for individual participants. Dashed lines, showing the model's predicted cumulative distances for each condition. The plot indicates alignment between observed and predicted trends, with some variability at higher cumulative distances. Predictions closely follow observed trajectories, demonstrating the model's accuracy in capturing cumulative distance patterns.}
    \label{fig:observed_and_predicted_cumulative_distances_by_version_m3}
\end{figure}

Our Bayesian model converged successfully, as indicated by an $\hat{R}$ value of 1 in the model summary. Figure~\ref{fig:observed_and_predicted_cumulative_distances_by_version_m3} presents the posterior predictive distribution for cumulative edit distance, demonstrating alignment between the predicted and observed data.

% TC:endignore

\end{document}